\documentclass[aps,prd,reprint,groupaddress,noshowpacs,noshowkeys]{revtex4-1}

\usepackage{amsmath,amssymb,mathrsfs}

\usepackage{graphicx}
\usepackage{dcolumn}
\usepackage{bm}
\usepackage{xcolor}
\usepackage[T1]{fontenc}
\usepackage[pdftex,
  hidelinks,
  colorlinks=true,
  linkcolor=blue,
  anchorcolor=black,
  citecolor=blue,
  urlcolor=blue,
  pdfauthor= {Michael Glinsky},
  pdfsubject = {A new theory of money},
  pdfdisplaydoctitle,
  bookmarks=true,
  bookmarksopen=true]{hyperref}

\begin{document}
\title{A new economic and financial theory of money}

\author{Michael E. Glinsky}
\affiliation{BNZ Energy Inc., Santa Fe, NM, USA}

\author{Sharon Sievert}
\affiliation{SCA Consulting, London, England, UK}

\begin{abstract}
This paper fundamentally reformulates economic and financial theory to include electronic currencies.  The inspiration for this work has been the advent of electronic currencies that span the gambit from blockchain based coins such as Bitcoin and Tether, to centralized currencies such as PayPal, Zelle, and Venmo.  The valuation of the electronic currencies will be based on macroeconomic theory and the Fundamental Equation of Monetary Policy, not the microeconomic theory of discounted cash flows.  Hence, the value of a potential financial investment will be its long-term contribution to the social aesthetic of sustainable economic activity, rather than the short-term exploitation of society via profits.  The view of electronic currency as a transactional equity associated with tangible assets of a sub-economy will be developed, in contrast to the view of stock as an equity associated mostly with intangible assets of a sub-economy.   The view will be developed of the electronic currency management firm as an entity responsible for coordinated monetary (electronic currency supply and value stabilization) and fiscal (investment and operational) policies of a substantial (for liquidity of the electronic currency) sub-economy.  The risk model used in the valuations and the decision-making will not be the ubiquitous, yet inappropriate, exponential risk model that leads to discount rates, but will be multi time scale models that capture the true risk.  These multi scale risk models will be used to make better investment decisions, operationally manage the businesses, and control the currency supply both over the short-term through arbitrage trading and over the longer term through investment.  The decision-making will be approached from the perspective of true systems control based on a system response function given by the multi scale risk model and system controllers that utilize the Deep Reinforcement Learning, Generative Pretrained Transformers, and other methods of Generative Artificial Intelligence (genAI).  This will be contrasted against current private financial investment, and uncoordinated governmental monetary and financial practices.  Finally, the sub-economy will be viewed as a nonlinear system with both stable equilibriums that are associated with short-term exploitation, and unstable equilibriums that need to be stabilized with active nonlinear control based on the multi scale system response functions and genAI.  The later is associated with long-term maximization of social aesthetic enabled by electronic currencies.  This will lead to the connection to religious philosophy.  There is a choice of the future (equilibrium or religious eskaton) of the sub-economy that is made by the electronic currency management firm by how it manages the sub-economy.  By choice of the electronic currency, an entity of the economy is choosing its future -- equilibrium, or eskaton; that is religion.  Therefore, there are two types of religions -- religions of maximum social aesthetic which require active nonlinear genAI control using electronic currencies, and the current religions of maximum exploitation which require resistive control using interest.
\end{abstract}

\maketitle

\section{Introduction}
\label{introduction.sec}
The recent ubiquitous adoption of electronic currency, whether that be centralized currencies like Zelle, PayPal and Venmo, or peer-to-peer currencies like Bitcoin and Tether, has created an opportunity to reform our economic, banking and financial systems.  It is an opportunity to reform them from ones based on short-term greedy, profit motivated exploitation;  to ones based on long-term virtuous, economic activity motivated benevolence (that is improvement of the social aesthetic \footnote{This paper coins the phrase ``social aesthetic''.  What is meant by the social aesthetic is the beauty or desirability of the society. This encompasses the society having a vibrant economy with full employment, to the society having a beautiful sustainable environment.} or good).  In order to realize this reformation, the theory of economics, banking and finance must be rehabilitated and extended to fundamentally include electronic currency.

\begin{figure}
\noindent\includegraphics[width=15pc]{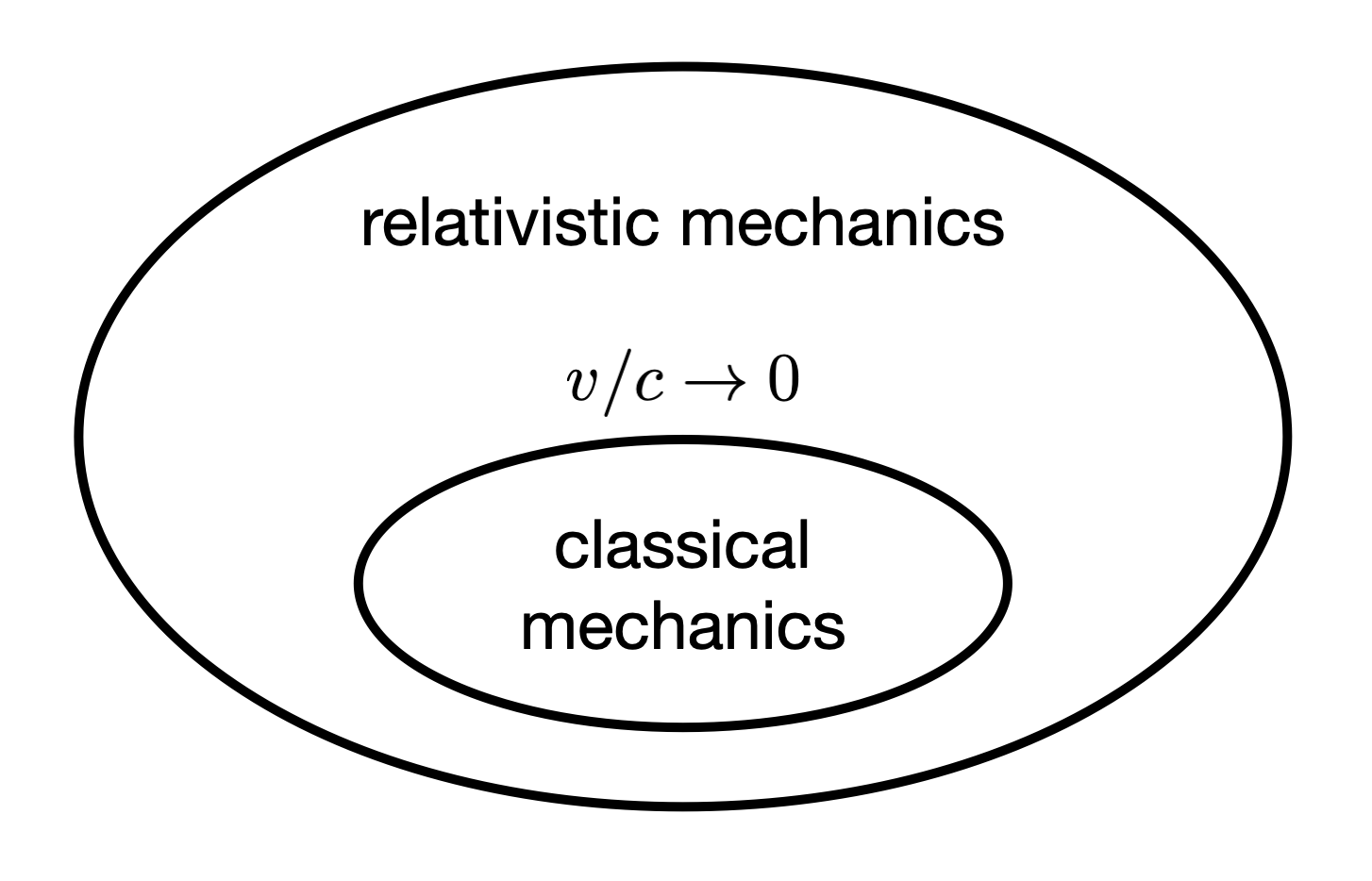}
\caption{\label{relativity.classical.fig} The relationship of the uber theory, relativistic mechanics to the subordinate theory, classical mechanics.  Relativistic mechanics reduces to classical mechanics in the limit $v/c \to 0$.}
\end{figure}

The approach that we will take to developing this economic theory, that fundamentally includes electronic currency, is borrowed from the physical sciences.  The example that we will build upon is the relationship of the ``uber theory'' of Einsteinian relativistic dynamics to the ``subordinate theory'' of Newtonian classical mechanics.  As is normally the case, Sir Issac Newton developed the subordinate theory of classical mechanics first in the late 17th century, inspired by an apple falling from a tree.  This was a very useful predictive theory that served the physics community well for over two hundred years.  In the later part of the 19th century, there emerged phenomenon and measurements that were not explained by classical mechanics.  In the early part of the 20th century, Albert Einstein developed relativistic dynamics, inspired by communication from a moving train.  It was shown that classical mechanics was obtained in the limit that the velocity was much less than the speed of light.  This condition is satisfied for a very large class of motions.  The problems with the theory came as the motion of photons and electrons about the nucleus were being examined in more detail.  The relationship of these two theories are shown in Fig. \ref{relativity.classical.fig}.

Let us now examine the current state of economic theory.  There is first the subordinate theory of microeconomics.  This is based on a local view of individuals and corporations, both private entities, that are motivated by profit.  This is called capitalism.  The Chicago School of economics lead by Milton Friedman also advocated that the private entities should be free to pursue this maximization of profit with minimal governmental regulation, and that government should leave as much as possible up to the private sector.  The profit is measured by the Net Present Value (NPV) of exponentially Discounted Cash Flows (DCF) \citep{allen11}.  This can be written as
\begin{equation}
\label{npv.eqn}
    \text{NPV} = \int{\text{e}^{-\nu t} \, (R-E) \, dt} \equiv \text{DCF} - I,
\end{equation}
where $\nu$ is the discount rate, $R$ is the reward or cash revenue (technically, inbound cash flows less investment or abbreviated as simply revenue), $E$ are the costs or cash expenses (technically, outbound cash flows or abbreviated as expenses), and $(R-E)$ is the cash profit (technically, free cash flow or abbreviated as profit), so that NPV is the Discounted Cash Flow (DCF) less investment ($I$).  Do note that our use of the words revenue, expenses, and profit are different than those of accrual based accounting.  Although the terms are technically different, they are morally equivalent.  NPV can now be identified as the short-term profit.  Equation~\eqref{npv.eqn} is also the definition of Discounted Cash Flow as $\text{DCF}\equiv\text{NPV}+I$, where $I=\int{I(t) \, dt}$.  DCF is therefore the upper limit or constraint on investment $I<\text{DCF}=I_\text{max}$.  The limit will be dependant on the cash flow profiles $R(t)$ and $E(t)$, and the form of $I(t)$.  The scaling of DCF, in an informative limit, is given in Sec.~\ref{conclusions.sec} as Eq.~\eqref{npv.constraint.eqn}.

Meanwhile there was developed the uber theory of macroeconomics.  This is based on a global view of the economy.  It is underpinned by the Fundamental Equation of Monetary Policy given by \citep{baumol20}
\begin{equation}
\label{macro.eqn}
    \text{GDP} = M \, V,
\end{equation}
where GDP (Gross Domestic Product) is the amount of economic activity, $M$ is the money supply, and $V$ is the velocity of money.  From this it can by shown that the value of the currency, $P_\text{ec}$, is given by
\begin{equation}
\label{revenue.eqn}
    P_\text{ec} \sim m_e S_0 R_0
\end{equation}
where $S_0=1/V_0$ is the savings or temporal multiplier of the primary entities of the economy, and $R_0$ is the revenue of the primary entities of the economy, and $m_e$ is the effective economic network multiplier.  The goal of the sovereign should be to maximize the socially aesthetic economic activity.

Microeconomic theory and macroeconomic theory were developed independent of each other, and a weak attempt was made to understand the relationship between them, unlike the strong attempt made in the physical systems example shown in Fig. \ref{relativity.classical.fig}.  Inspired by the television series Stargate and Mr. Robot, we will proceed by identifying macroeconomics as the uber theory.  We propose that, not only the sovereign, but all sub-economies and entities should be motivated to maximize the value of currencies that are relevant to them (those currencies in which they economically transact).  With the advent of electronic currencies, there now can be currencies for each sub-economy whether that be regional, industry-based or both.  The relationship of these currencies to each other defines a graph of the economy, and the monetary and fiscal policy of the individual currencies give economic control knobs matched to the structure (mathematically, the topology) of the economy.  Instead of a constrained maximization of an entity's DCF or NPV (that is, short-term profit) with the constraint of $\text{NPV}=\text{DCF}-I>0$, the entity should be motivated to maximize the long-term virtuous economic activity of the whole economy $m_e S_0 R_0$ with the constraint of not inflating the value of the currency (that is $m_e S_0 R_0-I>0$).  This economic theory will be discussed in more detail in Sec.~\ref{currency.theory.sec}, Sec.~\ref{solve.hjb.sec}, Sec.~\ref{faser.sec}, Sec.~\ref{resistive.HJB.sec}, and Sec.~\ref{traditional.DRL.sec} with an example of how it works given in Sec.~\ref{valuation.sec} (details of which appear in App.~\ref{econ.model.app}).  The practical new Ubuntu business model resulting from this economic theory, replacing the conventional debt based business model, is described in Sec.~\ref{ubuntu.sec}.

\begin{figure}
\noindent\includegraphics[width=15pc]{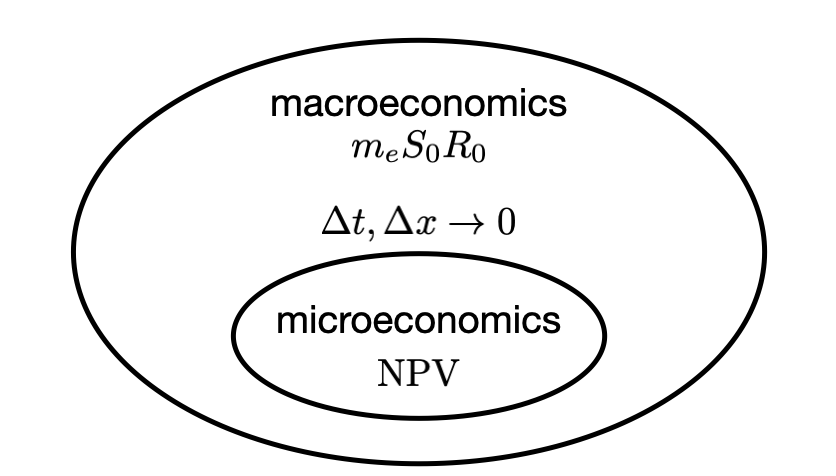}
\caption{\label{macro.micro.fig} The relationship of the uber theory, macroeconomics, which maximizes $m_e S_0 R_0$ to the subordinate theory, microeconomics, which maximizes NPV.  Macroeconomics reduces to microeconomics in the limit $\Delta t, \Delta x \to 0$, the random walk or diffusive limit with an exponential risk model.}
\end{figure}

It is very instructive to understand the relationship of the subordinate microeconomic theory to this uber macroeconomic theory.  It can be shown that if a local approximation in terms of both the business graph (direct transactions) and time (months to a few years) is made to the economic system, the financial dynamics can be reduced to that of a random walk or diffusion. This gives an exponential model of the financial dynamics (risk) that leads to a constrained maximization of DCF or NPV, replacing a constrained maximization of virtuous $m_e S_0 R_0$.  That is a maximization of
\begin{equation}
\label{local.approx.eqn}
    m_e S_0 R_0 \xrightarrow[\Delta t,\Delta x \to 0]{} \text{DCF or NPV}
\end{equation}
with the constraint
\begin{equation}
    m_e S_0 R_0 -I>0  \xrightarrow[\Delta t,\Delta x \to 0]{} \text{NPV}= \text{DCF}-I>0
\end{equation}
where $\Delta x$ is the random step size in the business graph, and $\Delta t$ is the time step.  This relationship is shown in Fig. \ref{macro.micro.fig}.  A more specific quantification of the approximation is given in Sec.~\ref{solve.hjb.sec}.

Unlike the approximation that leads to classical mechanics from relativistic mechanics, the local approximation of the financial dynamics is rarely satisfied, leading to catastrophic consequences.  This approximation is fundamentally introduced as an additional resistive term to control and stabilize the economic system, but dominates the true economic system response.  These consequences will be discussed in detail in Sec.~\ref{consequences.sec}.  This issue with the application of physical science and mathematics to the social sciences and economics was identified by Murray Gell-Mann, a Nobel prize winning physicist best known for his work on quantum chromodynamics and quarks.  During the latter portion of his career he became interested in complex systems, but quickly realized that there was a problem about how the theory was being developed.  \emph{``Science was so often badly applied.  Pretending that they could analyze and understand the most complex of processes, decision makers had embraced a kind of narrow rationality that took into account only things that are very easy to quantify.  Lost in the calculations were factors like beauty or diversity or the irreversibility of change.  The results have been disastrous.  With anything hard to quantify set equal to zero, a highway can be driven straight through a neighborhood or through a rare wilderness because there is no reliable quantitative measure of damage to set against the increased cost of running the road around the outside.''} \citep{gell.mann.00}  What is even worse in the case of financial dynamics is that what is quantified is not quantified correctly.

What has been set up is a classic game theory situation.  The local, short-term, greedy optimization of profit (that is NPV) is leading to a minimization of the global, long-term, virtuous economic activity (that is $m_e S_0 R_0$).  This is called the Nash Equilibrium.

We return now to the concept of economic conservatism embodied in the seminal work of Milton Friedman and summarized in the book ``Capitalism \& Freedom'' \citep{friedman.62}.  Like Friedman advocated, Government Inc.\ should be limited to matters of the common good like defense, education, social security, public infrastructure and resources, the arts, fundamental research and healthcare.  There should be freedom, not only of sub-economies, but of Government Inc., labor, intellectual property and law.  This will be discussed in more detail in Sec.~\ref{freedom.sec}.  Friedman was right that freedom is essential.  Freedom is how the leaders of the sub-economies are ultimately held accountable.  Free people will move from a sub-economy with a poor leader to a sub-economy with a good leader.  The problem comes with capitalism being based on the short-term maximization of profits.  The fact that it is left up to the private sector (the second ingredient of capitalism) is common to what we are proposing.  It would be best if Government Inc.\ is private with government executives and congressional representatives serving on the board.  The change that is being proposed is to move to ``Communism \& Freedom'', where we are talking about a private sector communism where the virtuous economic activity (that is the common good) is maximized.  It is very similar to existing economic cooperatives, but with a currency for the economic cooperative.

The rest of this paper is organized as follows.  The view of electronic currency as a transactional, stable, low risk equity (much like a checking account, associated with tangible assets), and of stock as a complementary, volatile, high risk equity (much like a savings account, associated with predominately the intangible assets) is discussed Sec.~\ref{tangibility.sec}.  The origin of the quantum field nature of financial systems, and the way the quantum field nature influences how the electronic currency management firm manages and observes the sub-economy, are discussed in Sec.~\ref{quantum.sec}.  How to manage the sub-economy is discussed in Sec.~\ref{management.sec}, how to tax the sub-economies for the public good is presented in Sec.~\ref{taxation.sec}, the view of the sub-economy as a nonlinear system is explored in Sec.~\ref{nonlinear.sec}, the resilience of this economic system to economic collapse is discussed in Sec.~\ref{resilience.sec}, what currencies currently exist and their characteristics are put forward in Sec.~\ref{existing.sec}, the interactions of sub-economies are explained in Sec.~\ref{interaction.sec}, the relationship of electronic currency to religious philosophy is explained in Sec.~\ref{religion.sec} answering the question ``why is crypto a religion?'', the inspiration for this work is given in Sec.~\ref{inspiration.sec}, and finally the conclusions are presented in Sec.~\ref{conclusions.sec}.

\section{Catastrophic consequences of the invalid economic localization}
\label{consequences.sec}
This section will start by examining the local diffusive approximation resulting in Eq.~\eqref{local.approx.eqn} and Eq.~\eqref{npv.eqn}, what it really means, how it becomes embedded in our economy, and why it is not a valid approximation for many cases, especially for infrastructure and other long-term capital assets.  The primary economic consequences of using this approximation will then be explored, followed by the secondary economic consequences, and finally the effects on the fabric of society.

The local diffusive approximation for financial dynamics is better known as the Fokker-Planck approximation in physical kinetics \citep{nicholson83,lifschitz83}.  It is also equivalent to a random walk or stochastic diffusive process.  In finance, it is the approximation that leads to the Black-Scholes equation \citep{black73,hull00}.  The underlying assumption is that the movement in the quantity, whether that be position, velocity or value, takes place by many small random steps.  To move an appreciable distance, many small steps need to be taken.  Specifically, to move a distance $L=N \Delta x$, where $\Delta x$ is the step size, $N^2$ steps need to be taken.  The evolution is governed by the Fokker-Planck equation
\begin{equation}
\label{fokker.planck.eqn}
    \frac{\partial f(J,t)}{\partial t} = \nu \frac{\partial}{\partial J} \Bigl( \sigma(J) \, f \Bigr) + \kappa \frac{\partial^2 f}{\partial J^2},
\end{equation}
where $f(J,t)$ is the probability, $\nu$ is the collision rate or dissipation rate or coefficient of dynamic friction or discount rate, $\sigma_c^2=\Delta x^2$ is the mean squared size of the step or the collisional cross sectional area, $\omega(J)=\omega_d$ is the frequency of the system as a function of $J$ or the drift frequency, $\sigma(J)= (\sigma_c/\nu) \, \omega(J)=\sigma_d$ is the square root of the drift cross section, $J$ is the quantity being diffused (e.g., action, economic activity, price, velocity, and position), and $\kappa=\left<\Delta J \Delta J/\Delta t \right>$ is the diffusion coefficient.  This equation evolves to the equilibrium distribution
\begin{equation}
\label{fp.equilibrium.eqn}
    f_{\text{eq}}(J) = \text{e}^{-H(J)/T},
\end{equation}
where $H(J)$ is the Hamiltonian or energy as a function of $J$ with units of $J/t$ (if $J$ is physical action, then $H(J)$ has units of energy), $\omega(J) = dH / dJ$, and $T$ is a constant temperature with units of $J/t$.  It will be shown in Sec.~\ref{solve.hjb.sec} that $\kappa = \nu \, \sigma_c^2 \sim T^{3/2}$, $\nu \sim \sqrt{T}$, and $\sigma_c^2 \sim T$.  So what is identified by the temperature $T$, or the random kinetic energy, of a heat bath or an external physical system, is proportional to the 2/3 root of the diffusion coefficient of a random walk or diffusive process.  For financial dynamics the diffusion coefficient, $\kappa=\left<\Delta P \Delta P/\Delta t \right>$, is called the volatility (which is the metric of risk), where $P$ is the price of the commodity.  Equation~\eqref{fokker.planck.eqn}, when $P$ is substituted for $J$, is the Black-Scholes equation of quantitative finance.

This assumption is made because it dramatically simplifies the general problem from the difficult to solve Hamilton-Jacobi-Bellman equation (Eq.~\eqref{hjb.eqn}), to an easily solvable partial differential equation (Eq.~\eqref{fokker.planck.eqn}) with a closed form, time independent (stationary), equilibrium solution given by Eq.~\eqref{fp.equilibrium.eqn}.  It is simple as foretold by Murray Gell-Mann.

While there are some physical systems that satisfy this assumption, that is not true of most economic systems.  This is especially true when it comes to major infrastructure projects, long-term capital assets like real estate, petroleum, mining, and energy.  These assets do not have legs and are not running around like drunken sailors.  The invalidity of this assumption was evidenced by the failure of Long-Term Capital Management (LTCM), the first trading firm to employ the Black-Scholes equation.

Unfortunately this assumption can not be escaped in our current financial systems.  It is fundamentally embedded in our economic, financial and banking systems.  It is baked-in by the US Federal Reserve Bank when it loans currency at a prime interest rate, by the US Treasury when it sells T-bills as bonds with coupon payments, by banks when they issue loans with interest, by stockholders when they demand dividends or simply value the stock using DCF analysis, and by corporations when they issue bonds.

The consequences of this invalid assumption are truly catastrophic.  The primary economic fallout is a significant under investment in the future economic growth and sustainability, businesses and economies not being well operated such that they dramatically under perform by factors of ten or more, and realized performance of businesses and economies that never meets projected expectations.  Secondary fallout includes the ``Research Valley of Death'' and the ``Innovator's Dilemma'' \citep{christensen.00}, as the long-term value of research is not captured, the natural development of exploitative and inefficient monopolies, and the failure of debt-leveraged private equity.  The failure of debt-leveraged private equity is the failure of the debt-based economy in miniature.  Private equity is not investing enough in their acquisitions, and are running their acquisitions inefficiently without enough inventory and savings, and consequently they find that their acquisitions do not meet their expectations of performance.

The effects on our social fabric are even more devastating.  The driving force of greed, that is short-term exploitation of others, pits identities against one another in a race to the bottom.  It leads to the rise of dictators, monarchs, fascism, authoritarianism, and monopolists; and the compensating fall of the general population and their savings (wealth).  The result is a social stratification that breeds animosity and a strong desire for retribution and recompense. The ultimate result is social revolution, discrimination, white supremacy, antisemitism, misogyny, and wokeness (political correctness and anti-discrimination)  -- a social war between the haves and the have-nots, amongst the haves, and amongst the have-nots.  People view everything as a zero sum game, so that if someone else is hurt that must be a personal gain even though that might be actually be a personal harm.  In general, it increases the mental stress on individuals.

These are the conditions that Frederick Engels described in 1845.  \emph{``The most important social issue in England is the condition of the working classes, who form the vast majority of the English people.  What is to become of these propertyless millions who own nothing and consume today what they earned yesterday?  The industrialists grow rich on the misery of the mass of wage earner, but prefer to ignore the distress of the workers.''} \citep{engels45,engels45a}  A more recent example, the transition of the United States from the large government economy started in the 1930s with the New Deal and reaching its zenith in the 1960s (based on a currency, with government spending based on the social good), to today's nearly pure, unregulated ``Capitalism \& Freedom'' starting in the 1970s and 1980s, is well chronicled by \citet{andersen21}.  This book describes increased economic inequality as evidenced by the Gini number.  The United States has moved from roughly equivalent to Canada in the more highly developed half of nations, to the most unequal rich country and to a place amongst the developing countries, better than the Congo and Uruguay but worse than Haiti and Morocco.  The average US worker now has almost no savings, living ``hand to mouth''.  As to quality of life, despite paying between two to three times as much for healthcare per capita than any other highly developed country, the average life expectancy is three to five years less.  Before 1980, the cost and life expectancy were both equivalent to other highly developed countries.   This is the exploitation of labor, a consequence of the invalid local risk assumption.

\section{Theory and valuation of electronic currencies}
\label{currency.theory.sec}
In this section, the economic theory forming the basis for monetary and fiscal policy of managing a sub-economy based on an electronic currency is developed.  Equivalently, this can be viewed as extending current macroeconomic theory, and the theory of banking and finance to include electronic currency.  Simply stated, this brings the global perspective of maximization of virtuous economic activity to the locality.  It will be built from scratch, starting with mathematical fundamentals.  The reduction of this theory to common microeconomic constructs based on Discounted Cash Flows (DCF) and Net Present Value (NPV) \citep{allen11} will be presented.  Finally, the practical implications of using this new theory will be contrasted to the use of current microeconomic theories of banking and finance \citep{ritter89}.

We start by assuming that a sub-economy saves its revenue at a constant rate, $V=1/S$, so that
\begin{equation}
\label{m.eqn}
    M = R \, \Delta t \sum_{n=0}^\infty{\left( 1-V \Delta t \right)^n} = \int_0^\infty{R \, \text{e}^{-Vt} \, dt},
\end{equation}
where $M$ is the amount of savings or the supply of the electronic currency, $R$ is the primary revenue of the sub-economy, $V$ is the currency velocity or the savings turnover rate, $S$ is the savings turnover time or savings multiplier, and
\begin{equation}
\label{r0.eqn}
    R_0 \equiv \lim_{T_0 \to \infty}{\frac{1}{T_0} \int_0^{T_0}{R(t) \, f[R(t)] \, d[R(t)] \, dt}}
\end{equation}
is the ensemble and long time average revenue of the primary sub-economy expressed as a functional integral over the distribution functional $f[R(t)]$ and time.  Equation~\eqref{m.eqn} can be summed or integrated to give
\begin{equation}
    M = \frac{R}{V} = S \, R
\end{equation}
which can be rewritten as $R=MV$, the fundamental equation of monetary policy.

We now recognize that the primary sub-economy will spend a large fraction of this revenue, and it will be become the revenue of the second level of the economy.  The second level of the sub-economy will spend a large fraction of its revenue, and so on.  The total amount of savings will be
\begin{equation}
    M = \sum_{i=0}^\infty{\frac{R_i}{V_i}},
\end{equation}
where $i$ is the level of the sub-economy.  Assume that the fraction of the revenue that is spent is a constant
\begin{equation}
    R_i = R_0 \, \left( 1-1/m \right)^i,
\end{equation}
where $m$ is the economic multiplier.  The total revenue can now be written as
\begin{equation}
    R = \sum_{i=0}^\infty{R_i} = m \, R_0.
\end{equation}
The total amount of savings can be rewritten as
\begin{equation}
    M \, V_0 = m \, R_0 \frac{\sum_{i=0}^\infty{R_i\frac{V_0}{V_i}}}{R} = m \, R_0 \, \left< \bar{S} \right>_R,
\end{equation}
where we define the effective dimensionless savings multiplier as
\begin{equation}
    \bar{S}_e \equiv \left< \bar{S} \right>_R \equiv \left< \frac{V_0}{V_i} \right>_R = \frac{\sum_{i=0}^\infty{R_i\frac{V_0}{V_i}}}{R}.
\end{equation}
This can be manipulated to give
\begin{equation}
\label{ec.value.eqn}
    \boxed{M = m \, S_0 \, \bar{S}_e \, R_0 = m S_e R_0 = m_e S_0 R_0,}
\end{equation}
where
\begin{equation}
    S_e \equiv \left< S \right> \equiv S_0 \left< \bar{S} \right>_R
\end{equation}
is the effective savings multiplier, and
\begin{equation}
    m_e \equiv \left< m \right> \equiv m \left< \bar{S} \right>_R
\end{equation}
is the effective economic multiplier.  Equation~\eqref{ec.value.eqn} is what we call the fundamental equation of electronic currency valuation.  The relationship to the value of the electronic currency can be seen by writing
\begin{equation}
    M = P_\text{ec} N_\text{ec},
\end{equation}
where $P_\text{ec}$ is the value of the electronic currency, and $N_\text{ec}$ are the number of units of the electronic currency in circulation.  So,
\begin{equation}
\label{ec.p.eqn}
    P_\text{ec} = \frac{1}{N_\text{ec}} m_e S_0 R_0 \sim m_e S_0 R_0,
\end{equation}
which is Eq.~\eqref{revenue.eqn} in Sec.~\ref{introduction.sec} with more precise definitions.

Let us now propose that the electronic currency management firm should maximize the value of the electronic currency, $P_\text{ec}$, as given by Eq.~\eqref{ec.p.eqn}.  This is in the interest of its stakeholders -- the holders of the electronic currency.  Equation~\eqref{ec.value.eqn} is made up of four parts.  

The first is the economic multiplier given by $m$.  This is a measure of how advanced and specialized the economy is (that is whether it is a first or third world economy, or how economically independent or how efficient individuals are).  Maximization of this term drives one to a first world, more efficient and dependant economy.  The electronic currency management firm, while it can influence this over the longer term, does not have a lot of direct influence on this parameter.  

It does have a very direct influence on the second term, the savings multiplier $S_0$, through the board level control of its subsidiaries.  The electronic currency management firm has a significant, yet indirect influence, on the third term, the dimensionless savings multiplier $\bar{S}_e$, as will be discussed in Sec.~\ref{management.sec}.  Given that $\forall i$, $V_i \gtrsim V_0$, this third term,
\begin{equation}
    \bar{S}_e = \left< \bar{S} \right>_R \lesssim 1,
\end{equation}
with the electronic currency management firm striving to make it as close as possible to 1.  The initial value is regionally and culturally dependant.  Currently it is much less than one since most businesses and people live ``hand to mouth'' with little savings.  

The increase in both directly and indirectly controlled savings has several benefits.  It enables a significant increase in the revenue as will be shown in App.~\ref{econ.model.app}, and eliminates supply chain issues like happened recently with the recovery from the COVID shutdown, by encouraging much higher levels of inventory (that is a form of savings).  It also allows leadership to focus on operational excellence not financing, employees to focus on performance not paying debt.  It eliminates cash flow problems, and reduces the stress on individuals.  The savings give financial inertia to the system, allowing the electronic currency management firm to more effectively stabilize the system, that is reduce the fluctuations, volatility and risk of the system.  Another way of looking at the benefits of savings is by recognizing that the financial system has individual fluctuations that are large.  The sub-economy needs enough savings to respond to these large fluctuations.  

The fourth term, the revenue, is a direct metric of economic activity.

If the electronic currency management firm needs to print new currency for investment to generate more virtuous economic activity, it should make that investment when 
\begin{equation}
\label{investment.eqn}
    m_e S_0 \Delta R_0 \ge \Delta I
\end{equation}
or
\begin{equation}
    \Delta M = m_e S_0 \Delta R_0 - \Delta I \ge 0,
\end{equation}
where $\Delta R_0$ is the increase in revenue coming from the investment financed by issuing $\Delta I$ in new electronic currency, and $\Delta M$ is the increase in monetary demand net of the investment, $\Delta M \equiv m_e S_0 \Delta R_0 - \Delta I$.  Note that the costs do not appear explicitly in this investment criteria (a small fraction of the costs appear in the required investment), unlike the $\text{NPV}>0$ constraint where all the costs appear explicitly as $E$ in Eq.~\eqref{npv.eqn}.  The revenue will be adjusted so that it includes all non-monetary value and/or mitigated costs to society.  This will be implemented as a tax on undesirable activities or a subsidy of desirable activities.  If the non-monetary value and/or mitigated costs are felt by members outside of the sub-economy, there should be a taxation of the members outside of the sub-economy for undertaking undesirable activities and investment of the proceeds from that taxation into the sub-economy, that is a subsidy of the desirable activities, by the government.  Note that if the tax is sufficiently large, the government or the currency management firm will never have to collect the tax.  The undesirable activities will not be undertaken.

We now turn your attention to the operational management of members of the sub-economy.  Pose this problem as a constrained optimization \citep{chiang84}.  Maximize the electronic currency demand, $m_e S_0 R(Q)$, subject to the constraint $\Delta M \ge \mu \ge 0$, where $\Delta I = (E(Q)-R(Q)) T_I$, $T_I$ is the time period of the investment, $R(Q)$ are the yearly revenues of the member of the sub-economy as a differentiable concave function of a parameter $Q$, and $E(Q)$ are the yearly expenses of the member of the sub-economy as a differentiable convex function of a parameter $Q$.  The Lagrangian is
\begin{equation}
    L(Q,\lambda) = m_e S_0 R(Q) + \lambda [-\mu - E(Q)+(m_e S_0 + T_I) R(Q)].
\end{equation}
The Kuhn-Tucker conditions are
\begin{equation}
    \frac{\partial L}{\partial Q}= m_e S_0 R'(q) - \lambda T_I E'(Q) + \lambda (m_e S_0 + T_I)R'(Q) \le 0
\end{equation}
and
\begin{equation}
    \frac{\partial L}{\partial \lambda}= - \mu - T_I E(Q) + (m_e S_0 + T_I)R(Q) \ge 0.
\end{equation}
These conditions give the following following equations for the optimum point $(Q^*,\lambda^*)$
\begin{equation}
    R'(Q^*) = \frac{\lambda^*}{m_e S_0 / T_I + \lambda^* (m_e S_0 / T_I + 1)} E'(Q^*)
\end{equation}
and
\begin{equation}
\label{mu.eqn}
    \mu = (m_e S_0 + T_I) R(Q^*) - T_I E(Q^*) \equiv \Delta M(Q^*),
\end{equation}
which can be solved for
\begin{equation}
    Q^* = \Delta M^{-1}(\mu)
\end{equation}
and
\begin{equation}
    \lambda^* = \frac{- (m_e S_0 / T_I) R'(Q^*)}{(m_e S_0 / T_I + 1) R'(Q^*) - E'(Q^*)}
\end{equation}
or
\begin{equation}
\label{lambda.eqn}
    \lambda^* = \frac{-m_e S_0 R'(Q^*)}{\Delta M'(Q^*)}.
\end{equation}
Now, identify the two boundaries given by the conditions $\Delta M'(Q_{\text{max}})=0$ and $R'(Q_{\text{min}})=0$.  Using Eq.~\eqref{lambda.eqn}, find that $\lambda_{\text{max}}=\infty$ and $\lambda_{\text{min}}=0$, and using Eq.~\eqref{mu.eqn}, find that $\mu_{\text{max}}=\Delta M(Q_{\text{max}})$ and $\mu_{\text{min}}=\Delta M(Q_{\text{min}})$.  Finally, summarize the solution for the different values of $\mu$.  It is not possible for $\mu$ to be greater than $\mu_{\text{max}}$.  If $\mu = \mu_{\text{max}}$
\begin{equation}
\label{r.max.eqn}
    \boxed{R' = \frac{1}{m_e S_0 / T_I +1} E'.}
\end{equation}
For $\mu_{\text{min}} \le \mu \le \mu_{\text{max}}$,
\begin{equation}
    R' = \frac{\lambda}{m_e S_0 / T_I + \lambda (m_e S_0 / T_I + 1)} E',
\end{equation}
where
\begin{equation}
    \lambda = \frac{-m_e S_0 R'(\Delta M^{-1}(\mu))}{\Delta M'(\Delta M^{-1}(\mu))}
\end{equation}
and for $0 \le \mu \le \mu_{\text{min}}$,
\begin{equation}
    R' = 0.
\end{equation}
The electronic currency management firm will want to maximize $\mu$ in order to maximize the value of the currency, so that $\mu = \mu_{\text{max}}$.  Therefore, the solution is given by Eq.~\eqref{r.max.eqn}.  For a well managed currency, $m_e S_0 / T_I \gg 1$.  For instance, the example given in App.~\ref{econ.model.app} has $T_I= 1 \, \text{yr}$, $S_0 = 2 \, \text{yr}$ and $m_e = 3.5$, so that $m_e S_0 / T_I = 7$ and $R'= (1/8) E'$.  This is close to being a revenue maximizing firm with $R'=0$ and $\mu = \mu_{\text{min}}$, which is not much less than $\mu_{\text{max}}$ since
\begin{equation}
    \frac{\mu_{\text{max}}-\mu_{\text{min}}}{\mu_{\text{max}}} = \frac{1}{m_e S_0 / T_I +1} \approx \frac{1}{8}
\end{equation}
for the example given in App.~\ref{econ.model.app}.  In contrast, if the local approximation is made, the value of the currency is still maximized by using Eq.~\eqref{r.max.eqn}, but the non-local business network and time effects are neglected. The effective economic multiplier $m_e$ will approach 1, and the savings multiplier $S_0$ will approach 0.  For Target Energy Inc., the typical firm operated with the local assumption discussed in App.~\ref{econ.model.app}, $m_e=2$ and $S_0=1/20 \, \text{yr}$.  This gives $m_e S_0 / T_I = 1/10$, so that $R'= (10/11) E' \approx E'$ -- a profit, not revenue, maximizing firm.

The functional integration over the distribution given by the functional $f[R(t)]$ in Eq.~\eqref{r0.eqn} is how the risk model enters this theory.  This is what could be approximated by the Generative Pretrained Transformer (GPT) \citep{radford.18,vaswani17,farimani23}, a Generative Adversarial Network (GAN) \citep{goodfellow16}, the Mallat Scattering Transformation (MST) \citep{glinsky23,glinsky.24c}, or the Heisenberg Scattering Transformation (HST) \citep{glinsky23c,glinsky.24e,glinsky23d,glinsky.24b,glinsky.24d}, all forms of Artificial Intelligence (AI) \citep{hastie09,sugiyama15,goodfellow16}.  In the case of the local approximation, it is approximated by solving the diffusive Black-Scholes equation given in Eq.~\eqref{fokker.planck.eqn}.  There are also more simple approximations that can be made that respect the low risk and long-term nature of many financial situations.

Starting with the equation for $R_0$, Eq.~\eqref{r0.eqn}, and substituting the solutions to the Black-Scholes equation, Eq.~\eqref{fokker.planck.eqn}, into it, one can derive the expression for NPV given in Eq.~\eqref{npv.eqn}.  It will be shown in Sec.~\ref{solve.hjb.sec} and Sec.~\ref{resistive.HJB.sec} how a constrained optimization of the expression for $R_0$ given in Eq.~\eqref{r0.eqn} leads to the optimization of NPV given in Eq.~\eqref{npv.eqn}, when a dominate diffusive risk model is superimposed on the conservative financial dynamics of a sub-economy.

\begin{figure}
\noindent\includegraphics[width=15pc]{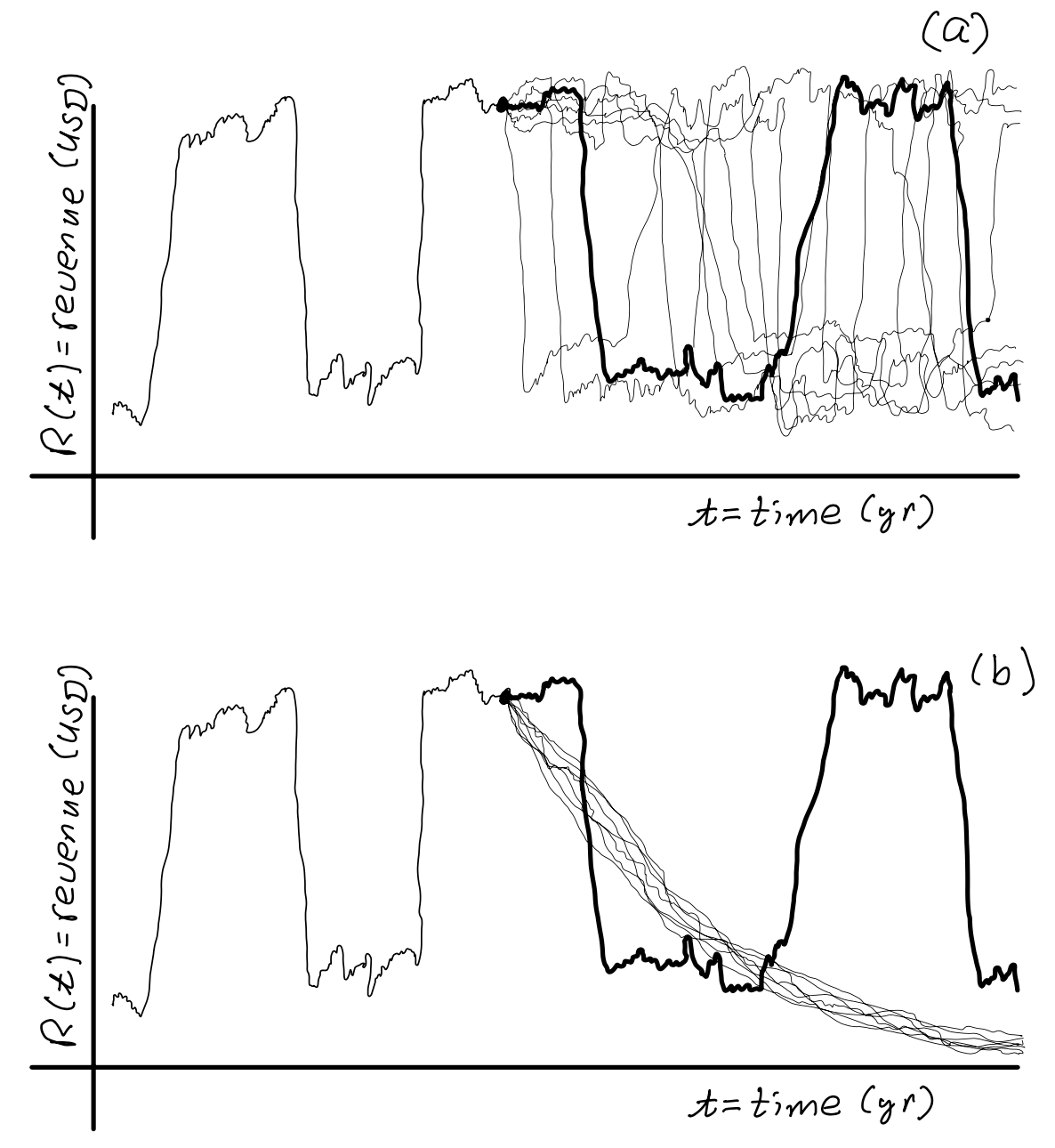}
\caption{\label{simulation.fig} Examples of conditional generation of price profile $R(t)$.  (a) using the distribution functional $f[R(t)]$ of Eq.~\eqref{r0.eqn} via the HST, (b) using the distribution function $f(R,t)$ of Eq.~\eqref{fokker.planck.eqn}.  The thick black line is the actual signal.  The thin black lines are the conditional realizations.}
\end{figure}

The dramatic difference between using the true distribution functional $f[R(t)]$ of Eq.~\eqref{r0.eqn} and the distribution function $f(J,t)$ of Eq.~\eqref{fokker.planck.eqn} is demonstrated in Fig. \ref{simulation.fig}.  In this figure the HST and the methods of AI are used to conditionally generate the distribution based on the historical behavior of the oil price (as will be explained in detail in Sec.~\ref{solve.hjb.sec}).  Note the discounting of the value of the next business cycle by the local approximation, so that no inventory would be amassed to take advantage of the next market upswing.  Also note how poor the simulations using the local approximations are -- that is how bad the approximation is.

In contrast to all of the advantages outlined in this section to optimizing $m_e S_0 R_0$, optimizing the NPV explicitly focuses on the short-term, due to the exponential term, and on profits, greed or exploitation due to the $(R-E)$ term.  When ancillary profit is made by the sub-economy optimizing $m_e S_0 R_0$, profit is not the objective.  The customer is paying forward, not being exploited.  The profit is re-invested to improve the product, so that the next time that the customer buys the product, it is better.

The preceding simplified example of constrained optimization demonstrates a stark difference between the status quo which uses the local approximation with the resulting exponential risk, and the proposed economic theory of this paper which optimizes sustainable economic activity.  Despite this, the practical situation demands a more sophisticated approach which includes the complete conservative stochastic system response.  The problem needs to be approached as a formal exercise in control of a complex system to optimize a functional given a model of the complex system.  Additionally, the system needs to be controlled to stabilize the optimal equilibrium and to minimize fluctuations about the equilibrium.  The method of choice is a solution of the Hamilton-Jacobi-Bellman equation \citep{goldstein80,kalman63}
\begin{equation}
\label{hjb.eqn}
    \frac{\partial V(q,t)}{\partial t} + \frac{\partial V}{\partial q} f(\partial V / \partial q,q) + R(q) = 0,
\end{equation}
where $R(q)$ is the reward given a state $q$ of the system, $f(p,q)=\dot{q}$ is the force equation or the constraint equation of the system given the state $q$ of the system and the co-state (canonically conjugate momentum or action) $p$ of the system, and 
\begin{equation}
    V(q_0,t) = V[q(t)] = \int_0^t{R(q(t)) \, dt}
\end{equation}
is the expected value of the system starting from state $q_0=q(t=0)$ at time $t$ (that is, the time integrated reward over the stochastic trajectory $q(t)$, or functional of $q(t)$). For the economic theory of this paper, the reward given the state of the sub-economy $q$ is $R(q)=m_e S_0 R_0(q)$, where $R_0$ is the primary revenue of the sub-economy under control per time, $m_e$ is the effective economic multiplier, and $S_0$ is the primary savings multiplier of the sub-economy under control.  This is the increase in the monetary demand $\Delta M$.

In Sec.~\ref{solve.hjb.sec}, it will be shown how to solve Eq.~\eqref{hjb.eqn} using the HST and the methods of AI based on either simulations of the system or observations of the system, driven by an external force $F_{\text{ext}}$.  The solution will be the fiscal (investment and operational) and monetary policy $\pi^*(q)$ that will locally maximize $V(q,t)$, but needs to be stabilized and cooled since local maximums are unstable equilibriums.  The reward that is optimized $R(q)$ can be modified to take into account other benefits or costs to society, such as sustainability and beauty, by applying a conservative economic control force $\Delta p / \Delta t = F_c(q)$.  Furthermore, the application of the economic force (arbitrage trading) to stabilize the system equilibrium $\pi^*(q)$ and minimize the fluctuations about the system equilibrium can be done in a similarly direct feedback manner by applying a force $F_\text{sf}(P)$ given in Eq.~\eqref{stable.feedback.eqn}.  It also can be done by applying a ponderomotive and diffusive force $F_\text{sp}(t)$, given in Eq.~\eqref{stable.pondermotive.eqn}, that is not dependent on knowing the state of the system.  More details of these control forces can be found in Sec.~\ref{solve.hjb.sec}.

If instead of the true model of the system, the diffusive model of Eq.~\eqref{fokker.planck.eqn} is used and applied to solve Eq.~\eqref{hjb.eqn}, it can be shown that the equilibrium optimum policy $\pi^*(q)$ is to optimize the NPV of Eq.~\eqref{npv.eqn} and to invest with terms of debt repayment.  This is equivalent to the profit maximization of the previous constrained optimization, instead of a revenue maximization.  It should be noted that the local NPV maximization leads to a significant reduction in the long-term economic activity and even long-term profits.  This greedy, local optimization when the system does not have exponential risk (that is diffusive dynamics), leads to very poor strategy (that is, policy).  It would be like playing chess with no regard for the long-term effects of the next move.  One would never sacrifice a piece to improve the long-term prospects of winning.  One would only consider what pieces could be captured or lost with the next move or two.  It also would be like playing PacMan with a strategy to swallow the next cherry as quickly as possible with no concern to whether that path leads to being consumed by a ghost in the near future.

\section{Tangibility of assets vis-a-vis equity}
\label{tangibility.sec}
We now examine where and how electronic currency appears on the balance sheet and what that means for it as a derivative security.  First we need to discuss the concept of tangible and intangible assets.

Tangible assets, as an accounting construct, are associated with the cash that has been spent to acquire or construct them.  For a building that would be the purchase price or cost of construction.  For intellectual property that would be the purchase price, licensing fee, or the cost of the research and development.  Since accounting is focused on the cash flows and valuation derived from discounted cash flows, the story ends here.

For the theory that we are presenting, the story must be extended to include intangible assets.  Before more recent accounting ```reforms'' this appeared on the accounting books as a goodwill asset associated with concepts such as brand value and value-in-place.  Value-in-place comes from the fact that the building is built, employees are on-the-job, properly educated and trained, and assimilated into the culture of the business, that is they know how to get things done.  For intellectual property it is the know-how and the show-how.  This value is subjective and is only valued when the business is sold, raises equity investment through a stock offering, or has a stock that is traded on a public exchange.  The intangible value is the difference between the book value (that is the cost of construction) and the market capitalisation or sales price.

\begin{figure}
\noindent\includegraphics[width=15pc]{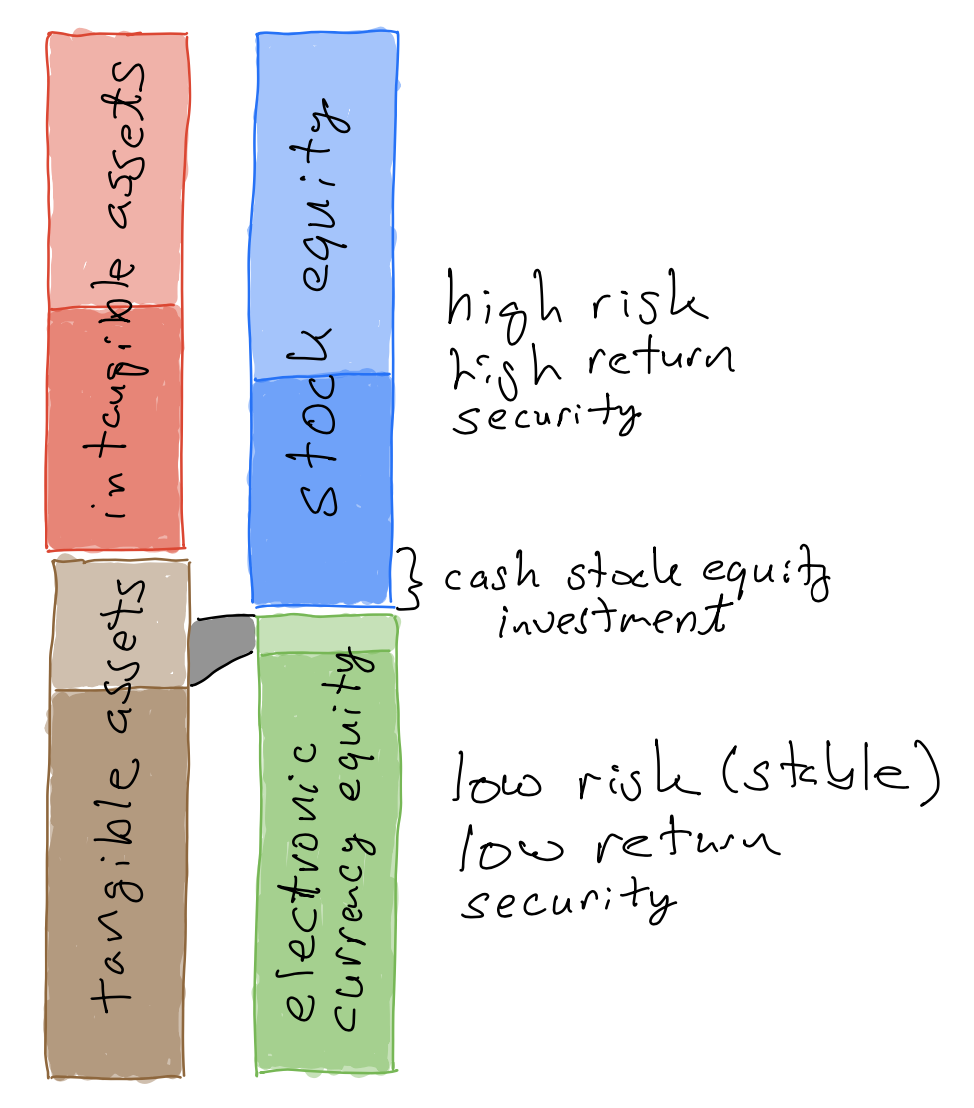}
\caption{\label{assets.equity.fig} The equity risk tranching into securities.  Shown on the left are the assets and on the right the associated fully backed securities.  The high risk intangible assets are shown by the red bar (the size of the fluctuations are shown by the light red area), and the low risk tangible assets are shown by the brown bar (the size of the fluctuations are shown by the light brown area).  The high risk stock equity is shown by the blue bar (with fluctuations shown by the light blue area).  The low risk electronic currency equity is shown by the green bar (with fluctuations shown by the light green area).  Note that the electronic currency equity is associated with the tangible assets and that the fluctuations are less than the fluctuations in the tangible assets due to the stabilization through arbitrage trading (shown by grey bar).  Also note the much larger fluctuations in the stock equity.  Since the number of shares of stock are roughly constant and the number of currency units scale with the size of the tangible assets, the value of the stock will increase proportionally with the growth of the intangible assets, while the value of the electronic currency will show little change.  The value of the currency will increase if the demand for the electronic currency is increased by more than the investment required to increase economic activity, but that increase will be much less than the increase in the price of the stock.}
\end{figure}

Electronic currency appears on the liability side of the balance sheet (see the example given in App.~\ref{econ.model.app}).  But, it is not a short-term liability like an accounts payable or a loan -- it is an equity, like owner's equity and retained earnings.  The terms of the loan are like that of cash raised through a stock offering.  It is repaid when, if, and how much it can be.  It also can be viewed, like stock, as an ownership of the business by the holders of the currency.  The difference is in the rights of the equity, and the resulting assets on which the value of the equity is based.  A close examination of the accounting example in App.~\ref{econ.model.app} finds that the electronic currency equity is associated only with tangible assets, while the stock equity is associated with predominately intangible assets.  (It only is associated with an amount of tangible assets equal to the amount of cash that was raised through stock offerings.)  Since the intangible assets are more speculative, ephemeral and simply fickle, they will exhibit larger fluctuations than the tangible assets and are therefore more risky.  The intangible assets will track the growth of the business, while the tangible assets will have a more modest growth rate determined by the efficiency of the investment in creating demand for the currency.  Therefore the currency equity will have less risk (made even lower by the active value control), and less return than the stock equity.  This is shown graphically in Fig. \ref{assets.equity.fig}.

The business can take advantage of a significant demand and resulting price increase in the stock price.  This is a way of the market telling the business that it needs to grow more quickly.  The business should increase its liquid reserves by raising cash through a stock offering, then issuing electronic currency and increasing the amount of investment into the sub-economy.

In summary, electronic currency is like a checking account where transactional savings, that is working capital should be kept.  Stock is like a savings account for long-term savings that will have larger returns, but also larger risk (that is fluctuations).

\section{Management of sub-economy}
\label{management.sec}
The electronic currency management firm needs to approach its role as a true fiduciary of the sub-economy.  The adoption and use of the electronic currency by the sub-economy for transaction and saving depends on: (1) transparency, (2) reserves, (3) accountability, and (4) stability.

First, with respect to \textbf{transparency}, the electronic currency will need to have online, real-time reporting.  This will include the full accounting books of the electronic currency management firm and its subsidiaries and companies that it holds a significant ownership interest.  Highlighted will be important financial metrics like reserves (cash, inventory, capital assets, and electronic currency), electronic currency supply, and investment valuations.  A part of this will be detailed stochastic financial models of projected business performance, including potential acquisitions.  Not only the models will be made available, but the software, so that investors can re-evaluate the financial models with their own assumptions.  A benefit of this will be crowd sourcing the financial models and the model assumptions.  Sunlight is the best disinfectant for fraud, and essential to hold the electronic currency management firm accountable and trusted -- ``trust but verify'' as arms negotiators say.

Second, the electronic currency management firm will need to commit to minimum cash and capital \textbf{reserves}.  This will include encouragement of savings by its subsidiaries, employees, suppliers and other members of the sub-economy.  There are several ways that this can be done.  They can range from explicit control of subsidiaries by investing sufficient electronic currency (then controlling use through the board of directors), paying suppliers and employees in electronic currency, establishing capital savings accounts for employees (that vest over time and can only be spent on capital assets like homes and education), paying off existing student debt and mortgages (vesting over time), having put options that are super-glued to electronic currency savings of the employee or supplier and cash reserves of the electronic currency management firm (effectively insured deposits of the electronic currency), operating a bank with both insured deposits of cash and electronic currency, and having taxation for public infrastructure programmed into the electronic currency as a transaction tax.

Via the prospectus that is part of an Initial Public Offering (IPO) of the electronic currency management firm, the commitment to reserves and savings can be made, along with the commitment to hyper-transparency.  Therefore, the electronic currency management firm will be held both criminally and civilly \textbf{accountable} if it does not live up to those commitments.  This is done to build investor confidence in both the stock of the electronic currency management firm and the electronic currency.

The last attribute of \textbf{stability} leads to a very rich discussion that will occur in Sec.~\ref{nonlinear.sec}.  This is based on a Generative Artificial Intelligence (genAI) control system incorporating a multi scale, topological model of risk (i.e., the financial system response).  This will be done via arbitrage trading by the electronic currency management firm (that is, buying and selling of the electronic currency on the open market using its cash reserves).  On the longer term, the electronic currency management firm will buy companies, resources, and labor in recessionary periods and sell companies and resources in inflationary periods.  The current practice of using bonds is a poor control system that immediately reduces the currency supply (anti-inflation) but is a commitment to future inflationary coupon payments.  The same is true of central bank loans that immediately increase the currency supply (anti-recession) but is a commitment to future recessionary loan repayments.  Both bonds and loans are exploitative in motivation and not matched to the risk of the sub-economy.

Without these safeguards, it is both very easy and tempting for the sovereign to focus on the local optimization of economic exploitation.  This manifests via a lack of transparency (through propaganda, destruction of education, censorship, Lügenpresse, destruction of freedom of speech and press), lack of ability to replace the leadership, lack of ability to emigrate to another country, and sole governmental control of currency.

The fiscal investment management of the sub-economy must be coordinated with the monetary management.  Investments need to evaluated with the advanced genAI multi-scale models of risk and with operational decision analysis, that is operational management of the sub-economy, done with the same genAI risk models and metric of virtuous economic activity.  This is not the case today.  Governments have uncoordinated fiscal investment policies that are based on political, not financial considerations.  Businesses are managed based on local metrics of financial exploitation.

The criteria on which investments should be made is whether the demand for the electronic currency that the investment will create is greater than the investment, as shown in Eq.~\eqref{investment.eqn}.  The electronic currency management firm can then electronic currency leverage the funds that it has in reserves by issuing new electronic currency to make the investment.  This ensures a future growth in the value of the currency, and that the sub-economy will not be paying for the investment through inflation, effectively an inflation tax.  This is in contrast to issuing debt, which exploits the sub-economy and leads to sub-optimal levels of investment and economic performance of the sub-economy.

The fact that there are many sub-economies matched to the structure (topology) of the economy (both regionally and industrially) and that the sub-economies are individually managed, leads to further optimization.  What is good for one sub-economy is not good for another.  Having more control knobs (degrees of freedom) allows for a much better optimization.  The result will be short-term stabilization and long-term growth of the economy as a whole.

\section{Sub-economy as a nonlinear system}
\label{nonlinear.sec}
\begin{figure}
\noindent\includegraphics[width=15pc]{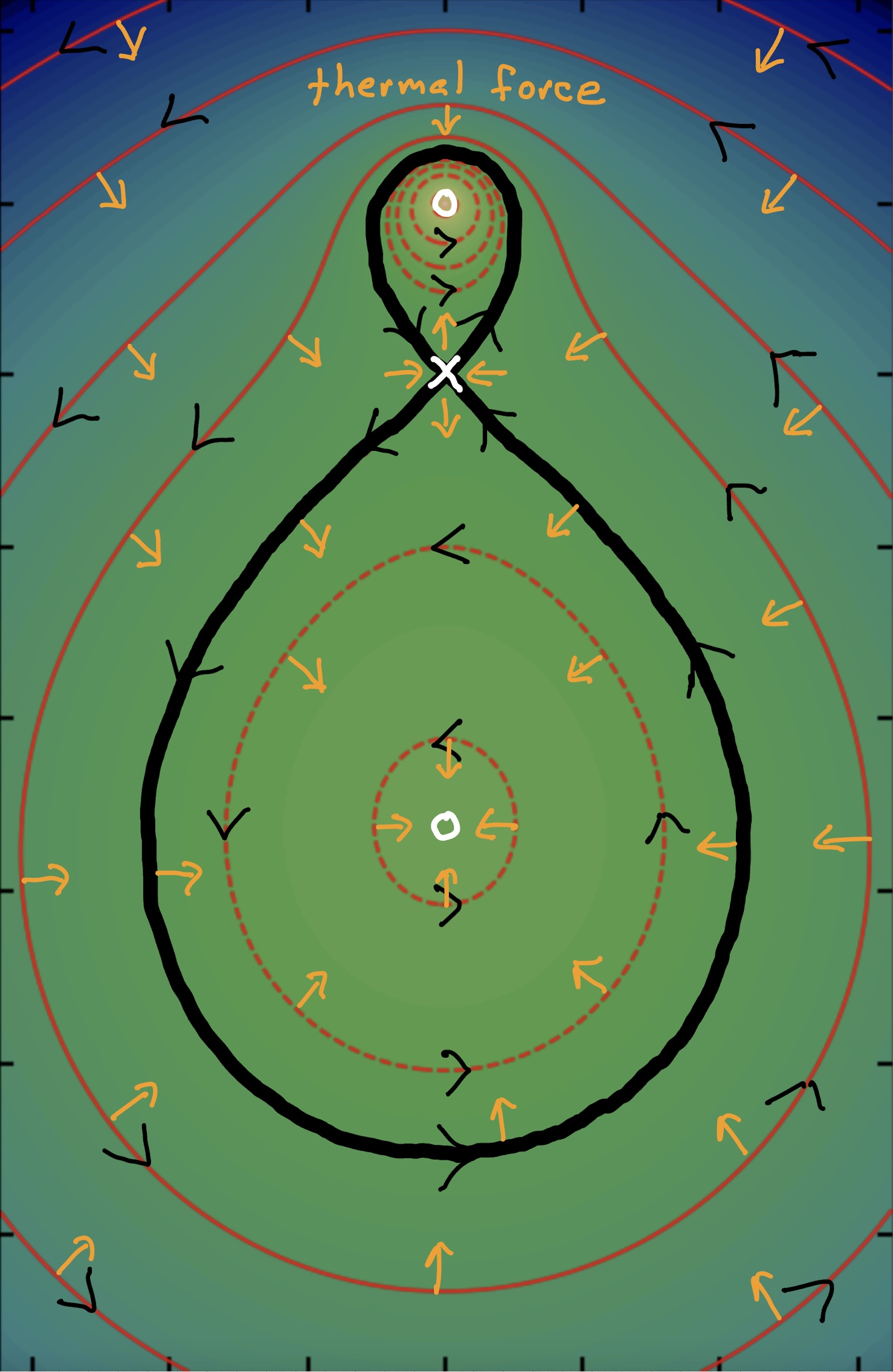}
\caption{\label{gca.dp.plot} Plot of particle trajectories shown by red lines with black arrows of an electron motion about an ion in a plane perpendicular to a strong magnetic field.  Note the boundary (thick black line) between two basins with the basin centers indicated by the white o-points, and the mountain heights with a mountain pass indicated by the white x-point.  The motion will circulate around the red lines and slowly relax in the directions shown by the orange arrows due to dissipative thermal forces.  The motion will relax from the mountain heights and eventually end up at the mountain pass (x-point), but this point is an unstable equilibrium and the motion (without control) will relax across the boundary into one of the two basins depending on how it approaches the mountain pass.  The motion will then continue to relax to the basin center at the o-point.  These are stable equilibriums.  For financial systems, the o-points are local minimums associated with exploitation, and the x-point is associated with an equilibrium of maximal sustainable economic activity that needs to be stabilized with economic control.}
\end{figure}

\begin{figure}
\noindent\includegraphics[width=15pc]{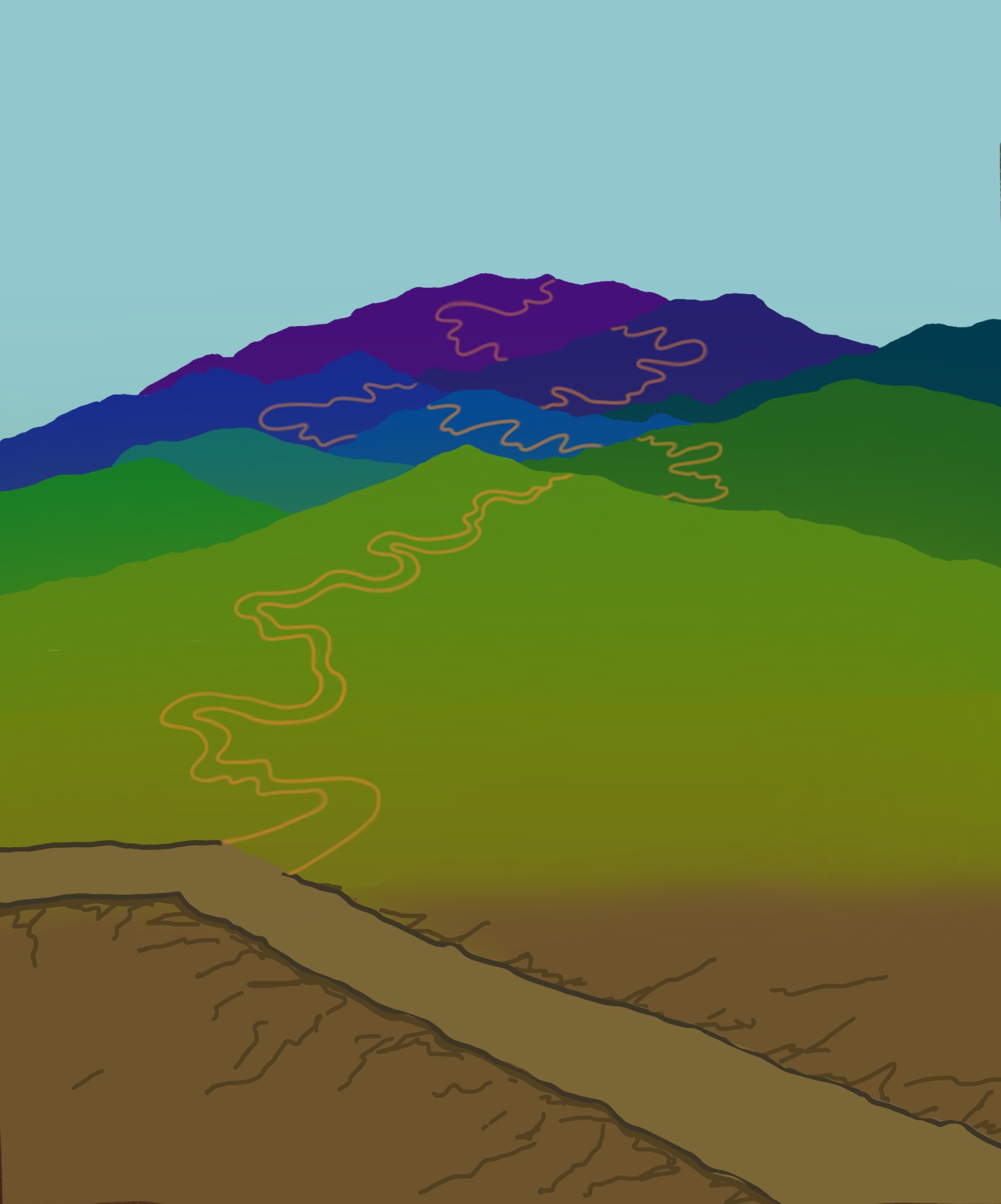}
\caption{\label{lexi.graphic} The choice at the economic crossroads.  Take the easy path down to the center of maximum exploitation and minimal sustainable economic activity, or take the difficult path up to the mountain pass of maximal sustainable economic activity.}
\end{figure}

\begin{figure}
\noindent\includegraphics[width=15pc]{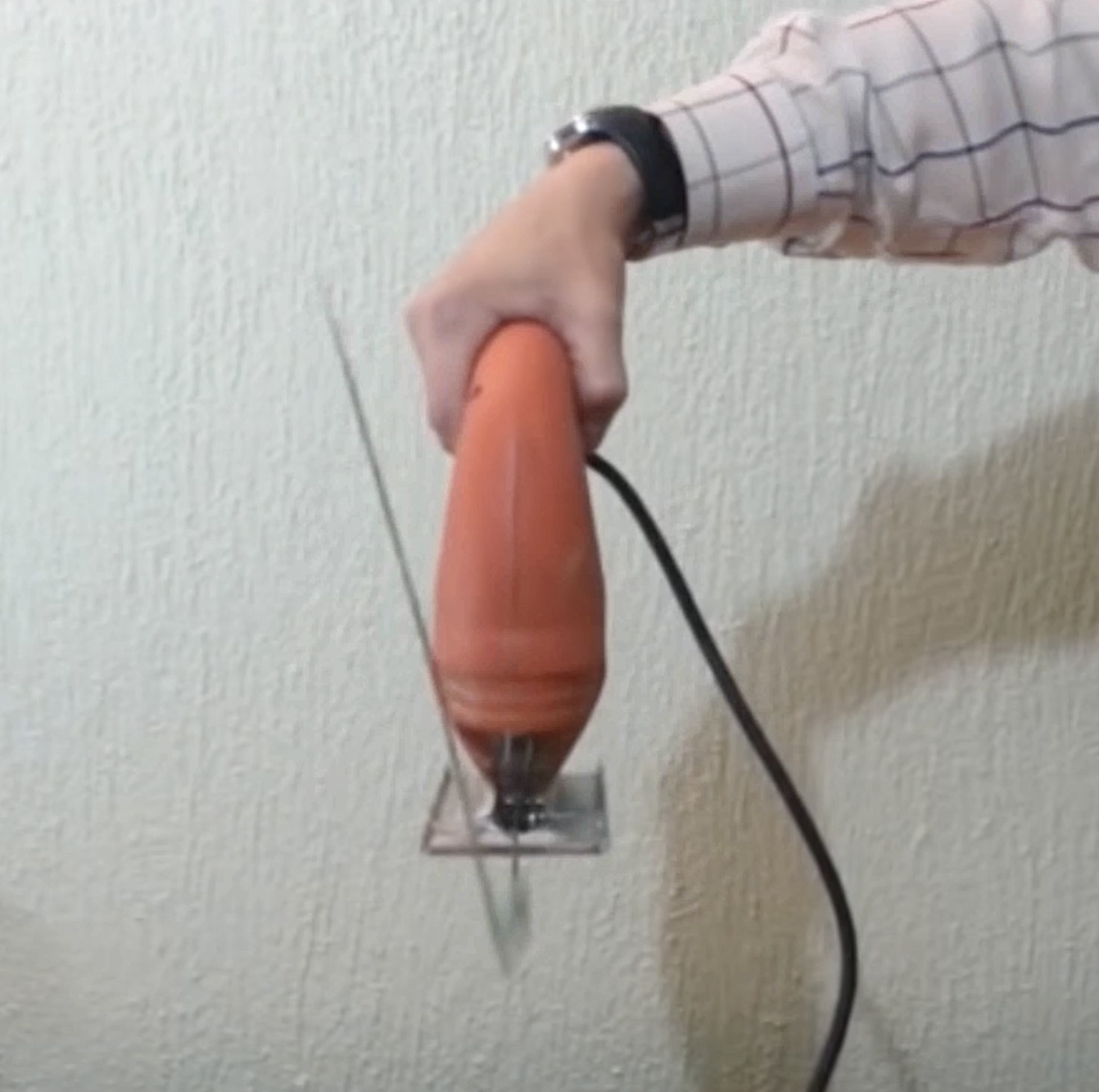}
\caption{\label{inverted.pendulum.fig} Stabilization of the pendulum about the unstable equilibrium (saddle point) by use of nonlinear control.  The full YouTube video can be found at this \href{https://youtu.be/cjGqxF79ITI}{Link}.}
\end{figure}

In this section, we will approach the understanding of a financial system as a nonlinear system like Complexity Economics \citep{farmer.24} does, and the financial and monetary policies as a problem in nonlinear systems control.  We start this discussion by referring to Fig. \ref{gca.dp.plot}, which is a representative contour plot for a physical system \citep{kuzmin04}.  It also can be looked at as a topography map.  This map has two basins with basin centers indicated by the o-points, and one saddle point (mountain pass) between the basins indicated by the x-point.  There are strong exploitative thermal forces that will take a sub-economy to the o-point once in the respective basin.  They are also what are called stable equilibriums.  The discipline of nonlinear dynamics refers to these o-points or stable equilibriums as attractors.  This is in contrast to the x-point, which is the point that the sub-economy descends to from the mountain, but it is not a stable equilibrium so that as it is approached the sub-economy will fall into one of the exploitative basins.  The discipline of nonlinear dynamics refers to these x-points or unstable/metastable equilibriums as semi-attractors (that is half attractor and half repeller).  Semi-attractors first attract the trajectory to them, but once reached repel the trajectory away from them.   These unstable equilibriums are the desirable states from a social aesthetic perspective as shown in Fig. \ref{lexi.graphic}.  They need an active control system, though, in order to stabilize them.  This is like stabilizing an inverted pendulum with a vibrating saw as shown in Fig. \ref{inverted.pendulum.fig} \citep{kapitza51,kapitza51a,landau76}.  Stabilization is the construction of a small alpine valley at the mountain pass where it is easiest to do.  The electronic currency management firm needs to vibrate the market with its arbitrage trading to stabilize the system.  The technical problem is in identifying the natural frequencies of the system.  For instance, when one flies a plane, one must take into account that it takes several seconds for the plane to respond.  If one tries to make corrections faster than this, one will over-correct and cause the plane to go out of control.  The issue with financial systems is that they have many natural frequencies, in fact an infinite number.  The electronic currency management firm needs to use innovative genAI technology that will control the financial system at all the natural frequencies.  Details of how this is done are given in the US Patent Application, ``Systems and Methods for Controlling Complex Systems'' \citep{glinsky23a} and discussed in Sec.~\ref{solve.hjb.sec}.  Current control systems only have a single time scale, and often have catastrophic phase lags built in that destabilize the system like proof-of-stake systems (e.g., New Ethereum) and the buying and selling of bonds by a central bank (e.g., US T-bills and British GILTs).

Note that the electronic currency management firm will want to buy the electronic currency, using its cash reserves, when the price is less than the equilibrium price to increase the demand and therefore the price, and sell the electronic currency when the price is more than the equilibrium price to increase the supply and therefore decrease the price.  That is, it will buy low and sell high -- a money making financial heat engine.  Mother nature rewards doing what she wants.  The essential tricks are sensing what the equilibrium is and doing it on multiple time scales.  It is all about topological discovery of the financial system response and equilibrium, then topological manipulation to control and stabilize the financial system \citep{glinsky.24b,glinsky.24d}.  This is not the case today.

\begin{figure}
\noindent\includegraphics[width=15pc]{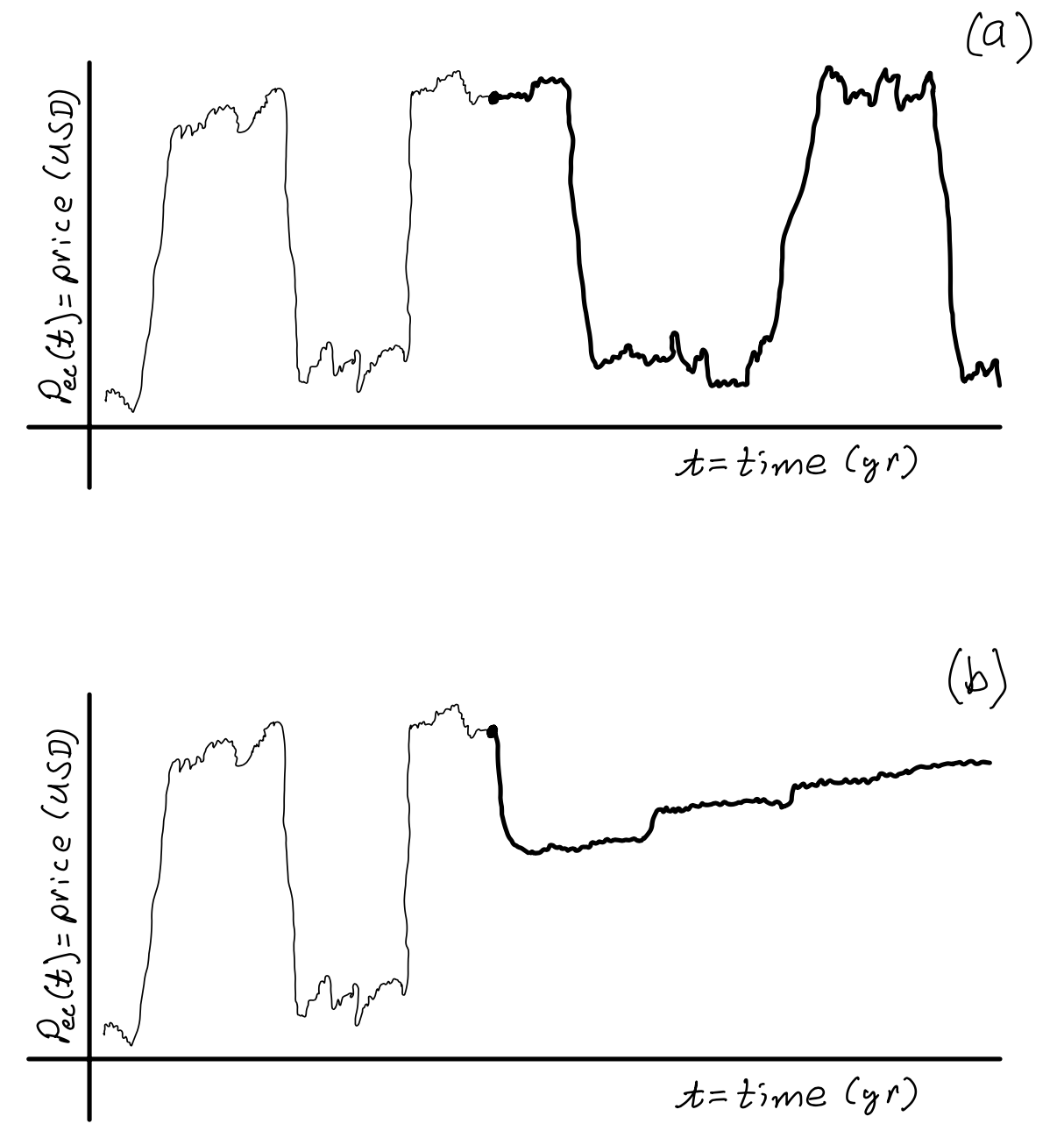}
\caption{\label{control.fig} Demonstration of control of a financial system.  (a)  the system response without control, (b) the system response with HST based control.}
\end{figure}

The effectiveness of this control system is demonstrated in Fig. \ref{control.fig}.  A realization of the uncontrolled oil price, as simulated by the HST, is compared to the same system controlled using the concepts discussed in this section based on the multi-scale topological understanding of the financial system dynamics.

\section{Solution of the Hamilton-Jacobi-Bellman equation}
\label{solve.hjb.sec}
The fundamental mathematical problem in economics is the constrained optimization (minimization) of an economic value or action functional \citep{chiang84}
\begin{equation}
    V[q(\tau)] = \int{R(q(\tau)) \, d\tau},
\end{equation}
where $R(q)$ is the reward or revenue or potential energy or potential economic activity given the state $q$ of the economic system, and $\tau$ is the evolution parameter, commonly time.  The optimization is constrained by a force equation
\begin{equation}
    \frac{dq}{d\tau} = f(q,dq/d\tau)=f(q,\dot{q}).
\end{equation}
This problem is normally approached using the method of Lagrange multipliers by forming the Lagrangian
\begin{equation}
    L(q, \dot{q}, p) = p(\dot{q}-f(q,\dot{q}))-R(q)
\end{equation}
where $p$ is the Lagrange multiplier or co-state of the system or action.  The Lagrangian can be re-written as
\begin{equation}
    L(q, \dot{q}) = L_0(q, \dot{q}) - R(q)
\end{equation}
where $p=\partial L_0 / \partial \dot{q} = g(q,\dot{q})$ so that 
\begin{equation}
    L_0(q,\dot{q}) = g(q,\dot{q}) \, [\dot{q}-f(g(q,\dot{q}),q)].
\end{equation}
The next step is to form the value or action functional
\begin{equation}
    V(q,\tau) = V[q(\tau)] = \int{L(q(\tau), \dot{q}(\tau)) \, d\tau}
\end{equation}
and use the calculus of variations to set $\delta V=0$ giving Lagrange's equation of motion for the path that minimizes the action
\begin{equation}
    \frac{d}{d\tau} \left( \frac{\partial L}{\partial \dot{q}} \right)- \frac{\partial L}{\partial q}=0.
\end{equation}
The system can also be analyzed from the Hamiltonian perspective by making the Legendre transformation
\begin{equation}
    H(p,q) = p \, \dot{q} - L(q,\dot{q}) = p \, f(p,q) + R(q)
\end{equation}
where $\dot{q} = g^{-1}(p,q) = \bar{f}(p,q)$ and $f(p,q) = f(q,\bar{f}(p,q))$.  The equations of motion are now Hamilton's equations of motion
\begin{equation}
    \frac{dq}{d\tau}=\frac{\partial H}{\partial p}
\end{equation}
and
\begin{equation}
    \frac{dp}{d\tau}=-\frac{\partial H}{\partial q}.
\end{equation}
If the motion is deterministic, the method of characteristics can be used, in what is commonly called Pontryagins Maximum Principal of Control Systems \citep{kalman63}.

The approach that we take is the canonical transformational approach \citep{arnold89,lichtenberg10} that results in the Hamilton-Jacobi-Bellman (HJB) equation \citep{goldstein80,kalman63}.  This method does not rely on the method of characteristics so that it can be applied to systems that are not integrable, that is stochastic.  This approach finds a canonical transformation generated by the value functional so that the transformed Hamiltonian is zero, giving transformed coordinates that are constants in $\tau$.  The resulting equation is
\begin{equation}
\label{hjb3.eqn}
    \frac{\partial V(q,P,\tau)}{\partial \tau} + H(\partial V / \partial q,q) = 0
\end{equation}
or more specifically
\begin{equation}
\label{hjb2.eqn}
    \boxed{\frac{\partial V(q,P,\tau)}{\partial \tau} + \frac{\partial V}{\partial q} f(\partial V / \partial q,q) + R(q) = 0,}
\end{equation}
giving
\begin{equation}
    p = \frac{\partial V}{\partial q}
\end{equation}
and
\begin{equation}
    Q = \frac{\partial V}{\partial P}.
\end{equation}
The value functional $V(q,P,\tau)$ is called Hamilton's Principal Function and can be written as
\begin{equation}
\label{sep.V.hjb.eqn}
    V(q,P,\tau) = \int{p \, dq - H \, d\tau}.
\end{equation}
Because the Hamiltonian $H(p,q)$ is not $\tau$ dependant, the value functional can be written as
\begin{equation}
    V(q,P,\tau) = W(q,P) - E(P) \, \tau 
\end{equation}
where $W(q,P)$ is called Hamilton's Characteristic Function and can be written as
\begin{equation}
    W(q,P) = \int{p \, dq},
\end{equation}
where
\begin{equation}
    p = \frac{\partial W(q,P)}{\partial q} \equiv \pi(q,P),
\end{equation}
\begin{equation}
    Q = \frac{\partial W(q,P)}{\partial P}
\end{equation}
and
\begin{equation}
    \omega_Q(P) \equiv \frac{\partial E(P)}{\partial P}.
\end{equation}
The equations of motion for the transformed coordinates are
\begin{equation}
    \frac{dP}{d\tau} = 0
\end{equation}
and
\begin{equation}
\label{wQ.hjb.eqn}
    \frac{dQ}{d\tau} = \frac{\partial E(P)}{\partial P} = \omega_Q
\end{equation}
with solution
\begin{equation}
\label{P.hjb.eqn}
    P(\tau)= P_0
\end{equation}
and
\begin{equation}
\label{Q.hjb.eqn}
    Q(\tau) = \omega_Q \, \tau + Q_0.
\end{equation}
To add external forces, that do not change the conservative nature of the system, we analytically continue the Hamiltonian and make the canonical transformation $\bar{q}=(q + \text{i} \, p)/\sqrt{2}$ and $\bar{p}=(p + \text{i} \, q)/\sqrt{2}$.  The complex analytic Hamiltonian $H(\beta)$ is now given by
\begin{equation}
    H(\beta) = H(\bar{p},\bar{q}) = H_{\text{Re}}(\bar{p}) + \text{i} \, H_{\text{Im}}(\bar{q})
\end{equation}
so that there are two orthogonal sets of motion, one for $H_{\text{Re}}$ (conservative motion generated by $H=H_{\text{Re}}=E$ and parameterized by group parameter $\tau$) with equations of motion
\begin{equation}
    \frac{dq}{d\tau} =\frac{\partial H_{\text{Re}}}{\partial p}
\end{equation}
and
\begin{equation}
    \frac{dp}{d\tau} = -\frac{\partial H_{\text{Re}}}{\partial q},
\end{equation}
and one for $H_{\text{Im}}$ (the motion generated by $\text{Ad}(H)=\text{i} H_{\text{Im}}=\text{i}\omega\tau=\text{i}\theta$ and parameterized by group parameter $\text{i}E/\omega=\text{i}J$) with equations of motion
\begin{equation}
    \frac{dq}{dJ}= \text{i} \, \frac{\partial (\text{i} H_\text{Im})}{\partial p}=-\frac{\partial H_{\text{Im}}}{\partial p}
\end{equation}
and
\begin{equation}
    \frac{dp}{dJ}= - \, \text{i} \, \frac{\partial (\text{i} H_\text{Im})}{\partial q}=\frac{\partial H_{\text{Im}}}{\partial q},
\end{equation}
where
\begin{equation}
    \Delta J = J -  J_0= \int_0^\tau{\frac{1}{\dot{q}} \frac{\partial H}{\partial \tau}d\tau}=\int{F_{\text{ext}} \, d\tau},
\end{equation}
so that 
\begin{equation}
    dE = \frac{\partial H}{\partial \tau} \, d \tau = \dot{q} \, dJ = \dot{q} \, F_\text{ext} \, d\tau = F_\text{ext} \, dq.
\end{equation}
When the Hamilton-Jacobi-Bellman equation given in Eq.~\eqref{hjb2.eqn} is solved in this analytically continued extended phase space, the transformed analytic Hamiltonian is given by
\begin{equation}
\label{Ha.eqn}
\begin{split}
    H(\beta) = H(P,Q) &=  E_P(P) + \, \text{i} \, \frac{\partial W(q(P,Q),P)}{\partial J}  \\
    &= E_P  + \, \text{i} \, \frac{\partial W}{\partial P} \, \frac{\partial P}{\partial J_P} \\
    &= E_P(P) + \, \text{i} \, Q \\
    &= E_P(P) + \, \text{i} \, \theta_Q(Q) \\
    &= H_\text{Re}(P) + \text{i} \, H_\text{Im}(Q)
\end{split}
\end{equation}
or
\begin{equation}
\label{Hpq.eqn}
    H(p,q) = E_P(P(q,p)) \, + \, \text{i} \, \frac{\partial W(q,P(p,q))}{\partial J},
\end{equation}
where 
\begin{equation}
    \theta_Q(Q) =  \omega\tau = Q
\end{equation}
and
\begin{equation}
    J_P(P) = E/\omega = P.
\end{equation}
The equations of motion for the conservative motion, with group parameter $\tau$, are
\begin{equation}
    \frac{dQ}{d \tau} = \frac{\partial E_P(P)}{\partial P} \equiv \omega_Q
\end{equation}
and
\begin{equation}
    \frac{dP}{d\tau} = 0,
\end{equation}
and the equations of motion due to an external force doing work on the system, with group parameter $\text{i}J$, are
\begin{equation}
\label{dQdE.eqn}
    \frac{dQ}{dJ} = 0
\end{equation}
and
\begin{equation}
    \frac{dP}{dJ} = 1,
\end{equation}
with differential solution
\begin{equation}
\label{dq.eqn}
    dQ = \omega_Q \, d\tau
\end{equation}
and
\begin{equation}
\label{dp.eqn}
    dP= \frac{dE}{\omega_Q}  = \omega_P \, dE = F_\text{ext} \, d\tau,
\end{equation}
where $\omega_P \equiv 1/\omega_Q$.  The finite solution is
\begin{equation}
    \Delta Q = Q - Q_0 = \int_0^\tau{\omega_Q \, d\tau}
\end{equation}
and
\begin{equation}
\begin{split}
    \Delta P &= P - P_0 \\
    &= \int_0^\tau{\omega_P \frac{\partial H}{\partial \tau} \, d\tau} = \int_0^{\Delta E}{\frac{dE}{\omega_Q}} \\
    &= \int_0^\tau{F_\text{ext} \, d\tau},
\end{split}
\end{equation}
where
\begin{equation}
    dE = \omega_Q \, dJ = \omega_Q \, F_\text{ext} \, d\tau = F_\text{ext} \, dq.
\end{equation}

It should be noted that after a significant amount of time has elapsed ($\omega_Q \, \tau \gg 1$), uncertainty in the value in $\omega_Q$ will cause the motion to become stochastic.  Not only will the value of $Q$ be not known, even the number of cycles of temporal period $\tau_0=2\pi/\omega_Q$ will not be known.  The value of $Q$ will simply be uniformly distributed from 0 to $2\pi$.

The form of the imaginary part of the Hamiltonian in Eq. ~\ref{Hpq.eqn}
\begin{equation}
\label{Hreal.eqn}
    H_\text{Im}(p,q) = \frac{\partial W(q,P(p,q))}{\partial J}
\end{equation}
is quite interesting.  Optimal control as done in genAI with Deep Q-Learning (DQN) \citep{mnih15} is based on parametric estimations of a Q-function
\begin{equation}
    \widetilde{Q}(s,a;\theta) = W(q,p;P) = W(q,P(p,q)),
\end{equation}
where $s=q$ is the state, $a=p$ are the actions to be taken, and $\theta=P$ are the parameters of the estimator.  What is at the kernel of DQN is the action $S_{\text{Ad}(H)}=\int{\tau \, dE}=\int{\theta_Q \, dJ_P}=\int{p \, dq}=W(q,p;P)$ of the $\text{Ad}(H)$ group with infinitesimal generating function $\text{Ad}(H)=\text{i}H_\text{Im}=\text{i}\omega\tau= \text{i}\theta_Q=\text{i}\,\partial W / \partial J$ and associated group parameter $\text{i}J=\text{i}E/\omega$.  This will be discussed in more detail in Sec. ~\ref{traditional.DRL.sec}.

Given this theory, we move on to the practical application of it to control the system.  This application will use the concepts of Artificial Intelligence \citep{hastie09,sugiyama15,goodfellow16}, as interpreted by \citet{glinsky.24b,glinsky.24d}.  First, construct a dataset by either doing an ensemble of simulations of the system or by observing the system.  It will be assumed that the system has a small number of dimensions.  Most of the systems of interest  present themselves in the domain of collective fields $f(x)$ that are elements of a Hilbert space not $q(\tau)$ with a small number of components.  How to de-convolute from the domain of the collective motion, the field $f(x)$, to the domain of the individual, $q(\tau)$,  (that is from a Hilbert space to $\mathbb{C}^n$) using the Heisenberg Scattering Transformation (HST) and a Principal Components Analysis (PCA) will be discussed at the end of this section.  This transformation can be done because the collective acts as one, because of the correlation or synchronization specified by the S-matrix $S_m$ given in Eq. ~\ref{smatrix.eqn}, which are the derivatives of the analytic Hamiltonian $H(\beta)$ of the individual.  

Start by doing an ensemble of simulations or measurements on the system of interest.  It is helpful to apply an external force $F_{\text{ext}}$ to the system being simulated or observed to sample phase space more efficiently.  A good choice would be a dissipation or a random diffusion which samples phase space well, as the system gradually relaxes to the stable equilibriums.  It is also good to apply an external force that is constructed to keep the dynamic trajectory in the vicinity of the unstable local maximums, that is stabilizes the unstable equilibriums.  The set of variables that should be recorded are $\tau$, in addition to variables that are related to the state $q(\tau)$ and the co-state $p(\tau)$.

\begin{figure}
\noindent\includegraphics[width=\columnwidth]{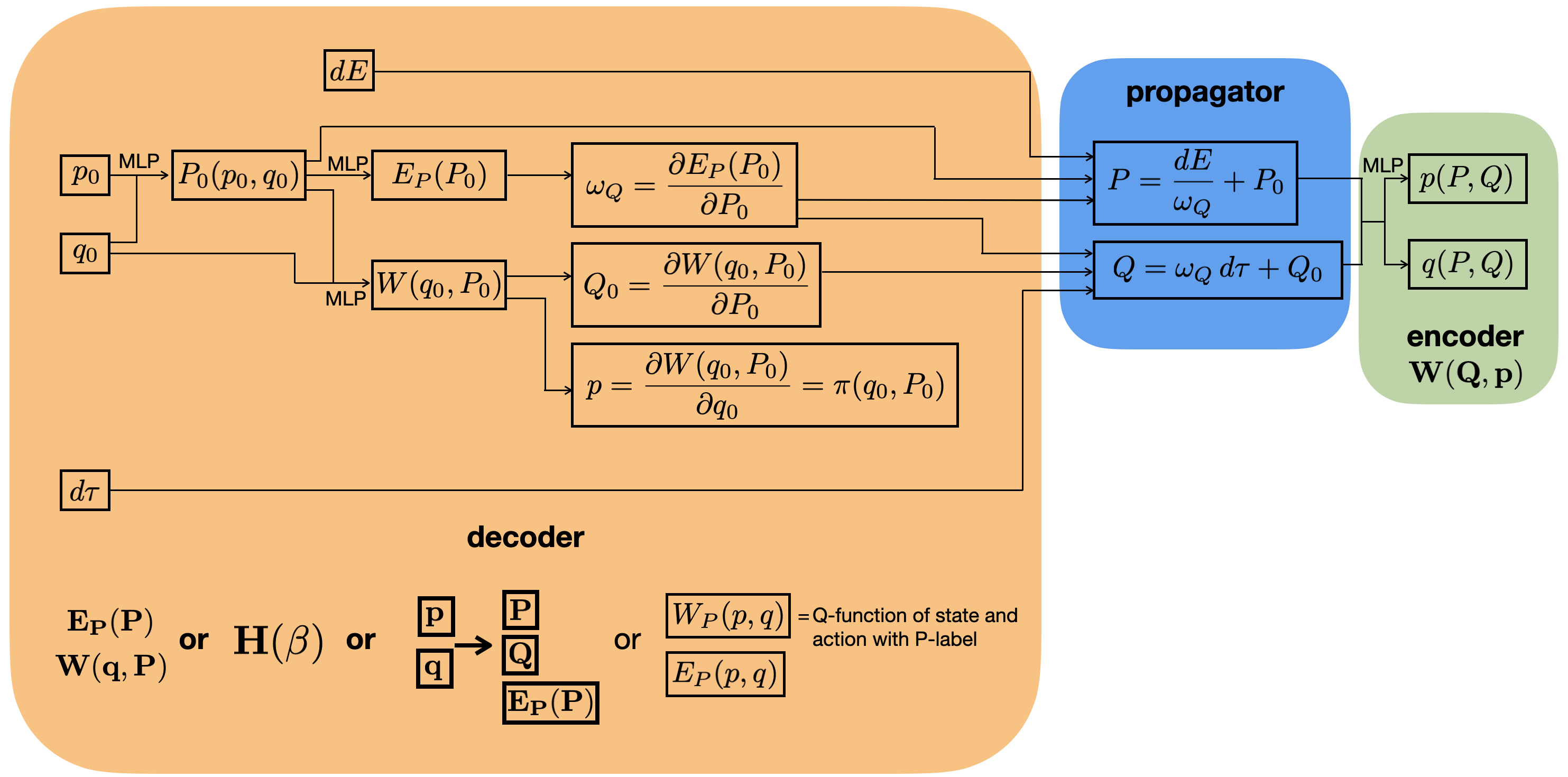}
\caption{\label{HJB_NN} The neural network architecture to estimate the solution of the Hamilton-Jacobi-Bellman equation.  It starts with a decoder from $(p,q)$ to $(P,Q)$ generated by the value function $W(q,P)$ related to the imaginary part of the analytic function $H(\beta)$ by Eq.~\eqref{Hreal.eqn}, a mapping to the real part of the analytic function $H(\beta)$ given by $E_P(P)$, the frequency $\omega_Q(P)$, the policy $\pi(q,P)$, the analytic mapping to advance $(P,Q)$, and a encoder from $(P,Q)$ to $(p,q)$.  The universal function approximator is a Multi Layer Perceptron (MLP).}
\end{figure}

Given the dataset, train a neural network with an architecture that matches the structure of the solution to the HJB equation to estimate:  (1) the decoding of the $p$ and $q$ coordinates into the $P(p,q)$ and $Q(p,q)$ coordinates that are the solution to the HJB equation as well as the encoding of $P$ and $Q$ to $p(P,Q)$ and $q(P,Q)$, (2) the value function $W(q,P)$ that generates these canonical transformations and is related to the imaginary part of the analytic Hamiltonian as shown in Eq.~\eqref{Hreal.eqn}, (3) the mapping of $P$ to the real part of the analytic Hamiltonian $E_P(P)$, (4) the frequency $\omega_Q(P)$, (5) the policy $\pi(q,P)$ and (6) the analytic advance of $P$ and $Q$ given in Eqns.~\eqref{dp.eqn} and \eqref{dq.eqn}.  Multi Layer Perceptrons (MLPs) \citep{goodfellow16} are used to approximate some of the functions.  The derivative functions are calculated by back propagating the MLPs.  This architecture is shown in Fig. \ref{HJB_NN}.  It is important that Rectified Linear Units (ReLUs) are used as activation functions in the MLPs because the MLPs are approximating analytic functions which are maximally flat but do have a limited number of singularities where the derivative is discontinuous.  MLPs with ReLUs are very good at doing this since they are universal piecewise linear approximators with discontinuities in the derivative.

\begin{figure}
\noindent\includegraphics[width=\columnwidth]{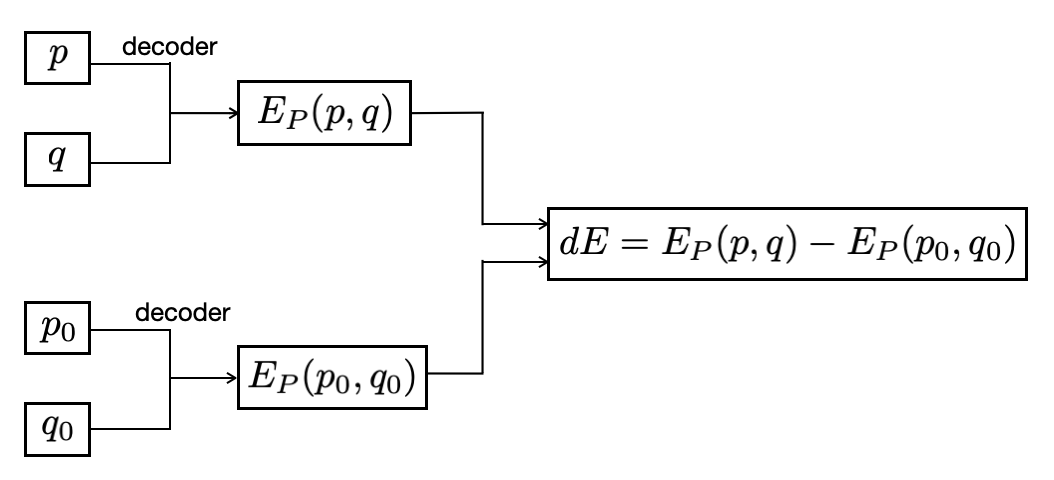}
\caption{\label{training.dE.fig} The addition to the neural network architecture to estimate $dE$ in the solution of the Hamilton-Jacobi-Bellman equation.}
\end{figure}

\begin{figure}
\noindent\includegraphics[width=\columnwidth]{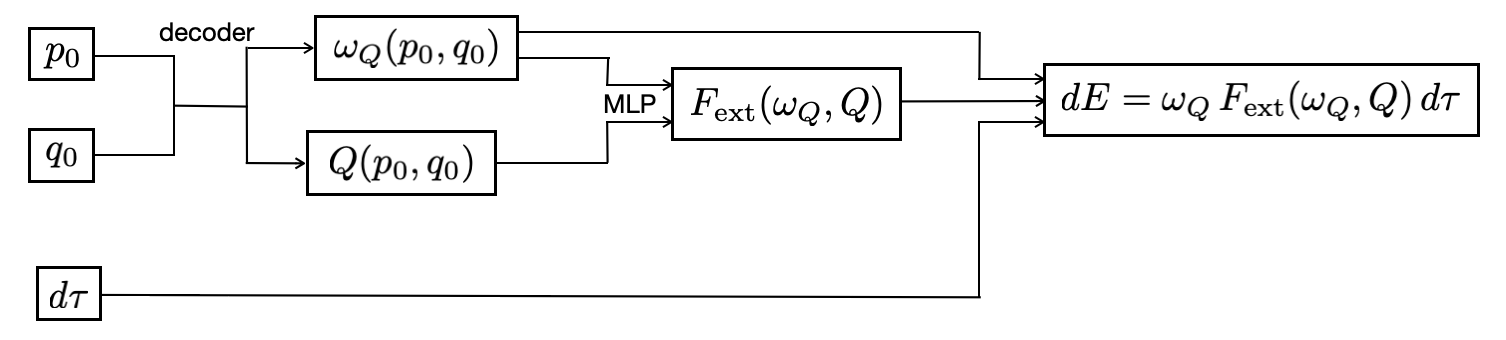}
\caption{\label{dissipation.dE.fig} The addition to the neural network architecture to estimate $F_\text{ext}(\omega_Q,Q)$ in the solution of the Hamilton-Jacobi-Bellman equation.  The universal function approximator is a Multi Layer Perceptron (MLP).}
\end{figure}

There is a non-trivial detail in this training step.  Although one has the inputs ($p_0$, $q_0$, $d\tau$) and outputs ($p$, $q$), what is $dE$?  For a conservative system with no external force being applied $dE=0$, but that is not the case with this dataset.  The solution is to use the decoder $E_P(p,q)$ to estimate $dE=E_P(p,q)-E_P(p_0,q_0)$, using the target outputs as an input to estimate $E_P(p,q)$, as shown in Fig. \ref{training.dE.fig}.  If this workflow is being used to train a surrogate where the external force is part of the dynamics that is being modeled, a model for the external force $F_\text{ext}(\omega_Q,Q)$ needs to be estimated using an MLP so that $dE= \omega_Q \, F_\text{ext}(\omega_Q,Q) \, d\tau$, as shown in Fig. \ref{dissipation.dE.fig}.  If the force is resistive, diffusive friction $F_\text{ext}=-\epsilon_P \, \omega_Q$, where $\epsilon_P \ll J_0$.  In this case, Rayleigh's Dissipation Function can be defined
\begin{equation}
    \mathscr{F} \equiv \frac{\epsilon_P}{2} \, \omega_Q^2
\end{equation}
so that
\begin{equation}
    \frac{dE}{d\tau} = -\epsilon_P \, \omega_Q^2 = -2 \mathscr{F}.
\end{equation}
You could view the estimation of $F_\text{ext}(\omega_Q,Q)$ as an estimation of Rayleigh's Dissipation Function where
\begin{equation}
    \mathscr{F}(\omega_Q,Q) = - \frac{\omega_Q}{2} \, F_\text{ext}(\omega_Q,Q).
\end{equation}

The solution of the system dynamics has the conservative force $F_0(q) = -\nabla R_0(q)$ of the uncontrolled system, where $R_0(q)$ is the reward optimized by the uncontrolled system.  The dynamics can be modified to optimize a desired reward $R(q)$.  In order to do this, calculate the conservative control force $F_c(q)$ (that is the incremental action, $\Delta p / \Delta \tau$) that needs to be applied to change the reward that is optimized. Estimate $F_0(q)$, then calculate the control force
\begin{equation}
\label{reward.force.eqn}
    \boxed{F_c(q) \equiv F(q)-F_0(q) = \Delta(dp/d\tau),}
\end{equation}
where $F(q)=-\nabla R(q)$ and
\begin{equation}
    F_0(q) = \frac{\partial E_P(p=0,q)}{\partial q}
\end{equation}
which is found by back propagating the derivatives in the MLP.  The system needs to be simulated or observed again, this time applying the control force, but not including that control force in the calculation of $\Delta E$.  The neural network needs to be fit again, with this new dataset.

One now has a solution of the HJB equation that has estimated the analytic Hamiltonian $H(\beta)$.  The next step is finding the equilibriums $\beta^*$ or $P^*$ where
\begin{equation}
    \frac{\partial E_P(P^*)}{\partial P}=0.
\end{equation}
Given the function $E_P(P)$ estimated in the workflow shown in Fig. \ref{HJB_NN}, $P^*$ can be found with a high performance root finder, both stable and unstable equilibriums.  The equilibrium policy can then be estimated as $\pi^*(q)=\pi(q,P^*)$ and the equilibrium value as $V^*(q)=W(q,P^*)$.  One now has estimated the $\beta^*$ which are viewed different ways by different technical disciplines.  These are: (1) the ground states of quantum field theory \citep{weinberg05}, (2) the attractive manifolds of nonlinear dynamics \citep{lichtenberg10}, (3) the emergent behaviours and self organizations of complex systems \citep{gros15}, (4) the Taylor relaxed states \citep{taylor86} and BGK modes \citep{bernstein57} of plasma physics, (5) the poles and branch cuts of control theory and complex analysis \citep{nehari12}, and fundamentally (6) the homology classes of the topology of the dynamic manifold or the geometry of the physics \citep{frankel12}.  The equilibrium values $\beta^*$ need not be points.  They can be manifolds with rich topography, that is algebraic structures, if $n>1$.

Knowing $\beta^*$ is equivalent to knowing the analytic function $H(\beta)$.  $H(\beta)$ is the solution of Laplace's equation given the boundary $\beta^*$.  The motion on the dynamical manifold is simply geodesic motion generated by the action $S(\beta)= \text{i} \int{H(\beta) \, d\beta}$ whose Taylor expansion coefficients are given by the S-matrix \citep{landau59,cutkosky60,chew61} 
\begin{equation}
\label{smatrix.eqn}
    S_m \equiv \frac{d^m S(\beta)}{d\beta^m}.
\end{equation}
The analytic function $H(\beta)$ specifies the geodesics $\text{Re}(H(\beta))=E$ of the motion generated by $H=H_\text{Re}=E_P$ with group parameter $\tau$, and the geodesics of the adjoint motion $\text{Im}(H(\beta))=\omega\tau=\theta$ generated by $\text{Ad}(H)=\text{i}H_\text{Im}=\text{i}\theta_Q$ with group parameter $\text{i}J=\text{i}E/\omega$.  The complete motion is generated by the Weyl-Heisenberg group $\mathbb{H}= H \otimes \text{Ad}(H)$ on extended phase space with Lagrangian or Poincaré one form $\lambda = p \, dq - H \, d\tau = \tau \, dE - E \, d\tau = \theta \, dJ - E \, d\tau$, symplectic metric or Poincaré two form $d\lambda=dp \wedge dq - dH \wedge d\tau=2 \, d\tau \wedge dE$, complex group parameter $\tau+ \text{i}J$, complex analytic Hamiltonian or complex group infinitesimal generating function $H(\beta)=E_P + \, \text{i} \, \theta_Q$, and group action or finite group generating function $S=\int{\lambda}=S_{\text{Ad}(H)}-S_H=W-\int{E\, d\tau}$, where $W=\int{\theta \, dJ}$ and $\theta=\partial W / \partial J$ -- a complex Lie Group.  Note that the complex finite group propagator is
\begin{equation}
\begin{split}
    U(\tau+\text{i}E/\omega) &= U_{\text{Ad}(H)}(E) \; U_H(\tau) = \text{e}^{\text{i}S} \\
    &= \text{e}^{\text{i}S_{\text{Ad}(H)}} \, \text{e}^{-\text{i}S_H} = \text{e}^{\text{i}\int{\tau \, dE}} \, \text{e}^{- \text{i}\int{E \, d\tau}} \\
    &= \text{e}^{\text{i}W} \, \text{e}^{- \text{i}\int{E \, d\tau}} \\
    &= \text{e}^{\text{i}W} \, \text{e}^{-\text{i}E\tau}, \; \text{ if } \frac{\partial E}{\partial \tau}=0.
\end{split}
\end{equation}
Therefore, if the motion is conservative, the propagators are
\begin{equation}
    U_{\text{Ad}(H)}(E) = \text{e}^{\text{i}W},
\end{equation}
and
\begin{equation}
    U_H(\tau) = \text{e}^{-\text{i}H\tau},
\end{equation}
the later being the well known field theory expression for the propagator.
If the motion is not conservative,
\begin{equation}
    U_{\text{Ad}(H)}(E) = \text{e}^{\text{i}W} = \text{e}^{\text{i}W_{P(E)}}
\end{equation}
is unchanged and
\begin{equation}
    U_H(\tau) = \text{e}^{-\text{i}\int{E(\tau)\,d\tau}} = \text{e}^{-\text{i}\int{H(p(\tau),q(\tau))\,d\tau}}.
\end{equation}
The important distinction to make is that $E(\tau)$ is changing with $\tau$, not the forms of $H(p,q)$ and $W_P(p,q)$.  Even if the motion is not conservative, it is still constrained to the manifold $\mathbb{H}= H \otimes \text{Ad}(H)$ with the algebra of $H(\beta)$ on $\mathbb{C}^n$.

It is interesting to note that at the equilibrium points $P^*$ the external forces can not change the system's energy because $\omega_Q(P^*)=0$ and $dE/d\tau=\omega_Q(P^*) \, F_\text{ext}=0$ for all $F_\text{ext}$.

Now to the solution to the general problem of control.  First, it might be necessary to change the reward or potential economic energy function from $R_0(q)$ to $R(q)$ using $F_c(q)$ of Eq.~\eqref{reward.force.eqn}.  This is improving the geometry by changing the location of $\beta^*$.  The trick is identifying $R_0(q)$.  This can be done by learning the solution to the HJB equation $E_P(P)$ and $W(q,P)=W_P(p,q)$, that is the energy or economic activity and the canonical generating function.  The solution can be harvested for $R_0(q)$, but more importantly for $\beta^*$.

\begin{figure}
\noindent\includegraphics[width=\columnwidth]{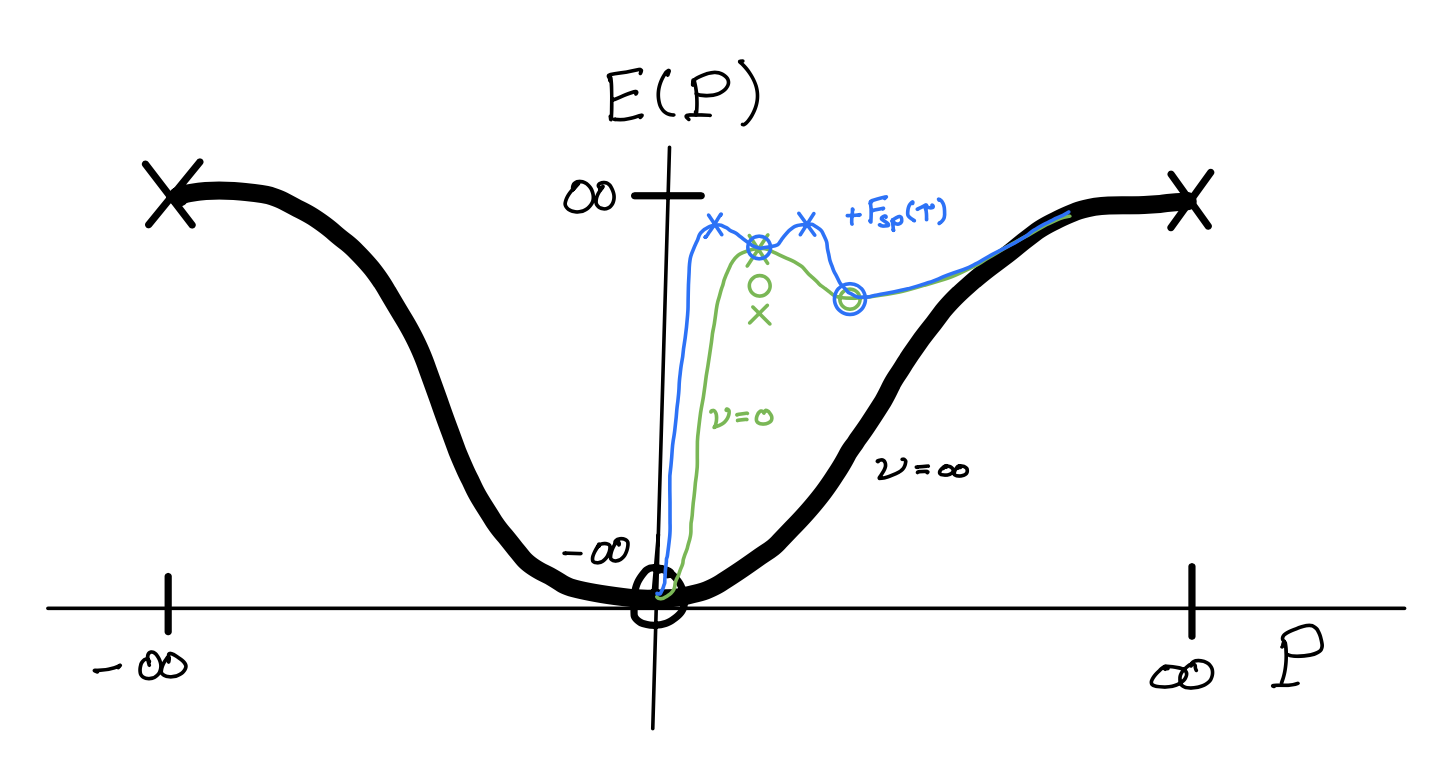}
\caption{\label{Ep.sp.fig} Graph showing the equilibriums where $\omega_Q(P^*)=0$.  Plotted is the energy $E(P)$ as a function of $P$.  The unstable local maximums are indicated by the x-points, and the stable local minimums (Nash Equilibriums) by the o-points.  The energy for the system with no dissipation $\nu=0$ is the thin green line.  How this appears in $(p,q)$ phase space can be found in Fig. \ref{gca.dp.plot}.  The heavily damped system $\nu \to \infty$ is the thick black line. The stabilized system (with $F_\text{sp}(\tau)$ or $F_\text{sf}(P)$) is the thin blue line.}
\end{figure}

There are two types of $\beta^*$, that is equilibriums where
\begin{equation}
    \omega_Q(P^*)= \frac{\partial E_P(P^*)}{\partial P}=0.
\end{equation}
Refer to Fig. \ref{Ep.sp.fig}.  Those that are stable equilibriums where
\begin{equation}
    \frac{\partial^2 E_P}{\partial P^2}=\frac{\partial \omega_Q}{\partial P}=\omega'_Q > 0,
\end{equation}
that is local minimums.  At some of these points $E_P \to -\infty$, $\omega_Q \to 0$, and $\omega'_Q \to \infty$.  Nothing further needs to be done in this case of a stable equilibrium.  Just put the system close to the stable $\beta^*$ and the system will oscillate about it gradually approaching $\beta^*$ due to naturally occurring interactions with an external heat bath, that is the economy external to the sub-economy.  These stable equilibrium points are referred to as attractors or o-points.  They are the Nash Equilibriums to which the system will naturally descend as it minimizes the value or action functional.  Do note that the local minimization of the economic action to accomplish the task (doing the task in the most efficient manner) is leading to a global minimization of the value.

There are also unstable equilibriums where $\omega'_Q < 0$, that is local maximums.  These unstable equilibriums are also known as saddle points or x-points or semi-attractors that first attract the trajectory to them with the system staying near them for a long time since $\omega_Q=0$ (a meta stable state).  But, eventually the system will start to move away from the semi-attractor, being repelled away from it.  The trajectory will eventually approach and be attracted back to the semi-attractor or might approach and be attracted to another semi-attractor (or several other semi-attractors) before being repelled by the last semi-attractor and returning to the first semi-attractor.  An example of a set of two semi-attractors could be a bull and a bear market.  These saddle points will need to be stabilized by application of an external force, then cooled by an external thermal force, otherwise the trajectory will evolve into the basin of attraction of a stable attractor and eventually relax to that attractor.  While small amounts of energy are needed to move from semi-attractor to semi-attractor, a significant amount of energy is needed to move from one attractor to another attractor or semi-attractor.  One needs to climb out of the basin of attraction or potential well, that is climb back up to the mountain pass (saddle point) out of the basin before descending into another basin.  When these semi-attractors are stabilized, there is still a local minimization of the economic action to efficiently accomplish the task, but the sub-economy is kept at a global sustainable maximum.

The stabilization and cooling of the saddle points can be done directly via an external feedback force
\begin{equation}
\label{stable.feedback.eqn}
    \boxed{F_\text{sf}(P) = -\omega_\text{sf}(P-P^*) - \epsilon_P \, \omega_Q(P) = \Delta(dP/d\tau),}
\end{equation}
where $\omega_\text{sf} \gtrsim \omega_0$ and $\epsilon_P \ll J_0$, and $\omega_0$ is the dominate spectral frequency or the ground state frequency $\omega_0=E_0/J_0$.  This is difficult to do because both $P$ (which is oscillating rapidly about $P^*$) and $P^*$ must be known or measured.  It is much better to apply the ponderomotive equivalent and random walk equivalent
\begin{equation}
\label{stable.pondermotive.eqn}
    \boxed{F_\text{sp}(\tau) = f_0 \, \text{e}^{\text{i} \, \omega_\text{sp} \tau} + \omega_0 \, \varepsilon_P = \Delta(dP/d\tau),}
\end{equation}
where $\omega_0 \ll \omega_\text{sp}$, $J_0 \, \omega_0 \lesssim f_0 \ll J_0 \, \omega_\text{sp}$ (so that the ponderomotive force is large but the motion is small) and $\varepsilon_P$ is a random $\Delta P$ of size $\epsilon_P \ll J_0$ taken every $2 \pi/\omega_0$.  This external $F_\text{sp}(\tau)$ force is not dependant on $P^*$ or $P$, just the time invariant mapping generated by $W(q,P)$ of $p(P,Q)$ and $q(P,Q)$, and the functional transformation iPCA+iHST of $f[p(\tau),q(\tau)](x)$ and $\pi[p(\tau),q(\tau)](x)$ or $f[p+\text{i}q](x)+\text{i}\pi[q+\text{i}p](x)$ as will be discussed later in this section.

The ponderomotive force can be intuitively understood.  The economic system is vibrated more when it evolves in a undesirable direction, and is vibrated less when it evolves in a desirable direction.  The system does not like to be vibrated, so a conservative ponderomotive potential is established that leads to a ponderomotive force away from the undesirable states and towards the desirable states.

\begin{figure}
\noindent\includegraphics[width=\columnwidth]{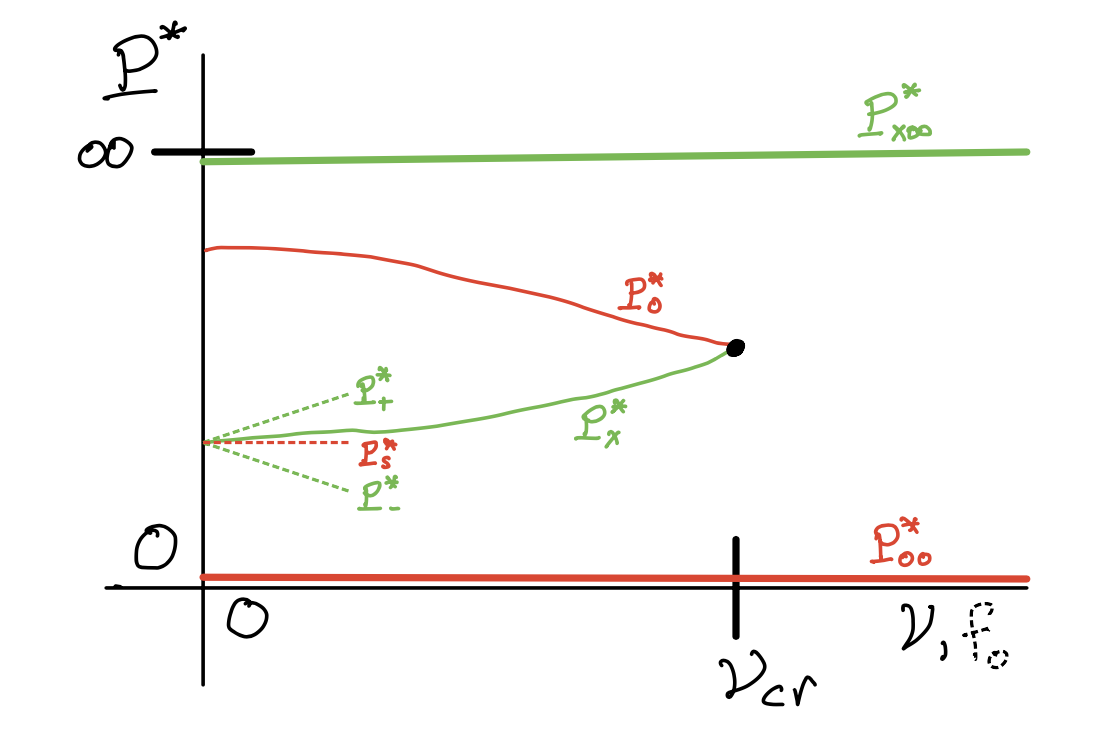}
\caption{\label{P.star.nu.fig} Graph showing the location of the equilibriums $P^*$, where $\omega_Q(P^*)=0$, as a function of the regularization or dissipation $\nu$ and the strength of the ponderomotive stabilization force $f_0$.  The thick green line shows the unstable equilibrium or x-point or local maximum at infinity $P^*_{\text{x} \infty}$.  The thick red line shows the stable equilibrium or o-point or Nash Equilibrium at zero $P^*_\text{o0}$.  This is the point of severe economic depression.  The thin green line shows the local maximum at $P^*_\text{x}$.  The thin red line shows the Nash Equilibrium at $P^*_\text{o}$.  This is the point of economic recession.  The thin dashed green lines show the local maximums of the stabilization force, $P^*_\text{+}$ and $P^*_{-}$.  The thin dashed red line shows the Nash Equilibrium of the stabilization force $P^*_\text{s}$.  This is the point of economic prosperity.  The upper limit on viscosity so that the singularities $P^*_\text{x}$ and $P^*_\text{o}$ are not destroyed is $\nu<\nu_\text{cr}$.}
\end{figure}

\begin{figure}
\noindent\includegraphics[width=\columnwidth]{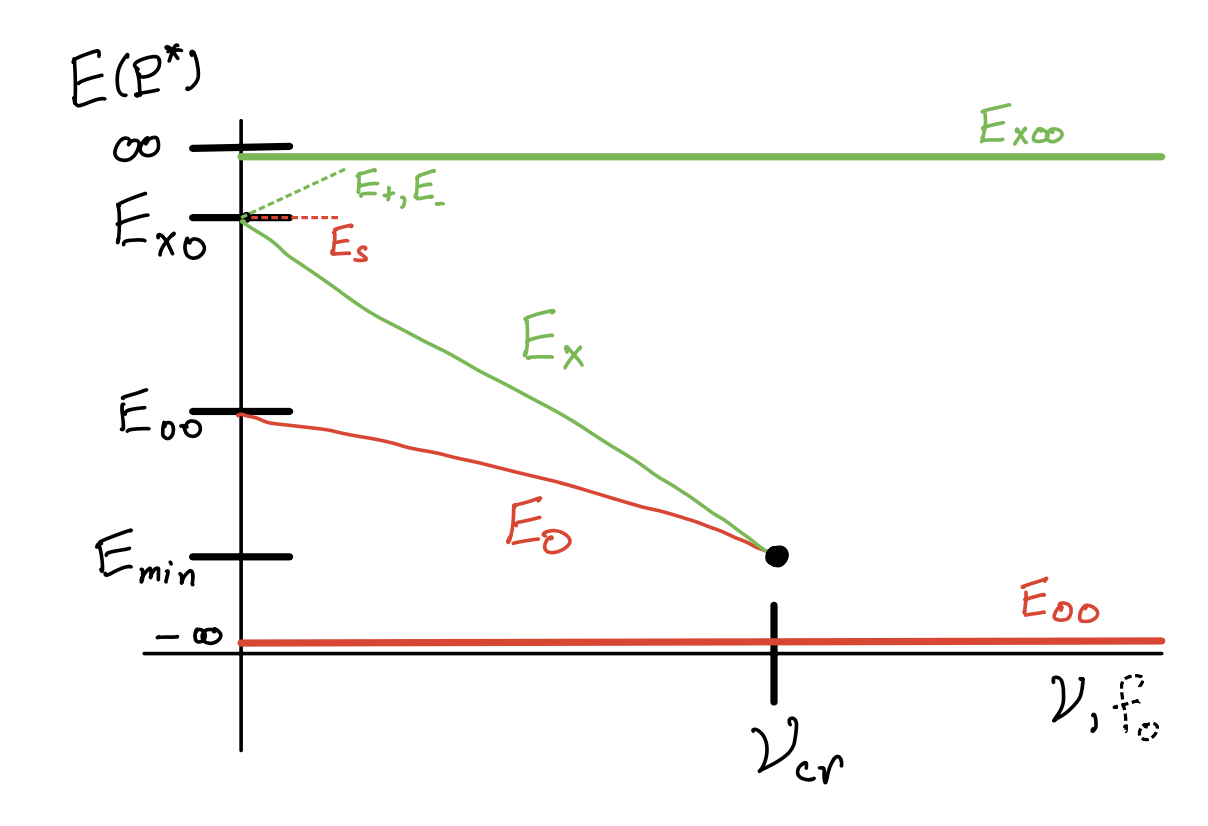}
\caption{\label{E.star.nu.fig} Graph showing the energies $E(P^*)$, where $\omega_Q(P^*)=0$, as a function of the regularization or dissipation $\nu$ and the strength of the ponderomotive stabilization force $f_0$.  The thick green line shows the unstable equilibrium or x-point or local maximum at infinity with energy $E_{\text{x} \infty}$.  The thick red line shows the stable equilibrium or o-point or Nash Equilibrium at zero with energy $E_\text{o0}$.  This is the point of severe economic depression.  The thin green line shows the local maximum at $P^*_\text{x}$ with energy $E_\text{x}$.  The thin red line shows the Nash Equilibrium at $P^*_\text{o}$ with energy $E_\text{o}$.  This is the point of economic recession.  The thin dashed green lines show the local maximums of the stabilization force with energies $E_\text{+}$ and $E_{-}$.  The thin dashed red line shows the Nash Equilibrium of the stabilization force with energy $E_\text{s}$.  This is the point of economic prosperity.  The upper limit on viscosity so that the singularities $P^*_\text{x}$ and $P^*_\text{o}$ are not destroyed is $\nu<\nu_\text{cr}$.}
\end{figure}

What the stabilization force has done is to modify and fortify the topology by adding one x-point and one o-point at the location of the original x-point.  It has left the o-point where the original x-point was and moved the two x-points out, sandwiching the o-point.  This has turned the mountain pass into a high mountain valley.  This is shown in Figs. \ref{Ep.sp.fig}, \ref{P.star.nu.fig}, and \ref{E.star.nu.fig}.  Note that the Nash Equilibrium at $P^*_{\text{o} 0}$ is the point of severe economic depression, the Nash Equilibrium at $P^*_\text{o}$ is the point of economic recession, and the unstable equilibrium at $P^*_\text{s}$ is the point of economic prosperity.  It is hard to stimulate an economy from an economic recession to economic prosperity since $E_\text{s} > E_\text{o}$, but it is very hard to stimulate an economy from a severe economic depression to economic prosperity since $E_\text{s} \gg E_{\text{o} 0}$.  Also note that if the economy is not stimulated enough, so that $E=E_\text{s}$, the economy will fall back into the recession or depression.

A very profound structure has been put on the dynamics by the sympletic, that is canonical, structure.  This symplectic structure can also be viewed as underlying toroidal topologies $T^2$ or cylindrical geometries of extended phase space or $\mathbb{C}$.  The underlying analytic function $H(\beta)$ will be specified by two types of singular points in the vector field: (1) o-points that are stable local potential energy (that is reward) minimums, and (2) x-points that are unstable local potential energy (that is reward) maximums.  The motion is geodesic motion with a metric of the value or action, that is to say the motion takes the path of minimum action (value or rewards or least economic activity or most efficient way to achieve the objective).  This asymmetry in stability induces a direction to time and a irreversibility to the motion.  The system will relax to states of minimum, not maximum, total energy (that is action or value).  This fact can not be altered.  What can be altered is the dynamics (topology) so that the state of maximum global (that is total) energy is a state of minimum global energy.  Since the Nash Equilibriums are the states of minimum global energy, the topology must be modified so that the global energy maximums are global energy minimums, that is Nash Equilibriums.  The topology is modified and fortified by application of the conservative stabilization force.  The control force of Eq.~\eqref{reward.force.eqn} enhances the geometry by changing the location of these new desirable Nash Equilibriums.  This theory recognizes that systems fundamentally minimize costs, but one person's costs are the another person's rewards.  In a game, to maximize personal rewards, the game must be played (that is modified) to maximize all players rewards.  If the game is played to greedily minimize personal costs, that is to minimize all player's rewards, personal rewards will be minimized -- the unmodified Nash Equilibrium.  This is viewing the system as a zero sum game instead of a ``rising tide floats all ships'' situation.  The same is true of negotiations.  The best result is a win-win solution, not a win-lose solution.  The win-lose situation is really a lose-lose situation.

From the perspective of the electronic currency management firm, the fiscal (investment and operations) and monetary policy has two parts.  The first is changing the economic potential or reward $R(q)$ to include all benefits and costs to society through application of a control force $F_c(q)$.  The second is stabilizing the economic equilibrium (that is fiscal and monetary policy $\pi^*(q)$) and reducing the economic fluctuations about the equilibrium by applying a stabilizing force, $F_\text{sf}(P)$ or more likely $F_\text{sp}(\tau)$, through arbitrage trading.

The stabilizing and cooling force is constructed to ``fine'' any malicious attempt to excite the economic system for financial gain such as a ``pump and dump'' scheme.  Since the controller knows $\pi^*(q)$, it will sell high as the malicious entity is pumping and will buy low as the malicious entity is dumping.

Another way of looking at this is that the controller has modified the dynamics to make the equilibrium a ground state.  An external system can only excite the conservative sub-system, putting energy into the sub-system.  The controller then de-excites the sub-system back into the ground state, taking energy out of the sub-system.  The net result is a flow of energy from the external system to the controller -- a heat pump of energy from the external system to the controller.  The concept of the ``pump and dump'' is a very interesting and deep subject that is discussed in detail in Sec.~\ref{faser.sec}.

The problem with current attempts to control complex systems is not knowing what reward $R_0(q)$ the system is naturally optimizing, and not knowing the equilibrium point $P^*$ of the system optimizing $R(q)$ (equivalently the equilibrium value $V^*(q)$ or the equilibrium policy $\pi^*(q)$).  Knowing both $F_0(q)$ and $\pi^*(q)$ are essential to controlling the system to optimize $R(q)$ and to be stable with minimum fluctuations about the equilibrium where the objective $V(q,P,\tau)$ is optimized.  The conservative force that must be applied is $F_c(q)$ and the external stabilizing and cooling force is $F_\text{sf}(P)$.  A simple way to state this is that the controller needs to know what to control about.  In this case, it is $F_0(q)$ and $\pi^*(q)$ or $P^*$.  For the ponderomotive control with $F_\text{sp}(\tau)$, current attempts at control do not know the canonical transformation generated by $W(q,P)$ or the characteristic spectrums $\left|\beta_i(z)\right>$ that need to be applied to the fields, as will be discussed next in this section.

\begin{figure}
\noindent\includegraphics[width=\columnwidth]{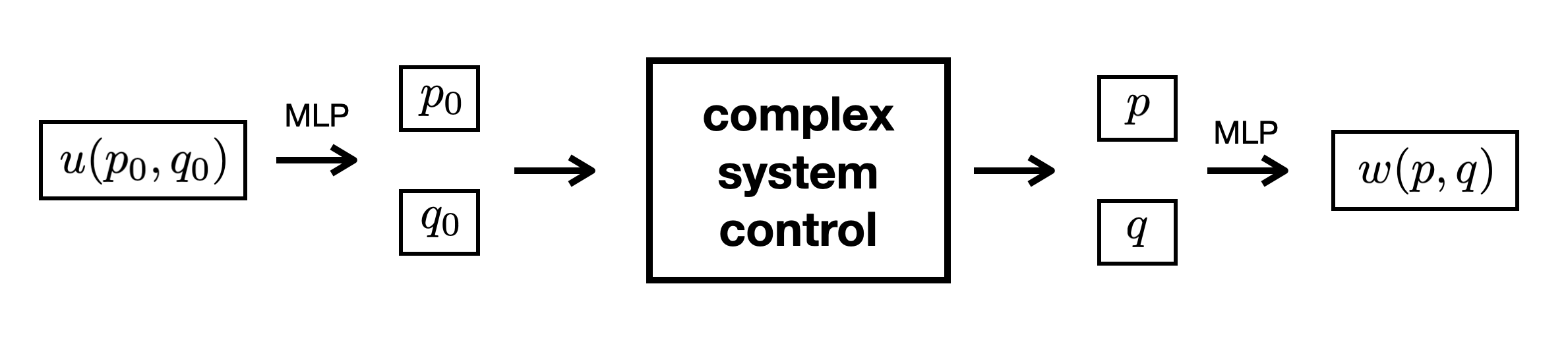}
\caption{\label{controls.outputs} Diagram of how to include input variables $u(p,q)$ (which includes control functions) and output variables $w(p,q)$ into the control system workflow. MLP is a Multi Layer Perceptron.}
\end{figure}

The inputs (which includes controls) and outputs of the control system may not be $q$ and $p$, but functions $u(p,q)$ for inputs and functions $w(p,q)$ for outputs.  A straight forward addition can be made to the workflow as shown in Fig. \ref{controls.outputs}.  An MLP that approximates the functions $p(u)$ and $q(u)$ should be added before the control system, and another MLP that approximates $w(p,q)$ should be added after the control system.

\begin{figure}
\noindent\includegraphics[width=\columnwidth]{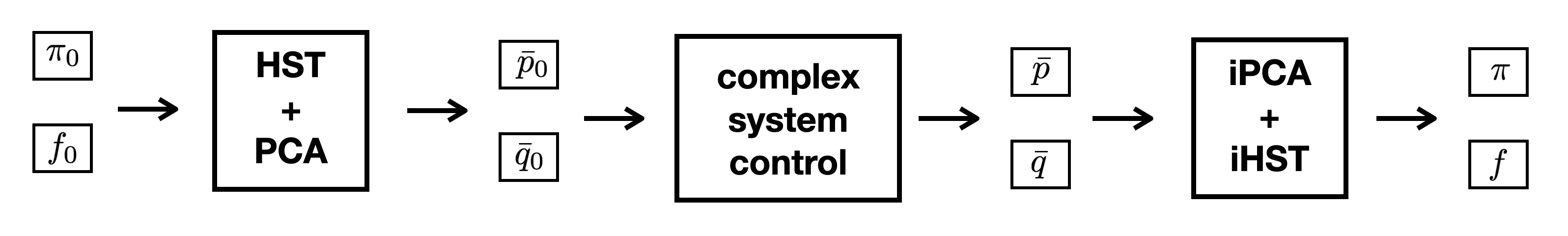}
\caption{\label{HST.app} Diagram of how the Heisenberg Scattering Theory (HST) and a Principle Components Analysis projection (PCA) are incorporated into the complex control system workflow.  The inverses are indicated as iHST and iPCA, respectively.}
\end{figure}

Now to the postponed, but important issue, that the system does not present itself in the low finite dimensional dynamical state $q$ and canonically conjugate momentum $p$ or co-state of the individual, but as an infinite dimensional field $f(x)$ and its canonically conjugate field momentum $\pi(x)$ of the collective \citep{glinsky.24b,glinsky.24d}, where $x$ is the base manifold or space with metric $\tau(x)$.  Another way of looking at this is that nature presents itself as a convoluted form of the dynamical variables that needs to be de-convoluted or decoded into the dynamical variables.  It is very common that $x=t$ and $\tau(t)=t$, but could be as sophisticated as $x=(x,y,z,t)$ and $\tau(x,y,z,t)=\sqrt{t^2-(x^2+y^2+z^2)/c^2}$ (that is relativistic dynamics on space-time). The transformation from the infinite dimensional Hilbert space with coordinates $(\pi,f)$ to the finite dimensional $\mathbb{C}^n$ space with coordinates $(p,q)$, has been an ongoing challenge to physics and complex system analysis.  It has manifested itself as the renormalization challenge to physicists that was first addressed by Ken Wilson \citep{wilson71} and has been the subject of two Nobel prizes.  Paul Dirac felt that renormalization is ``not a logical mathematical process'' \citep{dirac.82}.  There have been recent developments on this subject that have approached renormalization as a logical mathematical process, culminating in the Heisenberg Scattering Transformation (HST).  The HST is a closed form, specified Convolutional Neural Network (CNN) or Wavelet Conditional Renormalization Group (WC-RG) \citep{marchand22} which is a transformation from canonical $(\pi,f)$-field phase space to $\mathbb{C}^n$ with coordinates $\beta$.  When combined with the specification of the analytic function $H(\beta)$ or the equilibrium points (topological homology classes) $\beta^*$, it is a Generative Pretrained Transformer (GPT) \citep{farimani23}, and when further deployed as a controller it is a Deep Reinforcement Learning (DRL) \citep{bertsekas96,sutton18,yoon21,wang21,mnih15}.  The HST will be used as a pre-processor, and the inverse HST (iHST) will be used as a post-processor to the previously discussed analysis, as shown in Fig. \ref{HST.app}.  A detailed discussion of the relationship of our optimal control methodology to DRL can be found in Sec.~\ref{traditional.DRL.sec}, where particular attention is paid to practical differences between our optimal control methodology and DRL.

Details of the HST can be found in \citet{glinsky23c}.  The equation for the HST is
\begin{equation}
\label{hst.eqn}
    \boxed{S_m[f(x)](z) = \phi_{px} \star \left( \prod_{k=0}^{m}{\text{i} \ln R_0 \psi_{p_k} \star} \right) \text{i} \ln R_0 f(x),}
\end{equation}
where
\begin{equation}
   R_0(z) \equiv \frac{1}{\text{i}} h^{-1}(2z/\pi),
\end{equation}
$h(z) \equiv (z+1/z)/2$, $z \equiv p+\text{i} \, x$ and
\begin{equation}
    \psi_p \star f(x) = \int{\psi_p(x') \, f(x-x') \, dx'}
\end{equation}
is the wavelet transform where $\psi_p(x)$ are a normalized, orthogonal, localized and harmonic (that is coherent states) such that
\begin{equation}
    \psi_p(x) \equiv p^2 \,\psi(px),
\end{equation}
and
\begin{equation}
    \phi_{px}(x') \equiv p^2 \, \phi(p(x'-x)).
\end{equation}
The functions $\psi(x)$ and $\phi(x)$ are the Mother and Father wavelets that satisfy the Littlewood-Pauley condition.  

The motion, at this point, has been transformed into the space of all possible local complex spectrums $\left|\beta(z)\right>$ (that is, all possible solutions to the renormalization group equations).  This expansion or domain can be interpreted several different ways in addition to as solutions to the renormalization group equations.  It can be interpreted as Heisenberg's S-matrix expansion, as the $m$-body scattering cross sections, as the $m$-body Green's functions, and the Mayer Cluster Expansion.  This is why it has been called the Heisenberg Scattering Transformation.  It captures the $m$-body correlation structure of the collective motion.  This correlation synchronizes the collective motion so that the collective acts as one.  Motion of the system (or variation in $H(z)$) will be confined to a $n$-dimensional complex linear subspace $\mathbb{C}^n$ of the individual motion with basis vectors $\left|\beta_i(z)\right>$ easily identified by a Principal Components Analysis (PCA).  The $\left|\beta_i(z)\right>$ are the solutions to the renormalization group equations for the field theory with Lagrangian functional $L[f(x),\dot{f}(x)]$.  They are also the Taylor expansion coefficients of the action, that is the S-matrix ($S_m$) given in Eq.~\eqref{smatrix.eqn}.  The motion is projected onto this complex linear sub-space with complex coordinates $\beta=(\{\beta_i\})$.

The connection of the HST detailed in Eq.~\eqref{hst.eqn} to CNNs can be seen by identifying the iterative deep convolutional structure of the product, the nonlinear activation function $\text{i} \, \ln R_0$, and the pooling operation $\phi_{px}$.  The HST followed by the PCA projection to $\beta$ also can be interpreted as the Wigner-Weyl transformation done right \citep{case08,wigner32,weyl50}.

It should be noted that the HST followed by the PCA projection to $\beta$, leads to an analytic $H(\beta)$ with a few discontinuities $\beta^*$, in contrast to the original field which is normally very discontinuous.  Not making the HST has led to a mathematical industry of diffusive regularizations of the HJB equations called viscosity solutions \citep{bardi97}.  Unfortunately, the viscosity solutions are based on regularizing the solution by superimposing a diffusive model of risk as will be shown in Sec.~\ref{resistive.HJB.sec}.  The result is an optimization with the diffusive risk plus the true conservative sub-system risk.  Since $H(\beta)$ is analytic, $q(\tau)$ and $V(q,P,\tau)$ will be very continuous, $\text{C}^\infty$ except for a few points $\beta^*$ -- no regularization is needed.  Regularization has been an important part of renormalization in the historical way that it has been done in physics, based on the original ideas of Ken Wilson.  The second Nobel Prize was for dimensional regularization by \citet{veltman72} (the first Nobel Prize went to Ken Wilson).  The use of the HST in the renormalization regularizes the solution as it reduces the dimension by collecting all the singularities at a few points $\beta^*$ of the analytic $H(\beta)$.

What is the solution to the HJB equation given by $W(q,P)$ and $E_P(P)$, or more practically by $P(q,p)$, $Q(q,p)$ and $E_P(P(q,p))$?  These $2n+1$ coordinates are \textbf{the} Reduced Order Model (ROM) of AI.

A fundamental confusion has been thinking that $f(x=t)$ is $q(\tau=t)$ when it is not.  Whereas $q(t)$ which typically has $n$ equal to 2 to 8 dimensions with stochastic (non-integrable or chaotic), yet differentiable, Hamiltonian (governed by $H(\beta)$) dynamics characterized by a few singularities ($\beta^*$ or topological homology classes), $f(t)$ is a convolutional projection of these Hamiltonian dynamics and singularities onto an infinite dimensional Hilbert space.  What are simple isolated singularities on $\mathbb{C}^n$ are spread throughout the Hilbert space.  This has lead to renormalization procedures that have had to be regularized to collect the singularities into the homology classes of the underlying topology.  A $q(t)$ has been formally identified in models such as the three dimensional Lorenz system \citep{lorenz63}, but despite this there has continued to be confusion on how a seemingly one scalar field model $f(t)$ can be so discontinuous.

\begin{figure}
\noindent\includegraphics[width=20pc]{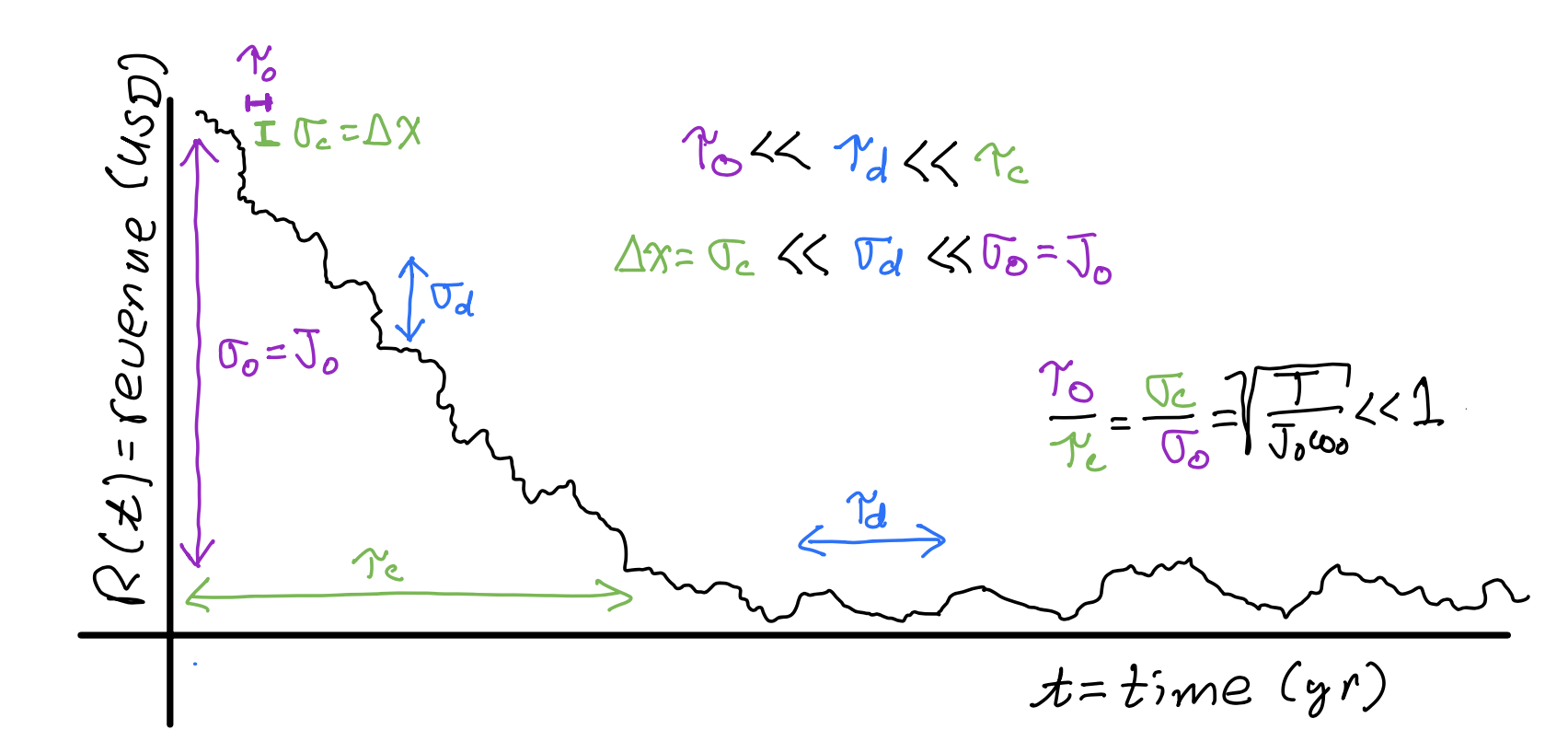}
\caption{\label{diffusion.fig} The three different diffusion scales (collisional=$c$=blue, drift=$d$=green, and ground state=$0$=magenta) are shown.}
\end{figure}

Given the theoretical structure developed in this section, the theoretical origins of the diffusive force can be elucidated.  First we need to take a closer look at the approximations that are made in the derivation of the Fokker-Planck equation.  There are three scales in the problem:  (1) the collision scale that will be identified by the subscript $c$, (2) the drift scale identified by the subscript $d$, and (3) the system ground state scale identified by the subscript $0$.  These scales are illustrated in Fig. \ref{diffusion.fig}.  The frequencies or times, and $J$ have the following ordering
\begin{equation}
    \nu_c \ll \omega_d \ll \omega_0
\end{equation}
or equivalently
\begin{equation}
    \tau_c \gg \tau_d \gg \tau_0
\end{equation}
and
\begin{equation}
    \Delta x = \sigma_c \ll \sigma_d \ll \sigma_0 = J_0 = E_0 / \omega_0,
\end{equation}
where $\tau_c \equiv 1/\nu_c$, $\tau_d=1/\omega_d$, $\tau_0=1/\omega_0$, $J_0$ is the action of the ground state, $\omega_0$ is the frequency of the ground state, and $E_0$ is the energy of the ground state.  The other variables have been previously defined in Sec.~\ref{consequences.sec}.  The fundamental assumption is that the collision is adiabatic, that is
\begin{equation}
    T \ll J_0 \omega_0 = E_0.
\end{equation}
Given this assumption, it can be shown that
\begin{equation}
    \frac{\nu_c}{\omega_0} = \frac{\sigma_c}{J_0} = \sqrt{\frac{T}{J_0 \, \omega_0}} \ll 1
\end{equation}
so that
\begin{equation}
    \kappa = \nu_c \omega_c^2 = \left( \frac{T}{J_0 \, \omega_0} \right)^{3/2} \omega_0 \, J_0^2,
\end{equation}
\begin{equation}
    \nu_c = \left( \frac{T}{J_0 \, \omega_0} \right)^{1/2} \omega_0,
\end{equation}
\begin{equation}
    \sigma_c = \left( \frac{T}{J_0 \, \omega_0} \right)^{1/2} J_0,
\end{equation}
\begin{equation}
    \sigma_d = \frac{\omega_d}{\omega_0} J_0,
\end{equation}
and
\begin{equation}
    \omega_d = \omega(J) = \frac{\partial H(J)}{\partial J}.
\end{equation}
In general, the inequality $\omega_d \ll \omega_0$ is satisfied.  Sometimes the gradients of $H(J)$ are so steep that $\omega_d \gtrsim \omega_0$.  In this case, the rate is limited so that $\omega_d = \omega_0$ and the dynamics is called ballistic.  This is a very common practice in high temperature plasma physics.  Finally, it is instructive to rewrite the Fokker-Planck equations as
\begin{equation}
\label{fp2.eqn}
    \frac{\partial f(J,t)}{\partial t} = \nu_c \frac{\partial}{\partial J} \Bigl( \sigma_d \, f \Bigr) + \nu_c \sigma_c^2 \frac{\partial^2 f}{\partial J^2}.
\end{equation}
The first term on the right represents the collisional slowing down of the system or mean reversion at a rate $\nu_c$ until it reaches the thermal equilibrium $f_\text{eq}(J)=\exp{(-H(J)/T)}$ or the mean.  The second term on the right represents the diffusion of the system.

Now, consider an external force exerted on a economic sub-system coming from an external economic heat bath of economic temperature $T_e \ll J_0 \, \omega_0$, where $J_0$ is the economic activity of the economic sub-system and $\omega_0$ is the billing frequency of the economic sub-system.  This leads to a diffusion in $P$ governed by the Black-Scholes or Fokker-Plank equation shown in Eq.~\eqref{fp2.eqn}, where $\nu_c$ is the rate at which steps are taken in the economic sub-system of size $\sigma_c$ due to the external interaction.  The equilibrium distribution will be
\begin{equation}
    f_{\text{eq}}(P) = \text{e}^{-(E_P(P)-E_P(P^*_s))/T_e}
\end{equation}
with the caveat that $P^*_s$ is a stable or stabilized equilibrium.  The status quo is that there is no stabilizing force $F_\text{sf}(P)$ or $F_\text{sp}(\tau)$ of unstable equilibriums or control force $F_c(q)$ to optimize the desired reward $R(q)$ so the economic sub-system relaxes to undesired stable equilibriums (local minimums) with $T_e$ giving the natural prime interest rate $\nu_\text{prime}=\sqrt{T_e/E_0} \, \omega_0$.  

Beyond the lack of stabilization, what is worse is that the conservative economic sub-system dynamics (natural business cycle) are modeled by increasing the economic temperature $T_e$ so that $\sigma_c=\sigma_0=J_0$.  This is done by setting the temperature to
\begin{equation}
    \frac{T}{\omega_0 \, J_0} = \left( \frac{T_e}{\omega_0 \, J_0} \right)^{1/3}
\end{equation}
so that
\begin{equation}
    \kappa = \left( \frac{T_e}{\omega_0 \, J_0} \right)^{1/2} \omega_0 \, J_0^2
\end{equation}
and
\begin{equation}
    \nu_c =  \left( \frac{T_e}{\omega_0 \, J_0} \right)^{1/6} \, \omega_0 \equiv \nu_b,
\end{equation}
where $\nu_b$ is the business cycle frequency and $\tau_b \equiv 1/ \nu_b$ is the business cycle time.  Not only is this improperly modelling the dynamics of the economic sub-system (risk), it violates one of the assumptions of the diffusive approximation (that is, $\sigma_c \ll J_0$) and double accounts (in a way that the diffusive force dominates the true dynamical force) for the risk of the economic sub-system.  These uses of the diffusive force are reinforced by central bank loans at a prime interest rate, bond financing, and the use of Discounted Cash Flow (DCF) analysis.  

What the control method of this section does is move the equilibriums from their uncontrolled natural positions to the desired positions according to the reward function $R(q)$, not $R_0(q)$, and stabilizes and cools a more desirable (with more economic activity) saddle point or semi-attractor.

We digress to note the present use of the prime interest rate $\nu_\text{prime}$ in economic control.  One of the ways that the money supply is traditionally controlled is by a central reserve bank loaning money at a set rate.  If this rate is lower than the natural prime interest rate, the money supply will increase as money is borrowed from the central reserve bank.  This is what is called quantitative easing and puts inflationary pressure on the economy.  The money supply will decrease if the set rate is greater than the prime rate.  This reduction in money supply will counter inflation.  As will be discussed in Sec.~\ref{resilience.sec}, this is a poor control system since the loaning of money causes a long-term consistent reduction in the money supply due to the coupon payments, countering the original increase in the money supply.

Even though a detailed parameterization of the fields $f(x)$ is beyond the scope of this paper, it is valuable to consider the fact that one or more of these fields will be the savings of the sub-economy, whether that be in electronic currency or inventory.  These can be viewed as the potential energy of the sub-economy where the kinetic energy of the economic activity of the sub-economy is stored.  As such, the savings rate should be $1/S_0 \sim \nu_b$, where $\nu_b$ is the frequency of the business cycle of the sub-economy and the maximum amount of savings will be
\begin{equation}
    E_0 \, \tau_b = J_0 \; \omega_0 \tau_b = J_0 \, \left( \frac{T_e}{\omega_0 \, J_0} \right)^{-1/6} \gg J_0
\end{equation}  
which is economic activity over the business cycle time $\tau_b$.  Typical business cycles are between 2-20 years.  For the example of App.~\ref{econ.model.app}, $S_0$ varies between 2-5 years.  The sub-economy of this example in the appendix is sampled every 2 years which is the nyquist frequency, implying a business cycle time of 2 years.

In contrast, if the dynamics of the sub-economy is improperly modelled by increasing the economic temperature, the savings rate should be $1/S_0 \sim \omega_0$, where $\omega_0$ is the billing frequency of the sub-economy.  The maximum amount of savings will be $J_0$, the economic activity over the billing cycle time $\tau_0$.   Typical business billing cycles are 15-60 days with much pressure being put on reduction of this cycle time (that is just-in-time inventory and fast collection of accounts receivable) by modern business practices.  For Target Energy Inc.\ of the example of App.~\ref{econ.model.app}, their savings is 15 days.  The correct savings rate in steady state is $\omega_0$, but if a significant investment is needed to reach equilibrium (that is steady state) the savings rate should by $1/ \tau_e$, where $\tau_e$ is the time to reach equilibrium.

It is interesting to note that when the collision rate is artificially increased so that $\sigma_c=J_0$, business cycles of 2-20 years would imply discount rates of 50\%-5\%.  For petroleum companies, their field lifetimes vary between 10 to 20 years implying discount rates of 10\% to 5\% -- typical Weighted Average Cost of Capital (WACC) for petroleum companies.  For Venture Capital funded startups, the average time to exit and the average life of a fund is about 5 years implying a discount rate of 20\% -- a typical required rate of return for a VC funded startup.  For New Energy Inc., the focus of the example of App.~\ref{econ.model.app}, has a business cycle time of 2 years implying a discount rate of 50\%.  The modelled rate of return of New Energy Inc.\ is 40\%.

\begin{figure}
\noindent\includegraphics[width=\columnwidth]{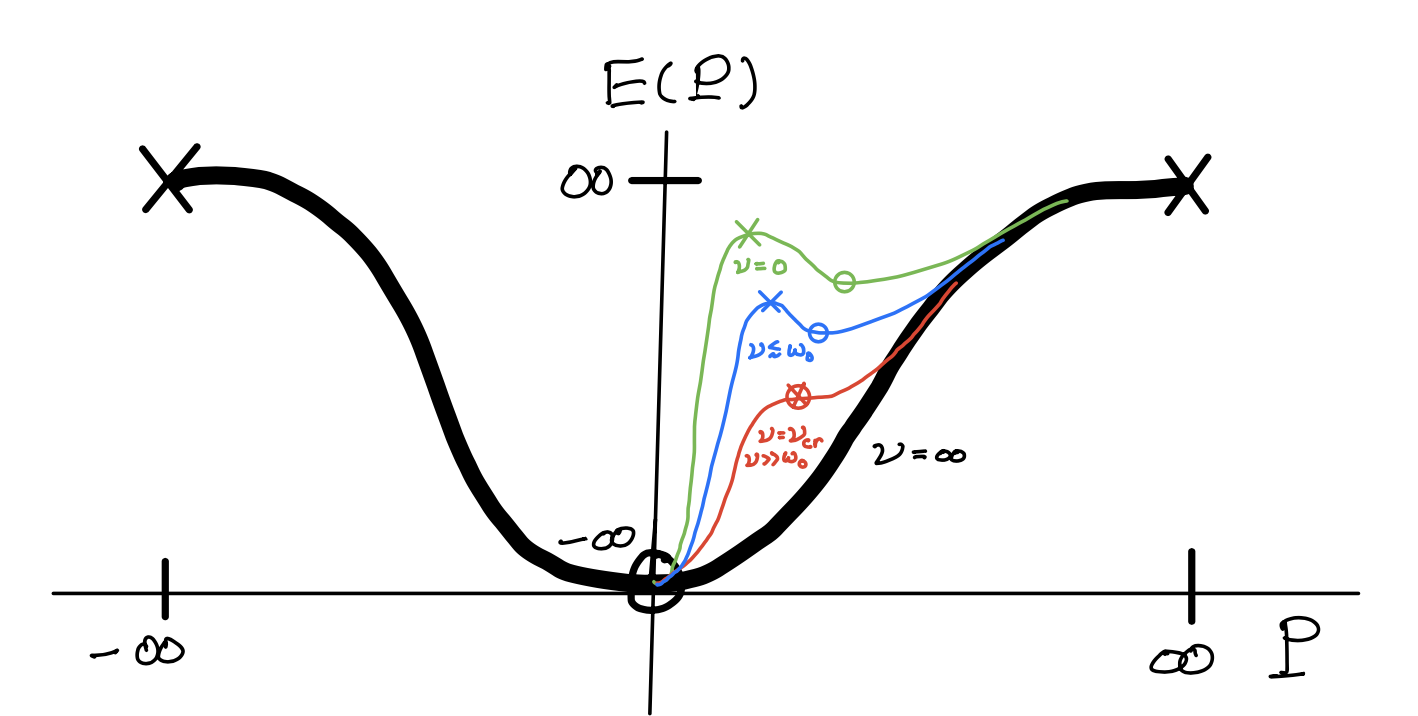}
\caption{\label{Ep.nu.fig} Graph showing the equilibriums where $\omega_Q(P^*)=0$.  Plotted is the energy $E(P)$ as a function of $P$.  The unstable local maximums are indicated by the x-points, and the stable local minimums or Nash Equilibriums by the o-points.  The energy for the system with no dissipation $\nu=0$ is the thin green line.  The heavily damped system $\nu \to \infty$ is the thick black line.  The system with modest dissipation $\nu \lesssim \omega_0$ is the thin blue line.  The system with the critical amount of dissipation that destroys the topology of the system $\nu = \nu_\text{cr}$ is the thin red line.}
\end{figure}

In order to understand all the implications of embedding resistive non-conservative dynamics in the sub-economy, the HJB theory of this section needs to be extended to include time dependant Hamiltonians which is done in Sec.~\ref{resistive.HJB.sec}.  The implications of embedding the non-conservative dynamics are significant and grave.  While the effects of the external thermal force can be mitigated and controlled, the embedded resistive force significantly changes the topology of the sub-system.  It regularizes all of the equilibriums $\beta^*$, reducing the energy of the equilibriums $E_0$ if $\nu \lesssim \omega_0$, and removing them if $\nu > \nu_\text{cr} \gg \omega_0$, so that the topology only has one point of singularity at zero with zero energy.  All trajectories inevitably spiral to zero.  This is shown in Figs. \ref{Ep.nu.fig}, \ref{P.star.nu.fig}, and \ref{E.star.nu.fig}.

\section{FASER: the LASER of financial systems}
\label{faser.sec}
\begin{figure}
\noindent\includegraphics[width=\columnwidth]{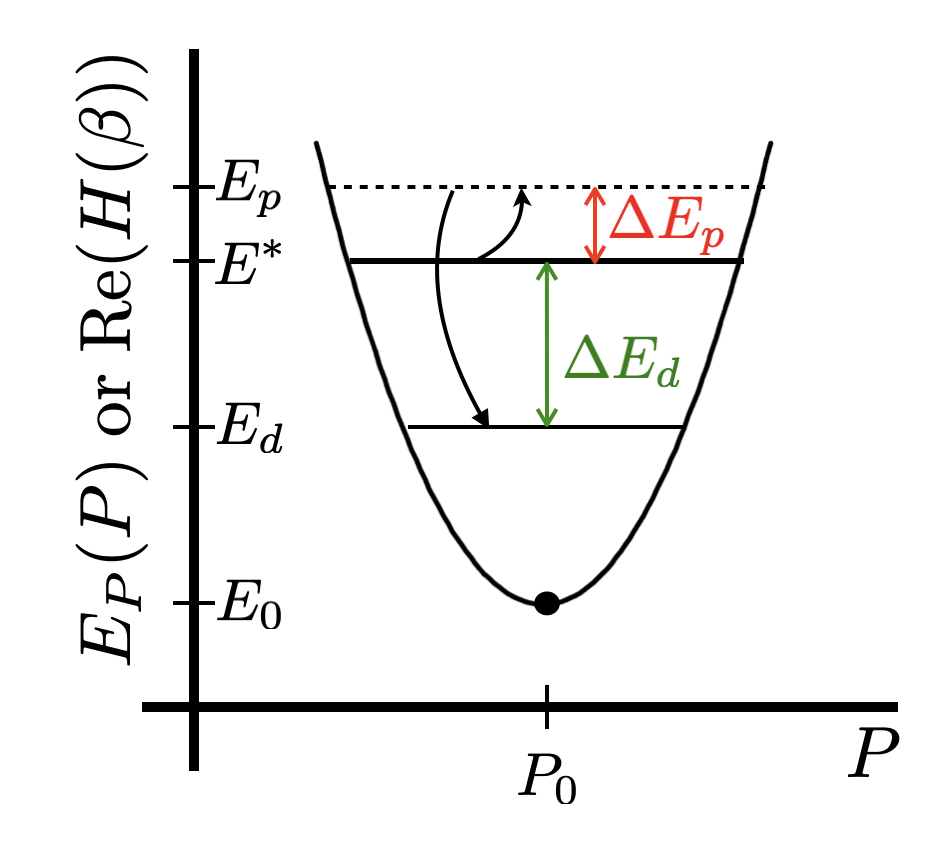}
\caption{\label{faser.fig} Shown is the architecture of a LASER or FASER (Financial Amplification by Stimulated Emission of Revenue).  Plotted is the potential energy or reward as a function of the dynamical coordinate $P$.  Displayed is the equilibrium coordinate $P_0$ and the corresponding ground state energy $E_0$.  The meta stable excited state energy is indicated by $E^*$, the energy to which the system is pumped by $E_p$, and the energy to which the system is dumped by $E_d$.}
\end{figure}

Given the detailed conservative dynamical perspective on financial systems presented in Sec.~\ref{solve.hjb.sec}, a very interesting interpretation can be made of the ``pump and dump'' financial scheme.  It is the financial equivalent of an optical LASER (Light Amplification by Stimulated Emission of Radiation) as shown in Fig. \ref{faser.fig}.  The ``pump and dump'' scheme is a Financial Amplification by Stimulated Emission of Revenue (FASER).

First, let us discuss how an optical LASER works.  Given a potential well around an atom, the ground state $P_0$ with ground state energy $E_0$ can be identified.  There exist long lived (meta stable) excited states with energy $E^*>E_0$.  The atom is put into the meta stable excited state.  The atom is then stimulated by the incident light into an even higher excited state with energy $E_p>E^*$, that is not meta stable, then the atom rapidly decays into a much lower energy state with energy $E_d<E^*$, usually the ground state.  This is a stimulated net emission of energy in light equal to the difference in energy between the initial meta stable state and the final state $\Delta E_d=E^*-E_d>0$.  The energy put into stimulating the system is the difference in energy between the higher excited state and less excited state $\Delta E_p = E_p - E^*$>0.  Finally the net gain in energy is
\begin{equation}
    E_\text{gain} = \Delta E_d - \Delta E_p.
\end{equation}
For a LASER, $\Delta E_p=0$ and $\Delta E_d = E^* - E_0$, so that $E_\text{gain} = E^*-E_0>0$.  If the LASER is not in an excited state, then $E_\text{gain}=-\Delta E_p<0$ so that there is negative gain and the LASER does not work.

The financial FASER works the same way.  The financial sub-system needs to be in a meta stable excited state, enjoying the irrational exuberance of a bull market.  This meta stable state is likely an unstable saddle point or semi-attractor that takes an infinite amount of time to approach since $\omega_Q^*=0$.  The market is then pumped into an even more excited bull state that is so over priced that it triggers a bear market and the market falls back into its ground (that is) stable state.  The financial gain (or profit/loss) is $E_\text{gain}$ which is positive for $\Delta E_p < \Delta E_d$.  If the system is already in the ground state $E_\text{gain} = - \Delta E_p<0$ -- a guaranteed money losing proposition pumping energy from the entity doing the ``pump and dump'' to the financial sub-system.

\section{Non-conservative HJB Equation}
\label{resistive.HJB.sec}
It is very common to embed a non-conservative force in the sub-system dynamics instead of applying an external force to the sub-system.  The effect on the dynamics of embedding a non-conservative force rather than applying an external force is dramatic.  Applying an external resistive force to a conservative sub-system simply relaxes the sub-system to the equilibriums of the conservative dynamics where the external system no longer has an influence on the dynamics and the sub-system follows the conservative dynamics.  The influence of the external resistive force, in fact any external force, can be both mitigated and controlled.  Embedding a non-conservative force fundamentally degrades the geometry, and eventually changes and destroys the topology of the dynamics.  If the embedded non-conservative force is resistive, the topology is regularized.  If the regularization is small $\nu \lesssim \omega_0$, then the topology is preserved, but the energy of the equilibriums $E_0$ are reduced as shown in Fig. \ref{E.star.nu.fig}.  If the regularization is large $\nu > \nu_\text{cr} \gg \omega_0$, the topology is reduced to one with a single homology class of a point so that trajectories inexorably spiral to the point at zero with zero energy.

Another way of looking at this is that:  but for the external force, the dynamics of the sub-system are conservative.  No matter how fast the deformation of the dynamical trajectory by the external force, the trajectory must still remain on the dynamical manifold $\mathbb{H}$ restricted by the topological obstructions $\beta^*$ -- $E$ changes not $H$.  By embedding the non-conservative force in the dynamics, the form of $H$ is being changed -- fundamentally changing the geometry and eventually the topology of the manifold $\mathbb{H}$.

To understand what is happening to the dynamics when the non-conservative force is embedded, the HJB equation needs to be modified to include this force.  The common way that this is done results in the following equation
\begin{equation}
\label{hjb.nu.eqn}
    \frac{\partial V(q,P,\tau)}{\partial \tau} + H(\partial V / \partial q,q) - \nu \, V = 0
\end{equation}
where $\gamma$ is the discount factor, and $\nu \equiv (1-\gamma)/\Delta \tau$ is the discount rate.  Another way this can be done is to make the Hamiltonian an explicit function of time yielding
\begin{equation}
    \frac{\partial V(q,P,\tau)}{\partial \tau} + H(\partial V / \partial q,q,\tau) = 0
\end{equation}
where $H(p,q,\tau)=p \, f(p,q) +\text{e}^{-\nu \tau} R(q)$.  The time-dependant Hamiltonian does not need to be limited to this form where $H(p,q,\tau)=p \, f(p,q) + g(\tau) \, R(q) \, $.  It can be any function $H(p,q,\tau)$.

It is interesting to note that the value functional corresponding to this diffusive form of $H(p,q,\tau)$ can be written as
\begin{equation}
    V[q(\tau)] = \int{\text{e}^{-\nu \tau} \, R(q(\tau)) \, d\tau}.
\end{equation}
A constrained optimization of this $V[q(\tau)]$ gives the expression for NPV found in Eq.~\eqref{npv.eqn}.

If $\nu=0$ the time-dependant HJB equation reduces to the separable Eq.~\eqref{hjb3.eqn}, with solution given by Eq.~\eqref{sep.V.hjb.eqn}, Eq.~\eqref{wQ.hjb.eqn}, Eq.~\eqref{P.hjb.eqn} and Eq.~\eqref{Q.hjb.eqn}.  The effect of the time dependency is to make the dynamics non-Markovian with the reward function being a functional of the path $H[q(\tau)]$, that is an explicit function of time $H(p,q,\tau)$.  The time-dependant HJB equation can still be solved for $V(q,P,\tau)$, Hamilton's principal function or action $\int{p \, dq - H \, d\tau}$, instead of $W(p,q)$ and $E(P)$.  The action $V(q,P,\tau)$ can be used to generate the canonical transformation to $P(p,q,\tau)$ and $Q(p,q,\tau)$ which are constants of the motion.  This can be done with the help of the relationships
\begin{equation}
    p=\frac{\partial V}{\partial q}=p(q,P,\tau)
\end{equation}
and
\begin{equation}
    Q=\frac{\partial V}{\partial P}=Q(q,P,\tau).
\end{equation}
The solution is then given by
\begin{equation}
    p=p(P_0,Q_0,\tau)
\end{equation}
and
\begin{equation}
    q=q(P_0,Q_0,\tau)
\end{equation}
where $P_0=P(p_0,q_0,\tau_0)$ and $Q_0=Q(p_0,q_0,\tau_0)$.

The inclusion of the dissipation term $-\nu J$ in the dynamics is regularizing (removing) all the singularities $\beta^*$, eventually, as $\nu \to \infty$, changing the topology to one with a single homology class of a point so that all trajectories spiral to zero with zero energy.

Embedding the time-dependant dissipation in the economic sub-system is practically done through resistive (non-conservative) debt financing.  This is equivalent to assuming that there is a probability $\alpha=1-\gamma$ that the game ends after each step, or that the reward is decreased by $\gamma$ for each step.  This is not the case.  The game always goes on.  If a business fails, the assets are transferred to another business, and the employees go to work for another business.  If a person dies, their offspring take their place in the work force.  When one game of chess finishes, another starts.  Time keeps marching on.  There is a future.

\section{Relationship to Deep Reinforcement Learning}
\label{traditional.DRL.sec}
Much of the theory and application of Deep Reinforcement Learning (DRL) and the very successful Deep Q-Learning (DQN) is based on the discrete parameterized Bellman equation
\begin{equation}
\label{bellman.eqn}
\begin{split}
    \widetilde{V}(s;\theta) &= \min_u{E[g(s,a,s') + \gamma \, \widetilde{V}(s';\theta)]} \\
    &= E_{a=\widetilde{\pi}(s;\theta)}[g(s,a,s') + \gamma \, \widetilde{V}(s';\theta)] \\
    &= g(s,a=\widetilde{\pi}(s;\theta),s') + \gamma \, \widetilde{V}(s';\theta),
\end{split}
\end{equation}
where $s$ is the state, $a$ is the action taken, $\theta$ are the approximation parameters, $\widetilde{V}(s;\theta)$ is the approximate value function, $\widetilde{\pi}(s;\theta)$ is the approximate policy, $\gamma$ is the discount factor, $E[g(x)]$ is the expected value of $g(x)$ which is equal to $\int{g(x) \, f(x) \, dx}$, and $g(s,a,s')$ is the reward of taking action $a$ and going from state $s$ to $s'$.  The approximate Q-function is
\begin{equation}
\label{Q.def.eqn}
    \widetilde{Q}(s,a;\theta) = E[g(s,a,s') + \gamma \, \widetilde{V}(s';\theta)],
\end{equation}
where
\begin{equation}
\label{V.exp.eqn}
    \widetilde{V}(s;\theta) = \min_u{\widetilde{Q}(s,a;\theta)=\widetilde{Q}(s,a=\widetilde{\pi}(s;\theta),\theta)}.
\end{equation}
To find the optimum value and policy, $\widetilde{V}(s;\theta)$ is relaxed to a minimum
\begin{equation}
    V^*(s) = \min_\theta{\widetilde{V}(s;\theta)} = \widetilde{V}(s;\theta^*)
\end{equation}
with the optimum policy $\pi^* = \widetilde{\pi}(s;\theta^*)$.  

The art is in how to relax the value and policy to the optimum at $\theta=\theta^*$.  The most successful strategy has been Q-learning.  This strategy is based on the following equation obtained by substituting the expression for $\widetilde{V}(s;\theta)$ in Eq.~\eqref{V.exp.eqn} into the definition of $\widetilde{Q}(s,a;\theta)$ in Eq.~\eqref{Q.def.eqn}
\begin{equation}
\begin{split}
    \widetilde{Q}(s,a;\theta) &= E[g(s,a,s') + \min_{a'}{\widetilde{Q}(s',a';\theta)}] \\
    &= \widetilde{Q}_\text{target}(s,a,;\theta).
\end{split}
\end{equation}
The value of $\theta$ is relaxed to the minimum by the following step
\begin{equation}
\label{Q.learning.eqn}
\begin{split}
    \theta_{k+1} = \theta_k - \alpha \nabla_\theta &E_{a=\widetilde{\pi}(s;\theta_k)+\varepsilon_a} \\
    &[(\widetilde{Q}(s,a;\theta) - \widetilde{Q}_\text{target}(s,a,;\theta))^2]
\end{split}
\end{equation}
where $\alpha<1$ is the learning rate, and $\varepsilon_a$ is a small random action.  The action is kept on policy by minimizing the approximate Q-function
\begin{equation}
    \widetilde{\pi}(s;\theta_k) = \arg \min_a{\widetilde{Q}(s,a;\theta_k)}.
\end{equation}
The small actions $\varepsilon_a$ are added so that the directions perpendicular to the policy (off-policy) are sampled so that the gradient can be approximated.  This algorithm is proceeded by a CNN that tries to calculate the fundamental dynamical variables (the states $s=q(\tau)$ and actions $a=p(\tau)$) as a functional transformation of fields $f(t)$ and canonically conjugate field momentums $\pi(t)$.  It is followed by a CNN that tries to calculate the fields and field momentums as a functional transformation of the states and actions.

Before the differences between this traditional approach and the methodology proposed in this paper for optimal control can be understood, the relationship of the discrete Bellman equation shown in Eq.~\eqref{bellman.eqn} to the continuous HJB shown in Eq.~\eqref{hjb.nu.eqn} needs to be established.  Start with the Bellman equation, Eq.~\eqref{bellman.eqn}, and rearrange terms to give
\begin{equation}
\begin{split}
    \widetilde{V}(s';\theta) - \widetilde{V}(s;\theta) &+ g(s,a=\widetilde{\pi}(s;\theta),s') \\
    &- (1-\gamma) \, \widetilde{V}(s';\theta)=0.
\end{split}
\end{equation}
Define 
\begin{equation}
\begin{split}
    \Delta \widetilde{V} &\equiv \widetilde{V}(s';\theta) - \widetilde{V}(s;\theta) \\
    &\equiv \Delta_\tau \widetilde{V} + \Delta_s \widetilde{V}, 
\end{split}
\end{equation}
$\Delta \tau \, R(s) \equiv g(s,a=\widetilde{\pi}(s;\theta),s')$, and $\alpha = (1-\gamma)=\nu\Delta \tau$ then rewrite the equation to give
\begin{equation}
    \Delta_\tau \widetilde{V} + \Delta_s \widetilde{V} + \Delta \tau \, R(s) - \alpha \, \widetilde{V}(s;\theta)=0.
\end{equation}
Now divide by $\Delta \tau$ to give
\begin{equation}
    \frac{\Delta_\tau \widetilde{V}}{\Delta \tau} + \frac{\Delta s}{\Delta \tau} \frac{\Delta_s \widetilde{V}}{\Delta s} + R(s) - \nu \, \widetilde{V}(s,\tau,; \theta) = 0,
\end{equation}
where
\begin{equation}
    \frac{\Delta s}{\Delta \tau} = f\left(s,a=\frac{\Delta_s \widetilde{V}}{\Delta s} \right).
\end{equation}
Identify $s=q$, $a=p$, $\theta=P$, $\widetilde{V}(s,\tau;\theta)=V(q,\tau;P)$, $\Delta_\tau \widetilde{V} / \Delta \tau = \partial V / \partial \tau$, and $\Delta_s \widetilde{V} / \Delta s = \partial V / \partial q$, then write the equation as
\begin{equation}
    \frac{\partial V(q,\tau;P)}{\partial \tau} + H\left(p=\frac{\partial V}{\partial q}, q \right) - \nu \, V = 0,
\end{equation}
the HJB equation with dissipation.  Here,
\begin{equation}
    H(p,q) = p \, f(p,q) + R(q).
\end{equation}
Now several things in DRL can be put in the context of the solution of the HJB outlined in this paper.  Identify,
\begin{equation}
    \widetilde{V}(s;\theta) = W(q,P),
\end{equation}
\begin{equation}
    \widetilde{Q}(s,a;\theta) = W(q,P(p,q)) \equiv W_P(p,q),
\end{equation}
\begin{equation}
    \widetilde{\pi}(s;\theta) = \frac{\partial W(q,P)}{\partial q} = \pi(q,P),
\end{equation}
and Eq.~\eqref{Q.learning.eqn} as
\begin{equation}
\label{P.grad.eqn}
    P_{k+1} = P_k \mp \alpha \, \frac{\partial E_P(P_k)}{\partial P}.
\end{equation}
where the sign of the update depends on whether the equilibrium is a minimum or maximum.  The optimum $P^*$ can be found more simply by solving for the roots of
\begin{equation}
    \frac{\partial E_P(P^*)}{\partial P}=0
\end{equation}
with a high performance root solver, rather than the much slower gradient decent of Eq.~\eqref{P.grad.eqn}.  Then, the optimal value function
\begin{equation}
    V^*(s) = \widetilde{V}(s,\theta^*) =  W(q, P^*) = V^*(q)
\end{equation}
and the optimal policy
\begin{equation}
    \pi^*(s) = \widetilde{\pi}(s;\theta^*) = \frac{\partial W(q,P^*)}{\partial q} = \pi(q,P^*) = \pi^*(q)
\end{equation}
can be explicitly calculated.

There are many ways that the optimal control methodology outlined in this paper is superior to DRL methodology.  The most significant is the functional transformation from the fields $f(x)$ and canonically conjugate field momentum $\pi(x)$ to the dynamical variables $q(\tau)$ and $p(\tau)$ by the HST+PCA.  This is a defined CNN with nothing to learn.  The PCA takes just seconds to calculate all that needs to be learned.  This is in contrast to searching the Banach space of all functionals.  This space is so large that an ansatz of the form of the CNN needs to be made that undoubtedly constrains the search away from the true form of the transformation.  Even with these ansatz, the CNNs have 10's of billions of parameters to learn or more.  Some of DeepMind's DRLs are taking 10's of millions of dollars to train or more.  Because of the constraint away from the complete solution, DRL does not fully renormalize the fields leaving significant amounts of singularities throughout the solution.  These remaining singularities make it very difficult to fit the functions with MLPs.  For this reason, dissipation is added to the Bellman equation to regularize the singularities, destroying the topology with grave consequences.

The MLPs and the structure of the workflow shown in Fig. \ref{HJB_NN} is matched to the structure of the solution to the HJB equation, finding surrogates of the important analytic functions such as $P(p,q)$, $Q(p,q)$, $W(q,P)$, $E_P(P)$, $\omega_Q(P)$, $\pi(q.P)$, and $F_\text{ext}(\omega_Q,Q)$.  The MLPs with ReLU activation functions are universal piece-wise linear function approximators that are very well adapted to fitting analytic functions that are maximally linear except for a limited numbers of singularities with cusp like discontinuities.  These ReLUs take 10's of seconds to train \citep{glinsky23}.  This is not done by DRL.  The approximation parameters of DRL are not matched to the dynamics in number or in form of the approximation parameters $P(p,q)$.  

While the value function $V(q,P)$ can be quickly calculated for any value of $P$ by our methodology, DRL only calculates it for $P^*$ that are stable local minimums or the unstable local maximums (saddle points).  The same is true of the policy $\pi(q,P)$, which is only calculated for $\pi^*(q)=\pi(q,P^*)$.  This enables our methodology to derive dynamical surrogates (including external forces) that can be very quickly evaluated with high fidelity.  This can not be done by current DRL.

Once these surrogates are formed, the equilibrium points $P^*$ can be efficiently found in 10's of seconds.  Not only are the stable equilibriums (that is local minimums, o-points, or attractors) found, but the more desirable unstable equilibriums (that is local maximums, x-points, saddle points, or semi-attractors) are also found.  These unstable equilibriums can then be stabilized and protected from external forces by the use of Eq.~\eqref{stable.pondermotive.eqn} or Eq.~\eqref{stable.feedback.eqn}. The reward function can also be modified as shown in Eq.~\eqref{reward.force.eqn}.  DRL uses a much slower gradient decent that needs to calculate two time-consuming minimums or maximums for each sample.

DRL finds the unstable local maximums by invoking ``CPT'' invariance.  It does this by reversing both time and ``charge'' (that is reward), then ``parity'' (that is minimums and maximums) will be reversed.  This is practically done by replacing the minimums with maximums and changing the sign of the reward or potential function, $R(q) \to -R(q)$.  While this changes the unstable local maximums to stable local minimums and vice versa, it fundamentally changes the dynamics of the system $W_{-}(q,P) \ne W_{+}(q,P) \equiv W(q,P)$.  While the unstable equilibriums $\pi^*(q)$ can now be found with the gradient decent or shrinkage method, a very different dynamical problem is being solved.  Time does not run in reverse with a reversal in reward.

DRL finds the optimum policy $\pi^*(q)$, not the control forces, $F_c(q)$, $F_\text{sf}(P)$, and $F_\text{sp}(\tau)$.  The optimum policy is neither single valued nor existing for all $q$.  This manifests itself in the DRL algorithm as a lack of convergence, lack of stationarity, and lack of IID (Independent and Identical Distribution).  DQN resolves these problems by an opaque and laborious procedure where
\begin{equation}
    \Delta \pi(q,P) = \frac{\partial W_{+}(q,P)}{\partial q} - \frac{\partial W_{-}(q,P)}{\partial q}
\end{equation}
the feedback stabilization force is learned, so that the force $\Delta p / \Delta \tau = - \omega_\text{sf} \Delta \pi(q,P)$ can be applied.  It should be noted that the inputs to DQN are both $f(x)$ and $\pi(x)$, not just $f(x)$.  This is done by inputting the field, effectively $q$, at different times so that $\dot{q}$, effectively the co-state $P$, can be estimated.

It is straight forward to construct training datasets using our methodology that sample phase space well and sufficiently explore off-policy.  This can be done by integrating trajectories with a small external dissipative force.  It is an exceptionally difficult task for DRL to keep most of the steps on-policy and simultaneously exploring off-policy, but not to far off-policy, for the determination of the adjoint motion.  The challenge can be seen in the form of Eq.~\eqref{Q.learning.eqn}, where $\widetilde{Q}_\text{target}(s,a,;\theta)$ is neither stationary nor IID.

The diffusion from the external force is modelled correctly with the $F_\text{ext}(\omega_Q,Q)$, retaining the geometry and topology of the conservative Hamiltonian system using our methodology.  It is not embedded in the dynamics as shown in Eq.~\eqref{hjb.nu.eqn}, first degrading the geometry, then destroying the topology of the conservative Hamiltonian system.  The regularization, if it is small $\nu \lesssim \omega_0$, leads to a geometric degradation, significantly reducing $E_0$.  If it is large $\nu > \nu_\text{cr} \gg \omega_0$, it leads to topological destruction and all trajectories spiraling to zero with zero energy.  The ubiquitous use of the discount factor $\gamma$ in DRL, to improve convergence and regularize the solution (needed because of the incomplete reduction to the dynamical variables $q(\tau)$) has these unintended effects.

\section{Quantum field nature of financial systems}
\label{quantum.sec}
The quantum field structure of financial systems comes about from real-time or space-like measurement limitations to those of a statistical nature, so that the Born Rule is applicable.  The Born Rule states that if a statistical measurement is made on a system, the measurement must exert a force on the system that will change the system dynamics -- the measurement is destructive.  Only statistical, that is destructive measurements, can be made on relativistic systems, systems that present themselves in the convoluted field domain of the collective, or systems where non-destructive observations are not available.  Practically, this comes about from the time that it takes to communicate, sum, or assimilate the measurements.  For instance, financial statements are only issued quarterly or annually, in a retrospective nature.  Even if they are reported real-time, they are still only retrospective.  The destructive measurement forces are most often non-conservative viscous forces, but can be conservative forces that erect a potential barrier that redirect, not extract energy from, the dynamical trajectory.

To measure, that is know, the state of a field, a measurement needs to be made of the future as well as the past.  That is not possible in real-time.  That is why all financial statements say that past performance is not a prediction of the future.  Technically, what they should say is that a measurement of the past is not a measurement of the future.  There is a predictability, that is a reality, to the financial dynamics, that is a predictability to the financial performance.  Although real-time non-destructive measurements can not be made of the state, so that feedback control is very difficult if not impossible, non-destructive retrospective observation of the system is possible.  From this the topology can be learned, that is $\beta^*$ or $H(\beta)$.  Given this knowledge of the topology $\beta^*$, the complex curvatures (that is, the S-matrix $S_m$) of the dynamical manifold, the complex financial system dynamics, or reality;  the topology can be modified by a ponderomotive force and the dynamics can be predicted to follow the geodesics of the dynamical manifold.  Philosophically, the ability to ponderomotively control a system is experimental proof that there is a reality.

In the physics literature, this is referred to as entanglement of the measurement with the geometry and ultimately the topology if the measurement force is large enough, and gives rise to the concept of Schrödinger's Cat.  Schrödinger's Cat is either alive or dead irregardless of whether it is being observed.  There is a reality.  A non-destructive retrospective observation can be made of whether it was alive.  From these non-destructive retrospective measurements, it can be learned that the cat has been alive for five years and that cats live for about another eight years if they have been alive for five years.  From this knowledge, a prediction can be made that it is likely that the cat will live for another eight years.  An example of a destructive statistical measurement that exerts a force on the system is stabbing the cat so that if the cat is alive it falls to the ground and dies.  If it was already dead nothing changes.  If the cat falls to the ground, the cat was alive.  If nothing changes, it was dead.  In either case, it now can be predicted with certainty that the cat will be dead in the future.  Any cat passing where the measurement is being made is killed.  Stabbing the cats has exerted a force on the system changing the dynamics, that is it has reduced the life span of cats.  The destructive measurement is not measuring the original reality, it is measuring an altered reality.  There is an entanglement of measurement and reality.

An example of a destructive measurement for a financial system would be to measure the money supply by significantly changing the money supply, then observing the change in economic activity.  The larger the change that is made to the money supply, the better it is known what the money supply will be.  The more times that the cat is stabbed, the more certain it is that it will be dead.

The counsel that this discussion gives the electronic currency management firm, is to centrally manage and carefully observe the electronic currency and sub-economy.  This precludes the use of distributed and un-observable (secret) crypto-currencies.  The use of a crypto-currency would preclude non-destructive observation of the system, and would require destructive statistical measurement.  

The destructive statistical measurement for `proof-of-work'' crypto-currencies is the physical work that needs to be done, that is energy dissipated in computers, to change the money supply.  For ``proof-of-stake'' crypto-currencies, it is the financial dissipation that needs to be paid to maintain the money supply.  Both of these statistical measurements change the system by embedding internal dissipation into the system.  The secrecy comes at a measurement cost, that adds significant friction to the system.  This cost is not alleviated by ``proof-of-stake'' crypto currencies.  The dissipation has simply been transferred from the real physical dissipation of energy in computers associated with ``proof-of-work'' currencies, to the financial dissipation of the interest payments due the ``proof-of-stake'' currencies.

The current banking system suffers from an inability to make detailed non-destructive observation of the system.  It results in many different money supplies $M$, that are all partial or indirect estimates of the true money supply, at best.  The true money supply must be destructively measured as discussed earlier in this section.   It is aggravated by the liquidity uncertainty generated by the banks not holding most of the balance of its customers savings accounts in cash.  This requires that the depositors treat their deposits statistically, and make statistical measurements.  The common statistical measurement of the depositors is requiring the bank to make regular interest payments.  In more extreme cases, it can result in the depositors making a very destructive measurement, a bank run where the depositors require the bank to return all their deposits, that results in bank failures, that is bank death, as discussed in Sec.~\ref{resilience.sec}.  In both cases, dissipation is added to the economy, degrading the economy's geometry and performance, and in extreme cases destroying the economy's topology resulting in severe economic recession.   This is a completely avoidable situation.  The bank should be printing the money that it is loaning, not reducing the amount in the savings accounts.

The current securities system also suffers from an inability to make detailed non-destructive observation of the systems.  Firms are only required to make very incomplete financial disclosures every quarter, with a bit more detailed disclosure every year.  This then requires that investors perform destructive measurements on the firm.  Practically, this is making significant changes in the amount of investment and seeing how the firm responds, or requiring regular dividend payments or bond payments or stock buybacks.  In the extreme case, if the reduction in investment or payments bankrupts the firm, then the investor definitively knows that more should have been invested or less payment required.  The problem is that the firm is now bankrupt.  In either case resistivity has been added to the economy and the performance of the economy has been degraded.  The problem with stabbing Schrödinger's Cat to see if it is alive, is that it can kill the cat, at the very least it will injure the cat and reduce its performance.  This is true of all investors, whether they be private equity, venture capital, banks, or holders of stocks and bonds.

Another way at looking at this is the following.  A business issues a yearly financial statement that says it generated a significant free cash flow and reinvested it in research.  The investors have no confidence in the truth of this financial statement.  Therefore, the next year they require that the free cash flow be returned to them as a dividend rather than reinvested in more research.  The larger the dividend that is demanded, the more certainty the investor has in the free cash flow.  After the dividend has been paid the following year, the investors have verified that the business does have that much free cash flow, but significant damage has been done to the future of the business by not doing the research.  The geometry of the dynamical manifold has been degraded.  If the dividend that was demanded was greater than the actual free cash flow, the business will have gone bankrupt.  The topology of the dynamical manifold has been destroyed.  The effect of demanding the dividend (the change in the geometry or topology of the dynamical manifold) will take time, a business cycle time, to be communicated to the investors as a reduction in free cash flow or a bankruptcy of the business.

An intuitive way of understanding a destructive statistical measurement is as ``kicking the tires'' of a car before you buy it.  When the tires are kicked a force is exerted on the car, and a fixed amount of work is done on the car.  If the car is well built, that is ``solid'', the change in the car's co-state, that is momentum, will be proportional to how ``solid'' the car is, that is $1/\omega_Q$ or the square root of the mass of the car.  The harder the tires are kicked the better the state, that is the condition of the car, will be known.  The problem is that kicking the tires too hard could do significant damage to the car if it subsequently crashes into another car as a result of having its tires kicked.  In any case, its dynamical trajectory has been altered by an external force, and the geometry of the dynamical manifold has been altered.  If the kicks are large enough, the topology of the dynamical manifold will be changed.

The Heisenberg Uncertainty Principle is simply a relationship between how large the change in the co-state needs to be, that is the force, in order to know the state of the system.  There is a fundamental lower limit on the product of the change in the co-state and uncertainty in the state given by $J_0=E_0/\omega_0$.   It is a limit coming from the finite communication speed, technically group velocity, of the system.  When there is a transient in the system, such as one starting to stab cats that pass by to see if they are alive, the topology of the dynamical manifold will be changed by the introduction of a singularity at the location where the cats are being stabbed.  This locally pulls on the dynamical manifold, causing a local change to the geometry, that is warping of the dynamical manifold.  This launches a wave, analogous to a gravitational wave in space time, that travels at the group velocity to the location of the passive non-destructive observer of manifold curvature.  After a travel time from the location of the destructive observer, who is stabbing cats, to the passive observer of dynamical manifold curvature, there will be a steady state change to the curvature of the dynamical manifold at the location of the passive observer.  It will take the time it takes the cats to walk from the destructive observer to the passive observer before the passive observer can know that the destructive observer is stabbing the cats, reducing their performance and life times.

The need to change the investment by a ``significant'' amount to be able to accurately know the state of the system is the origin of quantization.  The ``significant'' amount of energy is the ``quantum'' of energy $E_0$.  This is directly related to the system being stochastic so that only $P$ can be predicted, with $Q$ being uniformly distributed and periodic.  The periodic boundary condition on the probability (wave function $\left| \psi \right>$) in $Q$ induces the quantization of energy.

\section{Example of an electronic currency at work}
\label{valuation.sec}
In order to understand the potential for operating a sub-economy based on a metric of virtuous economic activity and how a coordinated monetary and fiscal investment policy would work using electronic currency leveraged financing, we present an example.  The example is based on the transition to sustainable energy, both sustainable hydrocarbon production with offsetting carbon sequestration and the development of sources of sustainable energy (such as blue and green hydrogen production, wind power production, solar power production, geothermal power production, and fail-safe nuclear fission power production with responsible waste disposal).

In this example, it is assumed that issuing of new Electronic Currency (EC\$) meets the demand so the exchange rate will be maintained at 1 EC\$ to 1 USD.  It is probable that the demand for the currency will be more than the supply, leading to appreciation in the value of the currency over time.  This appreciation could be enhanced by additional public offerings of stock in the currency management firm (which is called New Energy Inc.\ in this example) that will take advantage of the large appreciation in the stock and raise some of the capital needed for investment.  If the demand is different than projected because the actual economic multiplier, $m_e$, or the savings rate, $1/S_0$, being different than projected, the value of the EC\$ will just change so that the supply equals the demand.

The market capitalisation of New Energy Inc.\ will be determined by the value of the tangible assets of New Energy Inc.\ post IPO.  In actuality, the value of the stock is associated with predominately the intangible assets as shown in Fig. \ref{assets.equity.fig} and discussed in Sec.~\ref{tangibility.sec}.  It is expected that stock holders will perceive that they own the tangible assets and they will set a lower bound on the value of the intangible assets.  The value of those assets may be valued at several times that or possibly modestly less.  The reality is that the tangible assets of the sub-economy are practically controlled by the holders of the EC\$, that is the members of the sub-economy.  If members of the sub-economy leave the sub-economy, their part of the sub-economy leaves with them.  If the value of the New Energy Inc.\ stock rises well above this benchmark, it is an opportunity for New Energy Inc.\ to have a public offering of stock and raise capital at very favorable terms.

In order to simplify this example we have separated the development into four phases: (1) startup, (2) development of the energy transition for the state of New Mexico, (3) development of the energy transition for the United States, and (4) steady state.  For both Phase 2 and Phase 3, there are one NewCo and one BuyCo.  In practice, there will be a portfolio of companies.  It is assumed that \$20 million USD will be needed for Phase 1 and will be raised from the private markets at a valuation of \$60 million USD.  It will be used to set up the EC\$, to implement the on-line real time financial accounting, to obtain the necessary US SEC approvals, to establish the legal contracts for operating the sub-economy, and for governmental and public relations.  The beginning of Phase 2 will commence with an IPO to raise \$3 billion USD in cash reserves for the EC\$ at a New Energy Inc.\ valuation of \$15 billion USD.  The BuyCo is a typical size existing producer of unconventional gas in the Permian basin, and the NewCo is a production plant of blue hydrogen located next to the production reservoirs of the BuyCo.  Details of the transactions, including detailed balance sheets, can be found in App.~\ref{econ.model.app} along with other details.  An interesting thing to note on the summary of the balance sheets shown in Fig. \ref{balance.sheet.fig} is that the entries on the balance sheet for New Energy Inc.\ are reversed like a central bank with the EC\$ loans appearing as assets and the EC\$ as an equity (liability).

\begin{figure}
\noindent\includegraphics[width=\columnwidth]{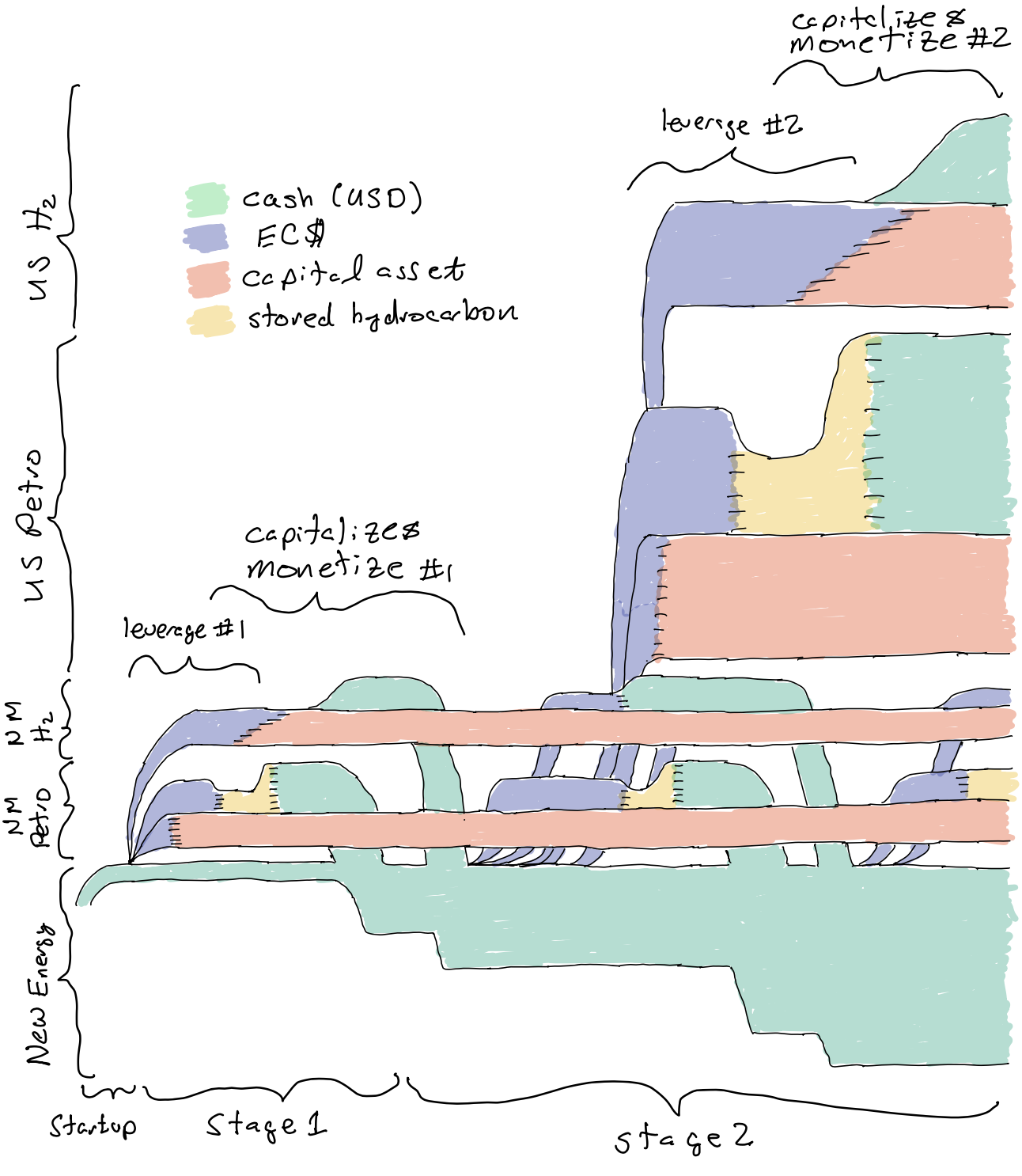}
\caption{\label{hydrodynamics.fig} Shown are the ``financial hydrodynamics''.  Note how the different types of assets are converted to one another by economic activity, and how the sub-economy displays exponential growth via the leverage of the electronic currency.}
\end{figure}

\begin{figure}
\noindent\includegraphics[width=\columnwidth]{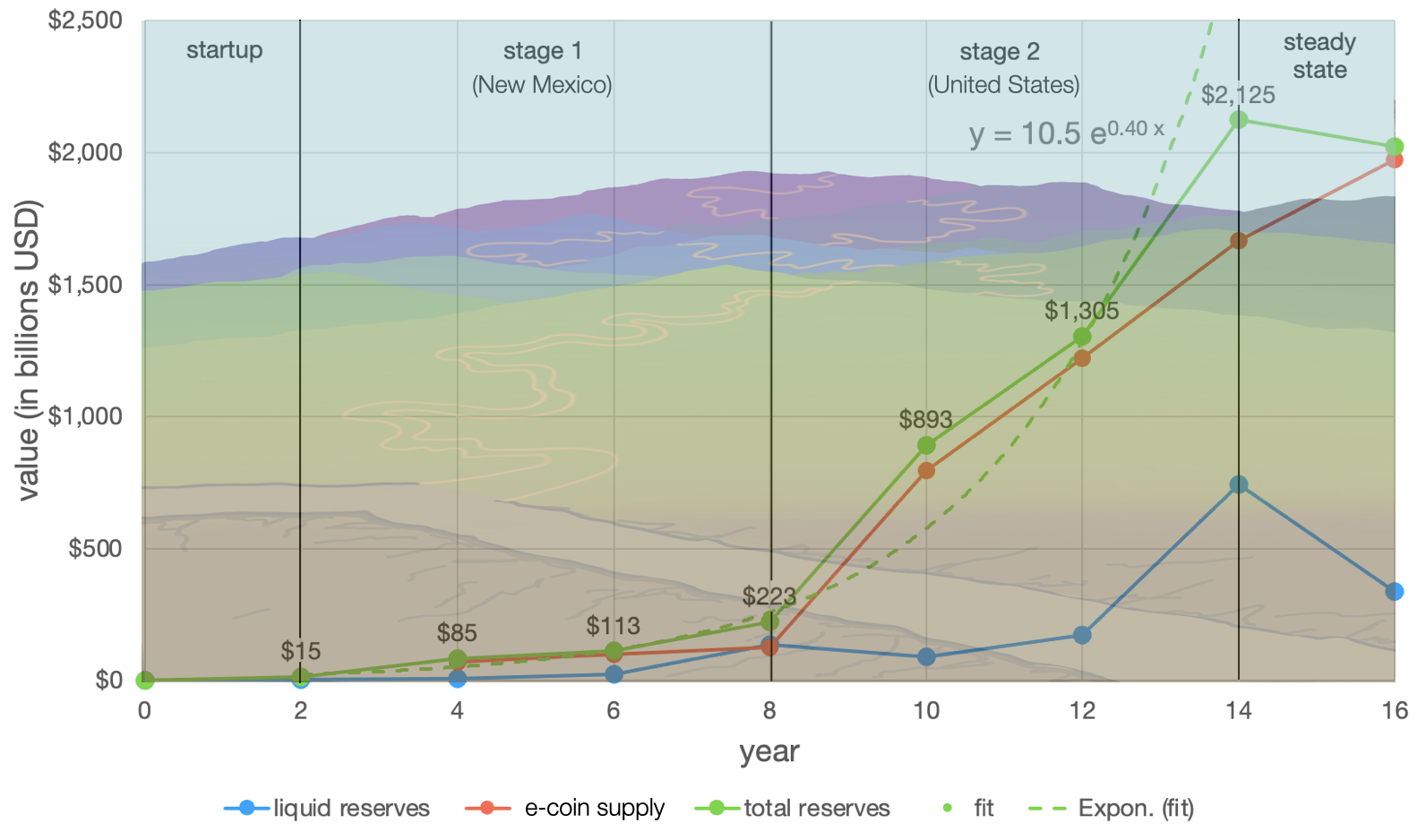}
\caption{\label{valuation.fig} Plot of the value and reserves of New Energy Inc.\ versus time.  The time between each event is roughly two years.  The ordinate is the $\text{USD}\$$ value.  Shown are the value of the: (1) liquid reserves as the blue circles, (2) EC\$ supply as red circles, (3) total reserves or the value of the sub-economy or equivalently liquid assets + capital assets (which includes the EC\$ savings of the sub-economy) as the green circles.  The different stages of growth are indicated, and the value of the sub-economy is fit to an exponential.  The fit has a slope (that is growth rate of the sub-economy) of 40\% per year.  The return on investment is therefore about 40\% per year, and the dividend is about 14\% per year when steady state is reached at a value of \$2 trillion USD. Some of the dividend is paid to the stockholders (4\%) and the rest of the dividend (10\%) is paid to society, the ultimate stakeholders in the sub-economy, in the form of $\text{CO}_2$ sequestration.  The value of the initial seed investment grows about 30,000x and the value of the IPO investment grows about 100x.  The sub-economy grows exponentially from the valley center of exploitation to the mountain pass of virtuous economic activity.}
\end{figure}

The ``financial hydrodynamics'' of how electronic currency leveraged financing works can be found in Fig. \ref{hydrodynamics.fig}.  First note how the original cash reserve raised by the IPO is multiplied by the issuing of EC\$, which are subsequently converted into capital and liquid assets by economic activity.  These assets then generate cash in excess of the original EC\$ investment.  This cash is returned to New Energy Inc.\ to significantly increase its cash reserve, allowing it to issue exponentially more EC\$ to invest in Phase III.  The process then repeats.  Note the exponential growth of capital and liquid assets coming from the leverage provided by the electronic currency.  The exponentially growing (at a rate of 40\%/year) capital and liquid assets, from the valley center of exploitation to the mountain pass of virtuous economic activity, is shown in Fig. \ref{valuation.fig}.  The value of New Energy Inc.\ grows from \$15 billion USD to more than \$2 trillion USD in 12 years.  The value of the initial seed investment grows about 30,000x and the value of the IPO investment grows about 100x.  New Energy Inc.\ is operated with no debt, and owns all the assets instead of renting or leasing them.  Both renting and leasing are effectively debt financing the assets.

In this example, note how New Energy Inc.\ maintains at least 10\% liquid reserves (in cash and New Energy Strategic Petroleum Reserve) and at least a 100\% total reserve ratio (in capital and liquid assets) at all times.  After the US assets are established, no further growth is anticipated in Phase IV.  Any further profits are returned to the investors as dividends or to society, the ultimate stakeholder, by sequestering $\text{CO}_2$.  It should be noted that the EC\$ loans of New Energy Inc.\ do not have repayment, foreclosure or ``call'' terms.  As discussed above a constant exchange rate is assumed for the EC\$ to USD.  If the value is not constant, New Energy Inc.\ will supplement the loans if the value goes down, and repossess the surplus if the value goes up.  It should be noted, that because of the high savings rates maintained in the New Energy sub-economy, there will not be any cash flow issues.  The primary EC\$ savings of the New Energy Inc.\ subsidiaries in steady state is 565 billion (about 2 years operating expenses on average) compared to the 2 trillion in circulation, giving a reasonable economic multiplier of 3.5.

It is also informative to note that when stock is issued an equity (liability) entry is made that is balanced by entries to cash and mostly to intangible assets.  But when electronic currency is issued an equity (liability) entry is made that is offset by entries to tangible assets.  This confirms the view shown in Fig. \ref{assets.equity.fig}.  When a company is purchased for electronic currency, the entries are identical to those made when a company is purchased for stock, the difference is in the type of equity that is issued.  Finally, when New Energy Inc.\ invests electronic currency in a subsidiary it is identical to the entries that would be made if it was investing cash in the subsidiary.

\section{Ubuntu business model}
\label{ubuntu.sec}
The example presented in Sec.~\ref{valuation.sec} deploys a new business model based on transactional equity.  This Ubuntu business model removes the dissipation that is embedded in the conventional debt based business model that uses borrow/loan financing  --  creating an economic superconductor free of resistance, that is debt.  Although the dissipation, that is interest, controls and stabilizes the economic system, it places a large constraint on the operations of the business reducing the revenue by 90\% or more, and robs the business of free cash flow needed for its sustenance and growth.  The dissipation does dominate the business performance forecasting, resulting in a simple exponential random walk or drift-diffusion to zero that is the solution to the easily solved Black-Scholes or Fokker-Planck equation.  In contrast, when the resistance is removed by use of the Ubuntu business model and replaced by genAI based economic control (forecasting for investment) and stabilization, the economy is put into motion by investment then is self sustaining without need for further investment, like a superconductor.

\begin{figure}
\noindent\includegraphics[width=12pc]{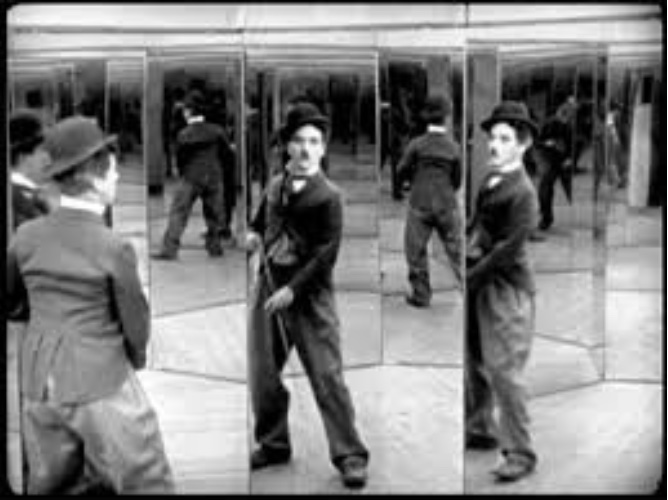}
\caption{\label{mirror.maze.fig} Demonstration of how the motion of the virtual individual is multiply reflected by a Hall of Mirrors to give the economic evolution of the collective. This is shown in the ``The Mirror Maze'' scene from the 1928 Charlie Chaplin movie ``The Circus'' (\href{https://youtu.be/G09dfRrUxUM}{ YouTube video}).}
\end{figure}

The Ubuntu business model recognizes that the economy is a collective of individuals, as such, there emerges a collective behavior or virtual individual that determines, that is controls, the collective behavior  \citep{glinsky.24b}.  This control emanates from a simply beautiful symmetry -- the coordination of the economically interacting individuals creating an economic force, and the economic force then coordinating the economic trade of the individuals.  Another way of looking at this is that the actions of the virtual individual, who follows a geodesic motion, are approximated by genAI.  The genAI then generates and multiply reflects the actions of the virtual individual through a Hall of Mirrors to obtain, that is forecast, the economic evolution of the collective.  This is demonstrated by ``The Mirror Maze'' scene from the 1928 Charlie Chaplin movie ``The Circus'' \citep{circus.28} and shown in Fig. \ref{mirror.maze.fig}.

Given how the actions of the virtual individual are multiply reflected through the Hall of Mirrors, the state of economic prosperity can be stabilized by genAI ponderomotive stabilization.  This method is similar to the way that a sheepdog herds sheep.  The sheepdog runs around the herd very fast, compared to movements of the herd, nipping at the heals of the sheep (vibrating them) if they wonder away from the mountain pass, that is the state of economic prosperity.  The sheepdog is effectively creating a small alpine valley at the mountain pass.  Practically, this is a high frequency arbitrage trading of the transactional equity that provides liquidity.

We use the name Ubuntu \citep{ubuntu.24,mandela.04} to describe the business model, the new unified economic theory, and the genAI because all three are based on the conservative interactions of collective members, that is the ``interconnectedness'' of the Spirit of Ubuntu.  The Spirit of Ubuntu can be viewed as the virtual individual of the collective.  We believe that we are ``bringing back the African Spirit of Ubuntu'' with this business model.

\begin{figure}
\noindent\includegraphics[width=\columnwidth]{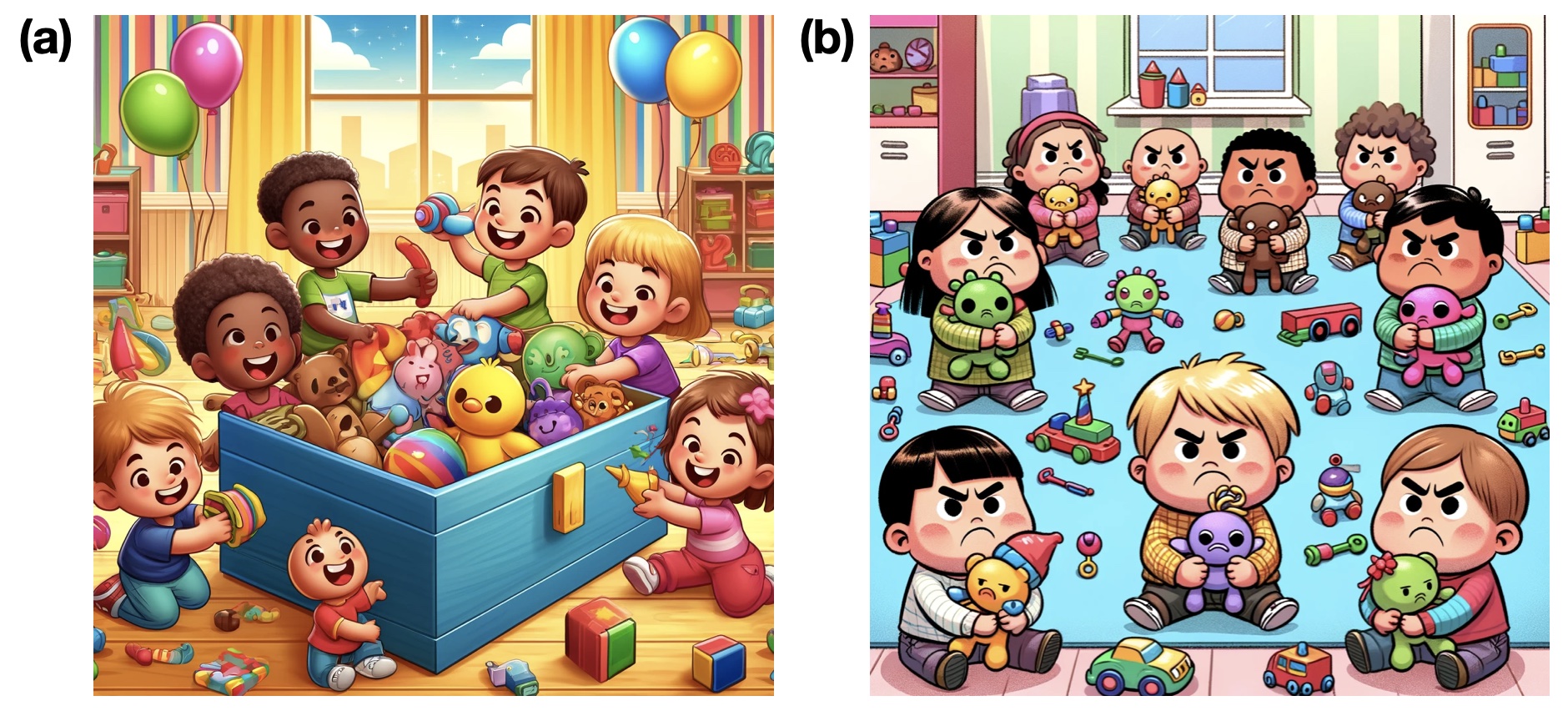}
\caption{\label{share.fig} The good of sharing. (a) Children putting toys in a community toy chest so that there are more toys from which to choose, leading to happiness.  (b) Children being greedy, keeping their toys to themselves, leading to sadness.  Images generated by OpenAI's DALL-E.}
\end{figure}

The Ubuntu way of ``mutuality'' is based on the concept of sharing, putting toys in a community toy chest so that there are more toys from which to choose as shown in Fig. \ref{share.fig}.  The community software chest is GitHub.  The community capital chest is Martin Luther’s Community Chest of the Protestant Reformation \citep{luther23}.  The electronic transaction equity version of the community capital chest is Ubuntu Financing.  This is related to the Economics of Mutuality \citep{mayer.21}, Modern Monetary Theory \citep{kelton20}, Marxian Economics \citep{wolff.87,desai.06,marxian.24,douglas.20}, Complexity Economics \citep{farmer.24}, and Special Purpose Acquisition Companies (SPACs) \citep{spac.24,ren.21}.

\begin{figure}
\noindent\includegraphics[width=12pc]{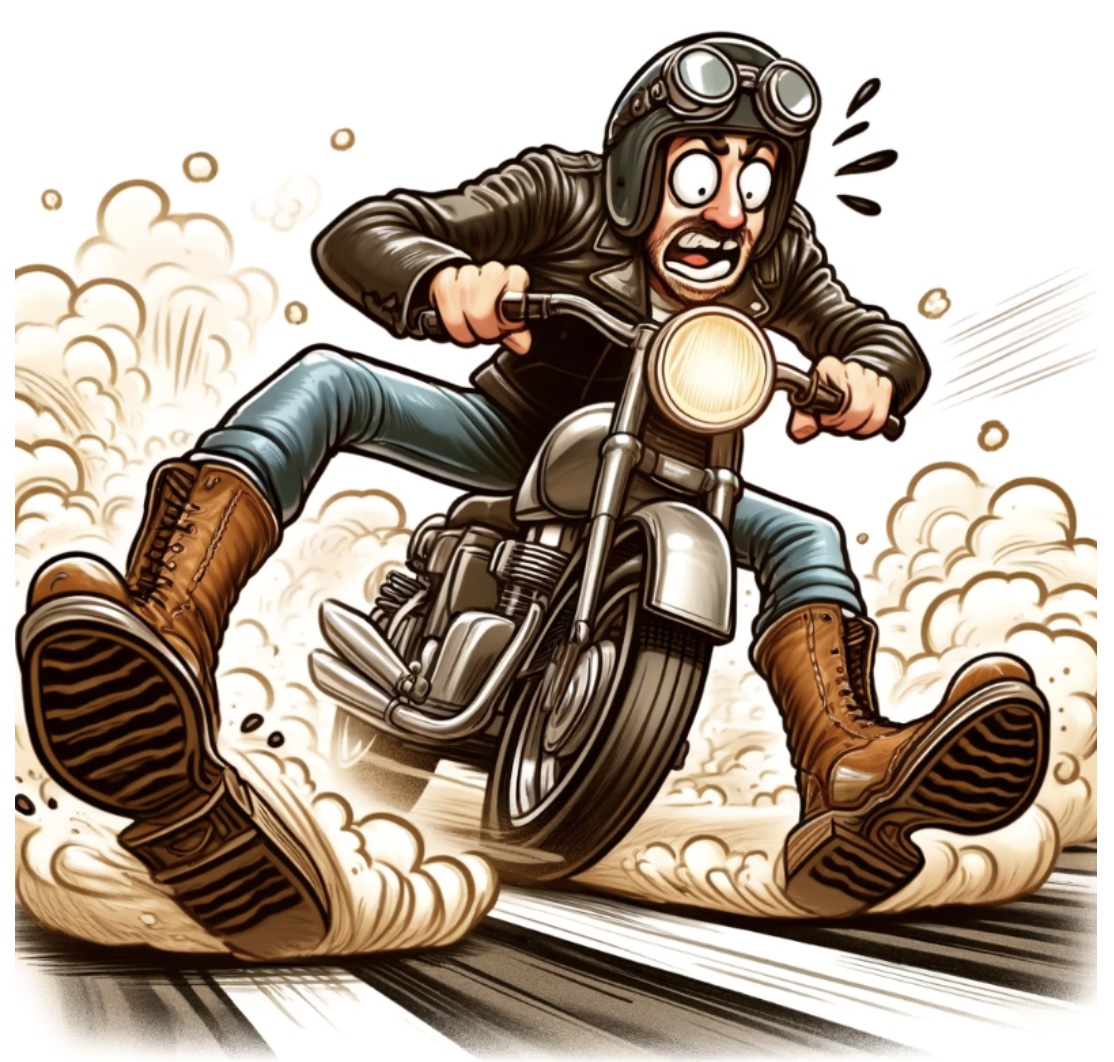}
\caption{\label{full.gas.fig} Illustration of resistive control.  Going ``full gas'' on a motorcycle and applying the brakes to change the speed.  Image generated by OpenAI's DALL-E.}
\end{figure}

\begin{figure}
\noindent\includegraphics[width=\columnwidth]{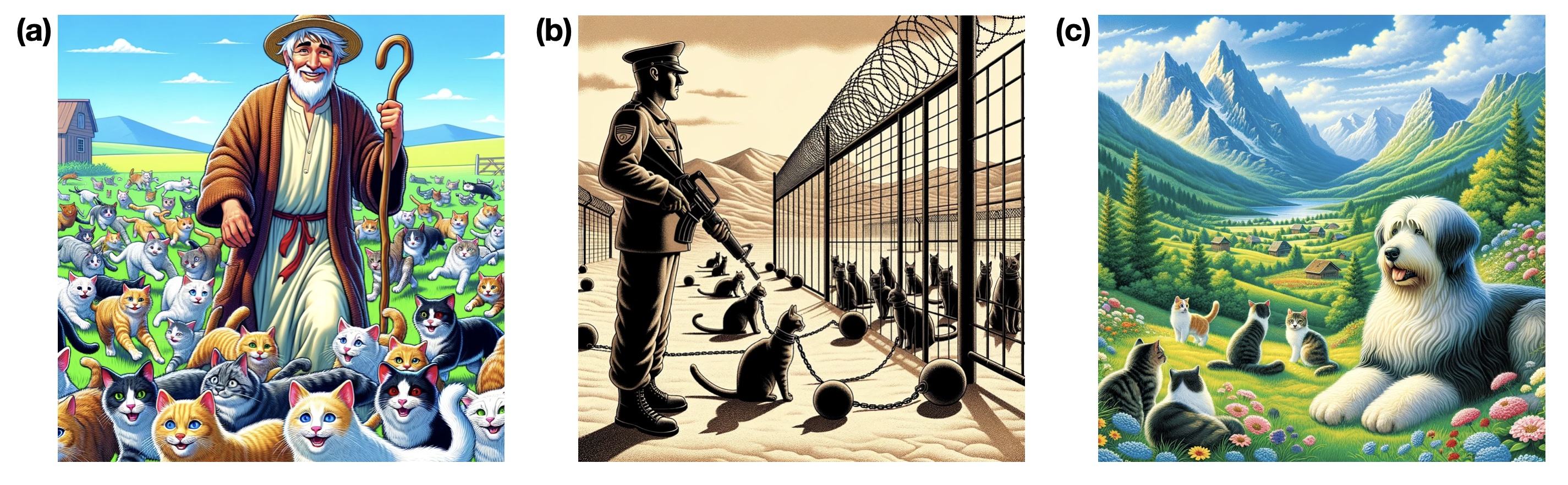}
\caption{\label{lexi.control.fig} Illustrations of different types of control.  (a) Jimmy Carter tries to shepherd the new free society (that is, The Great Society) emerging from the 1960's, which wanders out of control.  (b) The dystopian response is Lewis Powell and the cabal locking up society.  (c) Because of transactional equity and genAI, we are at the dawn of a new era of sustainable economic prosperity, in an alpine valley shepherded by genAI.  Images generated by OpenAI's DALL-E.}
\end{figure}

When the Ubuntu business model eliminates the dissipation from the business by the use of Ubuntu (print/invest) Financing with transactional equity, another method of control and forecasting must be found.  This control and forecasting is what is provided by the new Ubuntu genAI.  The difference between drift-diffusive forecasting and genAI forecasting is shown in Fig. \ref{simulation.fig}.  The resistive control, that is going ``full gas'' and applying the brakes to change the speed shown in Fig. \ref{full.gas.fig}, is replaced by the ponderomotive genAI control discussed in Sec.~\ref{solve.hjb.sec} and shown in Fig. \ref{inverted.pendulum.fig}.  The result of removing the dissipation without implementing ponderomotive control, the effect of resistive control, and the economic Nirvana created by ponderomotive genAI control are shown in Fig. \ref{lexi.control.fig}.

The \textbf{Ubuntu business model} is characterized by: (1) maximization of sustainable revenue, (2) business cycle (2 to 6 years) of transactional equity savings and inventory (both input and output), (3) forecast and control using genAI models, (4) Ubuntu (print/invest) Financing, (5) stakeholders (employees, suppliers and customers) being stockholders (owners), and (5) social responsibility (respecting the dignity of employees, suppliers and customers).  In contrast, the \textbf{conventional debt based business model} is characterized by: (1) maximization of short term profit, (2) transaction cycle (15 days or Just-In-Time) of cash savings and inventory, (2) forecast and control using using diffusive models (i.e., Discounted Cash Flow analysis or the Black-Scholes Equation for options and derivatives pricing), (3) debt based (borrow/loan) financing, (4) stockholders not being stakeholders, and (5) exploitation of society (that is, employees, suppliers, customers, and intellectual property).

The conventional debt based business model, without sufficient savings, has poor resiliency and performance.  Production, that is revenue, is constrained so that there is not full employment, supply is limited, and prices are artificially high to ration supply.  The economic collective is constrained so that there is never surplus supply, that is all production is immediately sold.  In contrast, the Ubuntu business model generates 10 times the revenue or value, makes 100 times the investment, has plentiful supply, is resilient to economic shocks, and has low prices.

Ubuntu Financing is based on the common stock equity and transactional equity model shown in Fig. \ref{assets.equity.fig}.  Transactional equity primarily securitizes the low risk tangible assets of the business, while the common stock equity securitizes the high risk intangible assets (things like value-in-place, goodwill, brand value, know-how, show-how, and intellectual property).  The risk (fluctuations) of the transactional equity are further reduced by the ponderomotive stabilization which transfers risk from the transactional equity to the common stock equity.  Such transactional equities are ubiquitous, whether that be Zelle, Venmo, casino chips, airline miles, StarBucks, gift cards, loyalty programs, subway tokens, REI dividends, or carnival tickets.

The Ubuntu firm will operate the Private Equity (PE) fund of the future.  The holders of the common stock equity take the place of the limited partners of the PE fund.  Conventional debt leveraging is replaced by Ubuntu Financing, that is the printing of transactional equity and the subsequent acquisitions or investments with that transactional equity.  This is equivalent to an all stock acquisition, for instance when Exxon printed \$60 billion in stock then bought Pioneer Resources with that stock.  The only difference with the Ubuntu firm is that one can tap-and-pay with transactional equity.  One does not first need to exchange the common stock for coin.  Transactional equity is a centrally accounted (not on a distributed ledger like the blockchain, that is it is not crypto) electronic coin.  This PE fund of the future does not have meddlesome limited partners, does not have banks that require regular coupon payments, can make much larger investments, and generates much more return, that is value.

\section{Government and taxation}
\label{taxation.sec}
We now turn our attention to the role of government and taxation.  The first role of government is to make the rules of the economic and social game, that is law, and to enforce them.  It is important that the rules be kept to a minimum and that the rules respect the freedoms discussed in Sec.~\ref{freedom.sec}.  Government must also keep the records on property rights.  In addition to these tasks, Government is responsible for the common physical and social infrastructure.  This includes:  (1) defense, (2) social security, (3) education, (4) arts, (5) public infrastructure (such as roads, railroads, and parks), (6) healthcare, and (7) fundamental and government related research. 

For games and economic systems, the system dynamics are determined and/or modified by rules and regulations.  These can either be by modification of the reward or conservative potential economic energy $R(q)$, or by application of an external force $f(q,\dot{q})$.  Even though the external force does work on the system and changes its energy, the system will remain conservative.  For this reason, it is very important that we digress at this time and closely examine what rules, regulations, and taxes will keep the system conservative, and which ones will not.  This also informs some of the freedoms that need to be protected to keep the system conservative.

As previously discussed, the Hamiltonian (real) is given by 
\begin{equation}
    H(p,q) = E(p,q) = p \, f(p,q) + R(q)
\end{equation}
and complex analytic Hamiltonian is given by
\begin{equation}
    H(\beta) = E(p,q) + \text{i} \, \omega\tau(p,q)
\end{equation}
with $\tau(p,q)$ the analytic continuation of $E(p,q)$ which can be found by a solution of the Cauchy-Riemann equations.  In order to remain conservative, the force $f(p,q)$ can be a function of both $p$ and $q$, but $R(q)$ can be a function of only $q$.  Because of the structure, when a time dependence is added to $f(p,q,\tau)$ and $R(q,\tau)$ they can be rewritten as
\begin{equation}
\begin{split}
    \bar{f}(p,q) &= f(p,q,\tau(p,q)) \\
    \bar{R}(p,q) &= R(q,\tau(p,q) \ne \bar{R}(q)
\end{split}
\end{equation}
so that $f$ can be time dependant, but $R$ can not and still remain a conservative system.  So, to remain conservative, the reward or potential must be a function of $q$ only, that is $R(q)$.  It can not be a function of either $p$ or $\tau$.  There is no restriction on the form of $f(p,q,\tau)$.  The reward can be identified by the energy $E$ when $p=a=0$, that is no action is being taken.  This energy or reward when there is no action $E(p=0,q)$ can not be a function of time.  The restriction of being a force on the system is that $\Delta R = \dot{q} \, f(p,q,\tau) \, d\tau$.  In other words, the reward is proportional to the rate of change in the state -- a tax on activity.

Examples of rules that are conservative are transaction and sales taxes, and equity and electronic currency funding which have repayment/issue terms that are conservative forces
\begin{equation}
    \Delta R = - \left( \frac{\partial R}{\partial q} - \frac{\partial R_0}{\partial q} \right) \, \dot{q} \, \Delta \tau.
\end{equation}
Examples of rules that are not conservative are wealth and property taxes, and loans with terms of debt $\Delta R = -\nu R \Delta \tau$ or $R(q,\tau)= R(q) \text{e}^{-\nu \tau}$.

The approach to management of government actions should be that of a private corporation with government appointed representatives forming the board of directors.  There should be separate corporations for the federal, state and local governments.  As usual, it is the board of directors that approves the fiscal policy (that is budget) of the corporation proposed by the executives of the corporation, Government Inc.  The budget could be subject to the approval of the respective elected legislative branch of government, and the key chief executives could be the elected members of the executive branch of government.  There should be an electronic currency for each corporation.  The monetary policy should be coordinated with the fiscal policy and implemented in the manner proposed by this paper.

There is one additional element to Government Inc. -- how the revenue is generated.  Since the infrastructure is common across all market segments of the economy, it needs to be generated through taxation.  We propose that this should be solely through a transaction tax on transfers of the electronic currency from one entity to another.  In general, the amount of the tax should be a common fixed rate.  For certain transfers that have significantly greater burdens on society, such as cigarettes and motorcycles that put a greater burden on the healthcare system, the tax rate should be greater.  These tax rates should be proposed by the executive of Government Inc. and approved by the board of directors.  The requirement of programming these taxes into the electronic currency is one of the important laws.

There are several reasons why we have proposed a transaction tax.  The first is that a transaction tax is easy to enforce since it is unavoidable.  It is programmed into the electronic currency and is paid at the point of the transaction when the currency is changing hands.  Instead of all the currency being transferred from the originator to the recipient some of the currency, programmatically, goes to the government.  There is need for little additional infrastructure or administration, unlike current methods of taxation like the income tax.  Second, current methods of taxation become nearly untenable when entities are dealing with multiple currencies.  For instance what currency is the tax accessed, and when is the tax accessed?  It is the international tax nightmare that expats and multinational corporations experience today, but much worse.  Third, a transaction tax discourages spending and encourages savings. Fourth, it maintains a conservative economy.  Finally, a transaction tax simplifies issues of multi-jurisdictional taxation.

\section{Freedom of labor, intellectual property, and law}
\label{freedom.sec}
We start by discussing the concept of ``Communism \& Freedom''.  Milton Friedman went out of his way in the early 1960's to cast capitalism as being superior to communism, but he made the point that it was essential that there be freedom so that the private sector could zealously pursue its maximization of profit.  If it was profitable, then it was in the social good.  This is fundamentally flawed.  Greed is not good, it is evil.  What is profitable is exploiting society.  The resistivity that the pursuit of profitability imposes on society, leads to oppression that opposes freedom \citep{glinsky.24b}.  Friedman's ideas formed the basis of Reagan and Thatcher conservatism of the 1980's.  

The problem with communism at the time, was the same issue that Friedman identified with capitalism, it needed to be free.  The problem with capitalism is that if the forces of greed are not constrained they naturally lead to political authoritarian dictatorships and business monopolies.  For communism, driven by economic activity maximization and benevolence, one is naturally led to a state of social beauty and economic communes.  Milton Friedman is right that freedom is essential to make this work.  Entities must be free to move, change employers, associate with another sub-economy, and use whatever currency is in their best interest.  This is what keeps the leaders of sub-economies motivated to do what is best for the sub-economy.  If they do not, they will lose their kingdom and with it their wealth.  By maximizing the economic activity of the sub-economy, they are maximizing their wealth.  If their subjects are not free, the leader can exploit them for their personal gain and the subjects can do nothing about it.  They are trapped.  The result is the existential angst that is prevalent in Latin literature.  This is the natural thing for the leader to do, since it is simplest to focus on the local here and now.  This is why ``Communism \& Freedom'', not ``Capitalism \& Freedom'' is the essential combination.  The point that is being made is that communism reinforces freedom, where capitalism naturally leads to subjugation and slavery, both economically and literally -- destroying freedom.

It should be noted that the combination of ``Communism \& Freedom'' that is being proposed is more economically conservative, more truly free or privatized, than existing \guillemotleft Communisme et Liberté\guillemotright\, in free societies such as France.  There is no one central government control that has a ``Robinhood'' policy of taking from the rich and giving to the poor, with an inescapable central government planning.  Instead there are multiple liberated private sub-economies that significantly raise the virtuous economic activity with some of the increase going to the wealthy, modestly increasing their wealth, but a much larger part of the increase going to the poor, significantly increasing their wealth and bringing them solidly into the middle class.  As we will discuss at the end of this section, this is a Democratic Socialism, not the centrally controlled Autocratic Socialism that Friedman criticized.

Restrictions lead to exploitation.  Therefore there needs to be complete freedom.  This starts with governments allowing their citizens to move about the country and even move to another country.  There should be freedom of speech including speech that criticizes the government, and political dissidents should not be imprisoned.

Labor should be free to work for whomever they want.  There should not be employment agreements with all-efforts clauses and non-compete clauses.

Intellectual property should be free to be used by whomever wants to use it.  Software should be open sourced, and all ideas should be put into the open literature.  There is no need for patents and copyrights since they lead to the enslavement and exploitation of the intellectual property.  Over the short-term intellectual property will still need to be patented to defend against patent trolls, so that the patents can be crossed licensed and made part of patent pools (that is patent communes).

Protection comes from the insular nature of the sub-economy, and the ability of the leader of the sub-economy to directly support (subsidize) the economic activity, effectively acting as a protective import duty.  For example, a leader of a sub-economy provides salaries in the currency of the sub-economy for artists based on how much of their art is consumed by the sub-economy.  The same is true of creators of intellectual property. As discussed in Sec.~\ref{taxation.sec}, some of this compensation and employment may be by Government Inc.\ with the revenue coming from a transactional tax.  Tokens of ownership (e.g., NFTs) or certification of action (e.g., carbon credits) are just digital certificates and should have nothing to do with a currency.

The law needs to be freed from case law and literal constitutional and statue interpretation.  The current lack of freedom leads to exploitation.  The basis of legal decisions should be on preamble, and the second reading of intent for social good.  Laws should be constitutional or not, based on whether they are in the social good or are exploitative of society.  Front line judges should be able to challenge whether a specific action is not in the social good like ``Jim Crow'' laws or is exploitative like laws that prohibit prostitution.  The burden of proof should be put on the state on appeal, not on the defendant who rarely have the resources to challenge the law.

Many of these freedoms also eliminate dissipation from the economic system which maintains the conservative dynamics of the economy, as discussed in Sec.~\ref{taxation.sec}.

The issue of freedom can be understood vis-a-vis the interaction of two things.  The first thing is the objective function that is optimized, whether that is DCF or GDP.  This is equivalent to the method of control, whether that is genAI or resistivity.  It is whether socialism is controlling the economic collective, or it is capitalism.  Whether it is controlled via profit maximizing capitalism, or it is controlled with revenue maximizing socialism.  

The second thing is the distribution of control.  Whether it is a distributed web of economic communes or a single economic commune centrally controlled by a Central Bank and a centralized government or Central Committee.  This is either autocracy or democracy. It is whether you end up with monopolies, both economic and political, or free-markets.  

\begin{figure}
\noindent\includegraphics[width=\columnwidth]{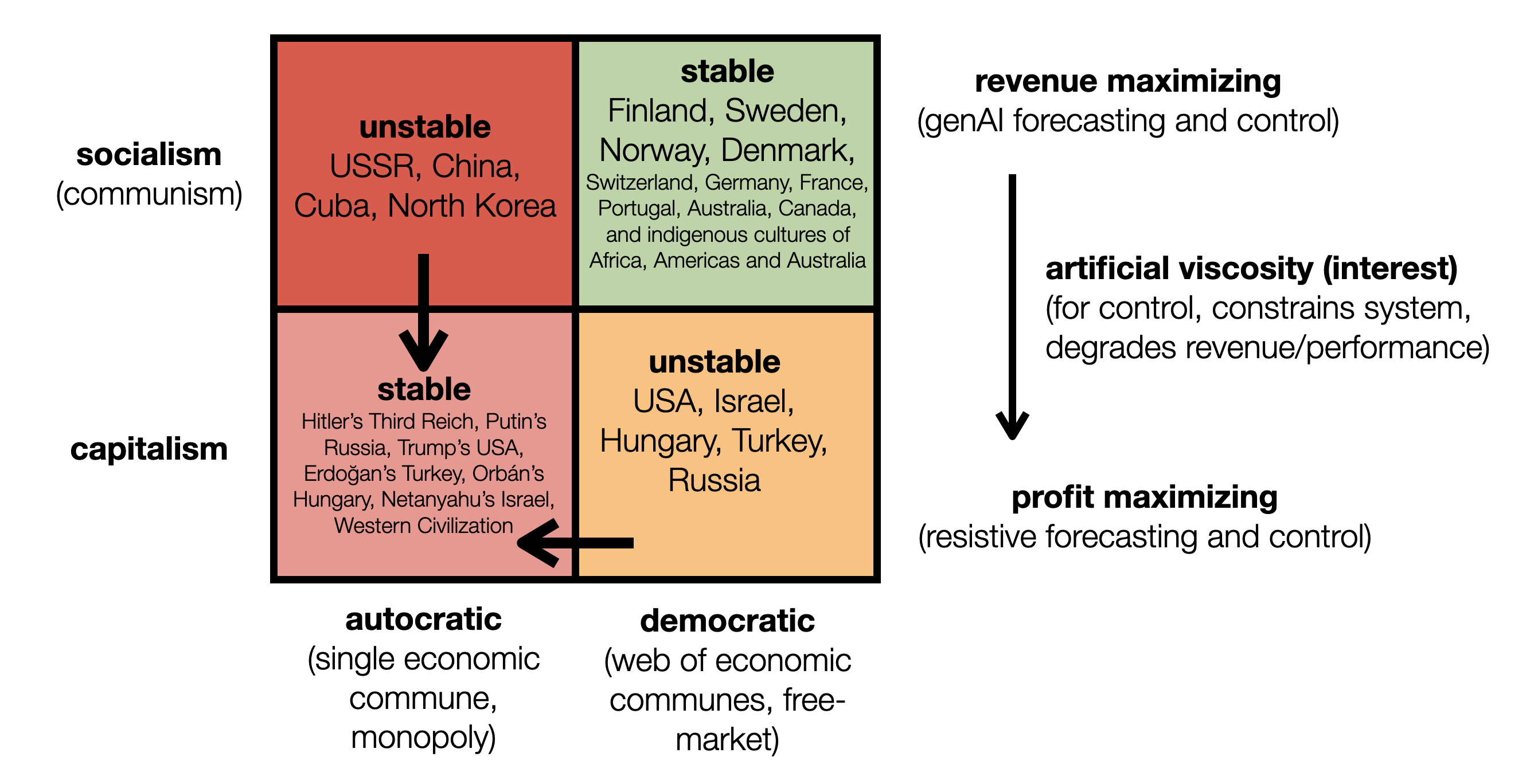}
\caption{\label{quad.chart.fig} Matrix of economic collective systems based on the objective and distribution of control.}
\end{figure}

Countries can now be put into quadrants of this matrix, shown in Fig. \ref{quad.chart.fig}.  First, there are the Socialist Autocracies like USSR, China, Cuba, and North Korea.  Second, there are the Socialist Democracies like the Scandinavian countries, to a lesser extent, France, Portugal, Australia, Canada, Switzerland, and Germany.  It is very interesting that most indigenous cultures of Africa, Americas and Australia are Socialist Democracies.  Third, there are the Capitalist Democracies like United States of the 1960’s, initially Israel, Hungary, Turkey, and Russia underneath Gorbachev and Yeltsin.  Finally the Capitalist Autocracies like Hitler’s Third Reich, Putin’s Russia, Trump’s USA, Erdoğan’s Turkey, Orbán’s Hungary, Netanyahu’s vision for Israel, and essentially most of the history of Western Civilization.  Note that Capitalist Autocracies are stable with poor economic performance.  Socialist Autocracies are unstable, mostly because of their very poor performance.  Capitalistic Democracies are also unstable with reasonable or modest performance.  The best and stable economies are Socialistic Democracies. 

We now go through these types of economic collective systems again.  \textbf{Democratic Socialism} is stable with high performance.  They are rated the happiest countries in the world.  This was successfully deployed in the American West from 1830 to 1860 \citep{galbraith.17}.  It was developed using multiple paper currencies, led by the Wells Fargo Bank, founded in 1852.  Each bank covered one day's horse ride and had its own promissory note or currency that it could print and grant.  During this time there was a two times decrease in prices due to increased productivity, that is deflation. There was a 100x increase in GDP of the American West over this time.  This movement continued after the US Civil War, as the Greenback Party (1874-1889).  A good example of an economic cooperative, that exists today, is the REI Co-op for recreational equipment.  It is 90 years old and counting.  It is the premier US retailer of camping gear.  It currently has \$4 billion per year in revenue and 15,000 extremely happy employees.  This leads to a distribution of wealth, power, and control that customizes the guidance, that is sharing, leading to and reinforcing a free market and democracy.  In contrast, there is \textbf{Autocratic Capitalism} that is stable with poor performance or \textbf{Autocratic Socialism} which is unstable with very poor performance.  The much studied one is \textbf{Democratic Capitalism}, which is familiar to most economics departments and business schools.  It is the capitalistic Monetarism (``Capitalism \& Freedom'') of Milton Friedman and the University of Chicago school of economics.  Note that commercial firms, under this method of control, have very short lifetimes, an average of about 15 years and shrinking.  This is because the diffusive model does not capture disruptions (does not model them), and the force of profit prevents the transition of firms through market disruptions.  Old firms, like Sears, fail and are replaced by new firms, like Amazon, when there is a market disruption.  The resistivity of 6.7\%, embedded in the economy via a typical AAA bond rate, constrains and grinds businesses to a halt in about 15 years.  The lifetime of the business is one over the interest rate, $t=1/\nu$.  The imposition of the interest rate is a forecast that businesses will go to zero revenue in 15 years, but imposition of the interest causes the business to have zero revenue in 15 years.  It is a self-fulfilling prophesy.  There is a strong natural force to concentrate wealth, power, control and guidance, so that there will be an evolution to autocracy and monopoly.  Greed will be an irresistible force, despite regulations and democratic constitutional constraints, that will prevail resulting in monopolies and autocracies despite one’s best efforts to prevent them.

The current lifespan of a business in the US (that is, a S\&P 500 company) is 18 years and shrinking, down from 35 years in 1965.  Only 50 companies are still on the Fortune 500 that were there 70 years ago.  With respect to the Dow Jones index, only one company (GE) is still on the index that was there in 1896 when it was formed.  There are four companies added in the late 1920’s and 1930’s (ExxonMobil 1928, Procter \& Gamble 1932, DuPont 1935, and United Technologies 1939).  The next oldest is 3M added in 1976.  Economists call this ``creative destruction'' that fuels economic prosperity.  Isn’t it rather corporate destruction driven by profit that destroys economic prosperity?  No business, other than IBM who has done it twice, has been able to survive a major disruption to their business model.  It is interesting that a lifetime of 15 years corresponds to a discount rate of 6.7\%.  This is not a coincidence.

\section{Resilience of economic system to economic collapse}
\label{resilience.sec}
There are three stages to economic collapse:  (1) a failure of the system of monetary control, (2) a risk squeeze, and (3) a liquidity squeeze.  In this section we will describe each of the stages, the cause, and how this system based on economic structure matched private electronic currencies, maximizing virtuous economic activity, remedy the economic pathologies.

The first stage is a \textbf{failure of the system of monetary control}.  Current systems of monetary control are dominated by the issuance of sovereign debt to decrease the money supply and loans from the central bank to increase the money supply.  Both of these have issues as a control system.  In order to go in one direction, let's say to the right today, there is a commitment to go in the other direction, let's say to the left, a bit for each of the next 20 years.  This phase shift in the control function causes the system to go out of control.  The economic manifestation is runaway inflation.  The dissipation has been increased so that the topology has been destroyed $\nu>\nu_\text{cr}$ and the system spirals into severe economic depression.  It could be a much more subtle failure of the control where the economy is at the point of economic prosperity and makes a small movement that is not controlled towards the basin of severe economic depression.  The fiscal investment policy is often an uncoordinated political process that leads to further issues with the control of the economy, exemplified by the recent British economy under Liz Truss. 

\begin{figure}
\noindent\includegraphics[width=15pc]{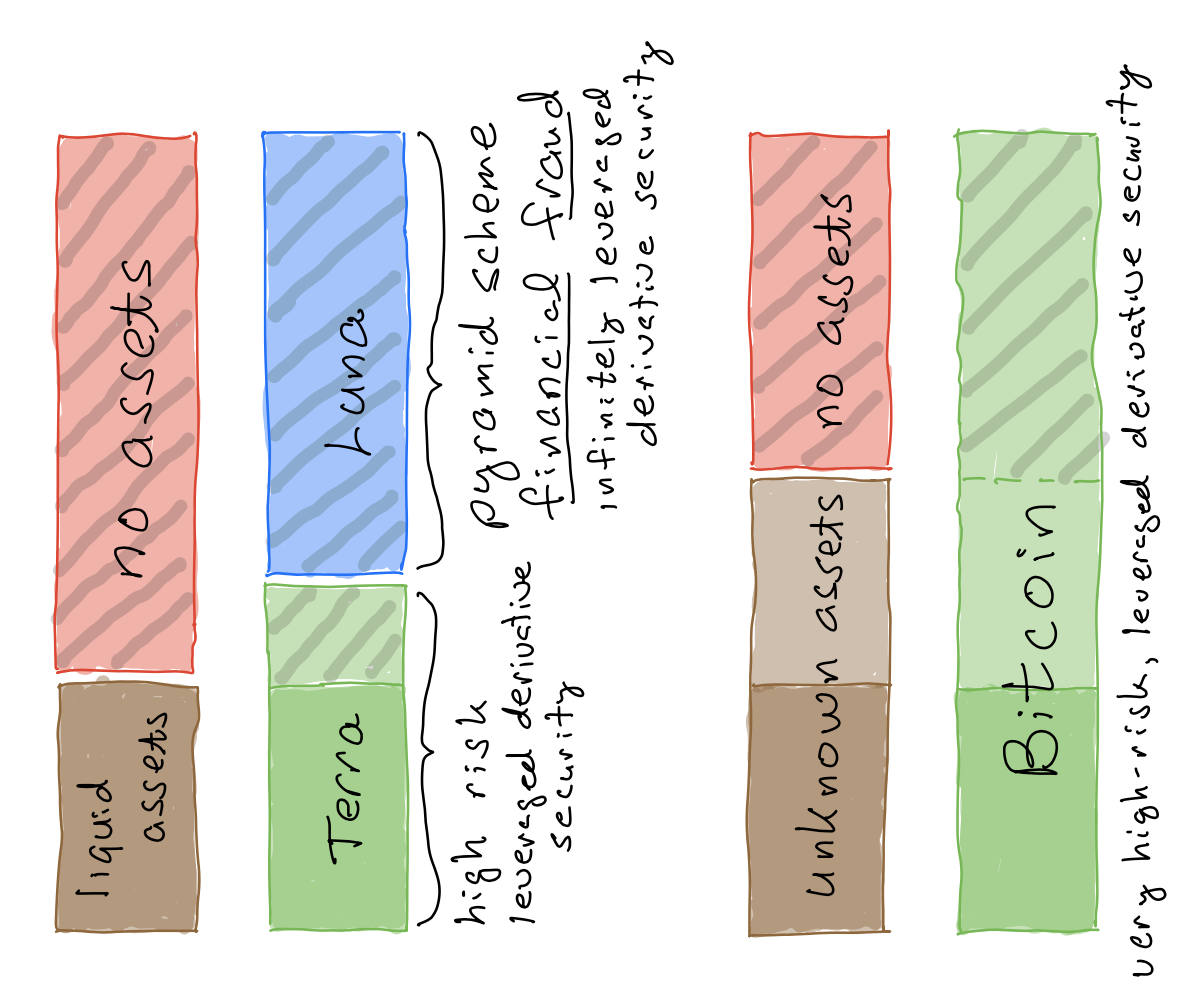}
\caption{\label{crypto.risk.fig} The risk tranching and leveraging of crypto currencies. (left) TerraLuna, (right) Bitcoin.  The foundational assets are shown by the brown bars (the size of the fluctuations are shown by the light brown bars).  The high risk assets (actually no assets, giving the ``leverage'') are shown as the red bars (the size of the fluctuations are shown by the light red bars, and the leverage by the diagonal grey stripes).  The lower risk leveraged derivative securities are shown by the green bars (the size of the fluctuations are shown by the light green bars, and the leverage by the diagonal grey stripes).  The higher risk leveraged derivative securities are shown by the blue bars (the size of the fluctuations are shown by the light blue bars, and the leverage by the diagonal grey stripes).  Note how both lower risk derivative securities are structured and leveraged so that they still have very significant risk.  In the case of Terra, it is associated with an infinitely leveraged pyramid scheme so that it will lose a large percentage of its value (break its peg) with the smallest decrease in the value of Luna.}
\end{figure}

Things are not much better with crypto currencies.  Proof-of-work crypto currencies, like Bitcoin, rely on a mining algorithm to control the creation of the currency.  There is no mechanism to reduce the money supply.  The creation of the currency is at a rate that varies depending on the time it takes to solve the elliptic problem on a computer.  Given that the economic system response has multiple time scales, and there will be times when the money supply should be decreased, this is again a poor system of system control.  The situation is even worse for proof-of-stake systems like the New Ethereum.  They have built-in, through the integration and debt-like coupon payments, a phase shift that leads to system instability.  There is also the large financial leverage that is built into the financial structure of these crypto currencies because of lack of both liquid and capital reserves, that will be discussed in Sec.~\ref{existing.sec} and is shown graphically in Fig. \ref{crypto.risk.fig}.  This can and has led to large volatility in the value of crypto currencies and their economic collapse.  There is rarely any mechanism for fiscal investment programmed into the crypto currencies.  When it is programmed into the currencies, it is not based on a rigorous economic evaluation of the investment opportunities.

The electronic currency that we are proposing is very different.  First of all it has significant levels of liquid assets and both tangible and intangible capital assets, as shown in Fig. \ref{assets.equity.fig}.  This provides a financial heat bath to give inertia to the electronic currency that will lead to a very stable value for the electronic currency.  In addition, the electronic currency management firm will use an advanced genAI control system, based on the multi-scale system response of the sub-economy, that will further reduce the fluctuations in the value of the currency and stabilize the economy.  The currency management firm will coordinate the fiscal investment policy with the monetary policy, issuing the currency needed for investment based on the projected demand that those investments will create.  This will lead to a modest long-term stable growth in the value of the currency.

The reserves are essential to the stability of the economy.  They give the dynamical financial system the inertia to change that stabilizes the economy.  As shown in Fig. \ref{assets.equity.fig}, the financial structure of the currency equity (fully backed security by tangible assets) and the stock equity (fully backed security by intangible assets) eliminates the super-charged risk of a highly leveraged derivative security as shown in Fig. \ref{crypto.risk.fig}.  First of all, the liquid reserves (whether that be gold, cash or inventory like stored hydrocarbons) give the assets that can be sold by the currency management firm to purchase the electronic currency as directed by the genAI, to stabilize the value of the electronic currency.  It is anticipated that less cash reserves are needed for a larger sub-economy since the relative size of the fluctuations are smaller.  Then there is the longer-term strategic monetary policy associated with execution of the fiscal investment policy and the control of longer term inflation and deflation.  In times of deflation (recession), this will involve issuance of new electronic currency to buy liquid assets, subsidize labor costs, buy capital assets, and invest in existing capital assets.  This will position the currency management firm to respond to periods of inflation by selling the liquid assets, terminating the labor subsidies, selling capital assets and reducing the investment in existing capital assets.  This buying and selling is in contrast to the loaning and borrowing of sovereign currencies.  Most governments do not have either the liquid assets (even if they do have a gold reserve like the US had at one time they do not actively trade it, it is only there to be a physical standard), or the capital assets to sell in a time of inflation.  They rely only on the issuance of debt.  A notable exception is the recent sale of the US petroleum reserve to combat the post COVID inflation.  There are additional ways that the reserves increase the resiliency that will be discussed in the following paragraphs of this section.

The second stage is the banking system being caught in a \textbf{``risk squeeze''}, resulting in banks being short on capital reserves when the system of monetary control fails.  Simply stated, the origin of this squeeze is the securitization of bank loans (many times on real estate) using the exponential risk model based on the local approximation previously discussed.  The assumes that a house has sprouted legs and is walking around like a drunken sailor random walking away in about 10 years or less.  The opposite is the norm.  A properly maintained and insured home will probably appreciate, possibly significantly.  This is a massive transfer of inflation risk from the homeowner to the bank.  The bank has not invested in real estate, but rather in a mortgage backed security by the terms they have set on the loan.  This security is valued as a long-term bond.  When interests rise, many times due to inflation, the value of these long-term bonds decreases significantly.  In the recent case of the Silicon Valley Bank (SVB) failure, they did not even have the fig leaf of creating the security by issuing the home mortgages (loans or bonds), SVB directly bought mortgage backed securities, making this squeeze obvious to depositors.  The bank becomes squeezed between the securitized exponential risk of the loan, and the riskless appreciating character of the real estate.  The result is that the bank finds itself in a situation where its capital reserves (that is the mortgage backed securities) are significantly less than its deposits.  This destroys depositor confidence and creates a run on the bank.  In the case of SVB, the depositor Peter Thiel lost confidence, and subsequently had every business he invested in withdrawal their deposits.  He also convinced many of his friends who were also investors in Silicon Valley to withdrawal their funds and the funds of every company they had invested in.  The result was the withdrawal of 42\% of SVB's deposits in one day.  This then leads to the terminal phase which immediately puts the bank out of business.

Before we proceed to that phase, let us examine why the system that we are proposing does not have a ``risk squeeze''.  The reason is simple.  This system does not make the local assumption and securitizes the loan using the real model of risk -- a multi-scale model of risk using genAI.

The third and final stage are banks being caught in a \textbf{``liquidity squeeze''}, when there is a run on the banks resulting from the banks being short on capital reserves.  This is an acute crisis that results in bank failures, the shutdown of the market for financing, and economic collapse and deep economic recession.  The origin of this squeeze is simply stated as the bank being squeezed between the representation of on-demand availability of deposits, as if they were kept in a vault, and the fact that the deposits have instead been invested in the longest term, lowest risk investment that the banks could find -- real estate.  When a significant number of depositors demand their deposits on the same day, there is no way that the banks can liquidate the loans (call or sell) to meet the demands of the depositors.  The result is immediate bank failure.

There are several ways that this system eliminates this squeeze.  The first is that different electronic currency is loaned out, than depositors keep in their electronic vaults (that is wallets).  They have the electronic currency securely kept in their electronic vault, and can always transfer it to someone else.  There are further protections against exchange to another currency.  The first and most important one is that there is no guarantee of exchange rate.  One must exchange at the going exchange rate.  This rate is controlled, as previously discussed, to be very stable and growing at a modest but constant rate.  If a manipulator like Peter Thiel does create a run on the bank, the value of the currency will quickly drop and most of the currency will be exchanged at a very low exchange rate.  Since the electronic currency is backed by the capital assets of the sub-economy, that practically means that most of the currency is held by entities that rely on it to transact on a daily basis (and many of them must hold and transact in the currency by loans and other contractual commitments), they will not want to sell low, even if they could.  The currency management firm, relying on its liquid reserves, will also issue insurance against such financial manipulations (such as the bank run started by Peter Theil), by exchanging at normal market rates the regular transactions.  This would only be for a couple days while the currency management firm is dealing with Peter Thiel.  The identification of these transactional exchanges will be aided by AI, and have case-by-case manual exceptions granted (like credit card fraud alerts).  Because transactions must be done in the electronic currency, a person like Peter Thiel must eventually exchange back into the electronic currency at a significantly above market price.  He has sold low and bought high, pumping money into the sub-economy and significantly increasing the cash reserves of the currency management firm.  Due to the currency value insurance, the members of the sub-economy are protected from this financial war, destined to be won by the currency management firm.  Just the threat of this happening will probably discourage the war taking place.

In summary, our system is protected from economic collapse by: (1) significant (about 10\%) liquid reserves, (2) complete capital reserves (that is the electronic currency being used for transactions and working capital), (3) genAI and non-debt based methods of monetary control, with insurance, matched to the structure of the economy, (4) a floating currency value, and (5) the use of electronic currency leveraged, not debt leveraged financing.

\section{Existing currencies}
\label{existing.sec}
Existing currencies are a grab bag of misfits that range from the mundane to fraud.  Let's start by examining the ubiquitous sovereign currency, that has been the only type of currency until recently.  Historical currencies have been based on a precious metal (gold or silver in most cases) liquid reserve that has literally been attached as a standard to the coin as the material from which it was fabricated.  In more recent history, paper notes were issued and pledged that there were gold or silver reserves in the vault of the sovereign.  In the most recent history, sovereigns have floated the currency, eliminating the gold reserves and backing the currency with the tangible and intangible assets of the sovereignty, with negligible liquid reserves in many cases.  Monetary policy has been controlled by the literal minting of coins and printing of currency, then having the central bank loan to commercial banks at a prime interest rate.  To reduce the currency in circulation the sovereign issues debt (i.e., US T-bills and British GILTs).  Rarely has the sovereign bought and sold  liquid reserves as an method of implementing monetary policy.  A recent exception is the selling of the US Petroleum Reserve by the US Government as a measure to combat the post-COVID supply-chain induced inflation.  As discussed previously in Sec.~\ref{resilience.sec}, this is a poor method of system control at best and is unstable in many cases.  In addition, there is usually an uncoordinated fiscal investment policy that is driven by politics, not economics.

\begin{figure}
\noindent\includegraphics[width=15pc]{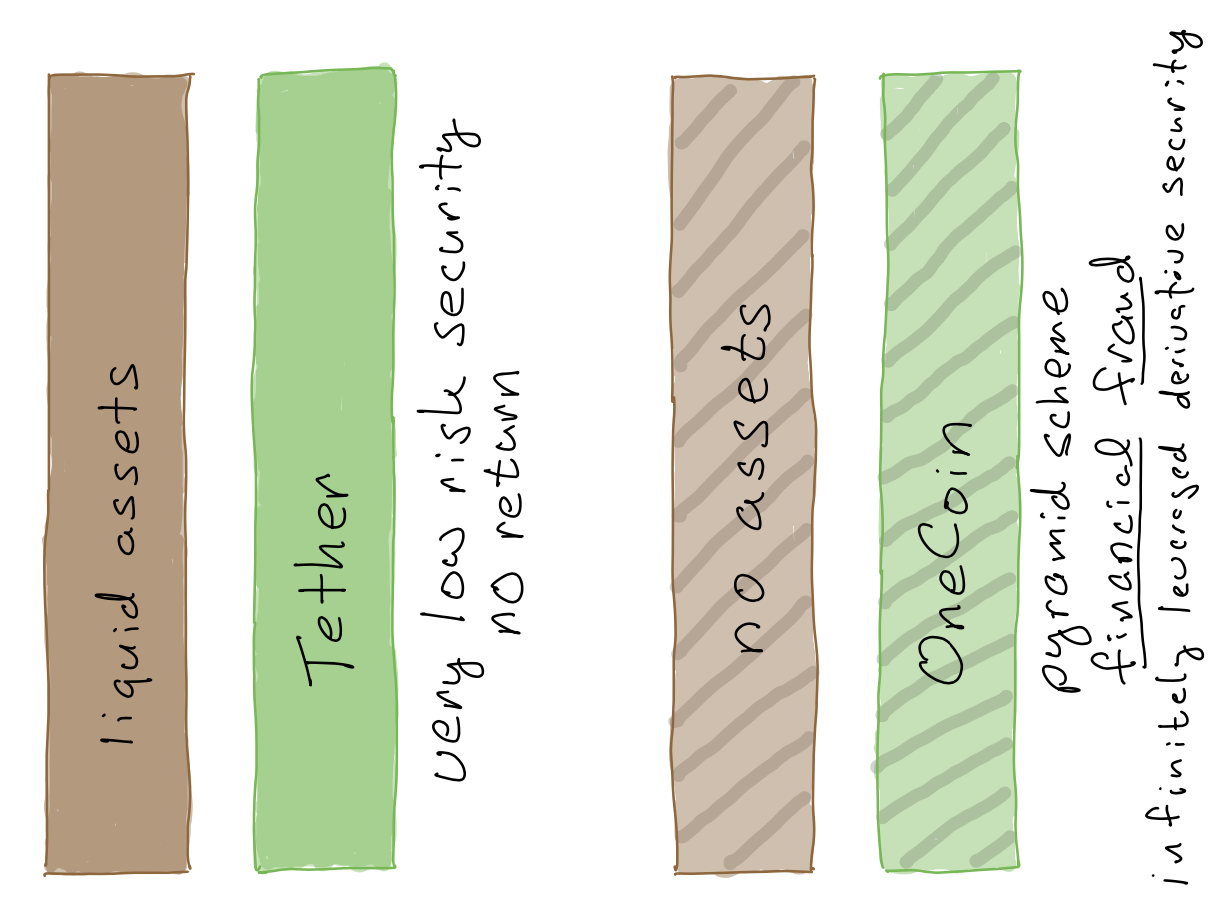}
\caption{\label{stable.fraud.fig} The risk tranching and leveraging of crypto currencies. (left) Tether, (right) OneCoin.  The foundational assets are shown by the brown bars (the size of the fluctuations are shown by the light brown bars, and the leverage by the diagonal grey stripes).  The lower risk leveraged derivative securities are shown by the green bars (the size of the fluctuations are shown by the light green bars, and the leverage by the diagonal grey stripes).}
\end{figure}

Over the last decade there has been the emergence of electronic currencies.  These have been of two types:  (1) centralized and (2) peer-to-peer, most times referred to as crypto-currencies.  Popular centralized currencies are PayPal, Venmo, and Zelle.  They all are 100\% backed by liquid reserves.  For the most part, they are backed by the USD.  These are very low risk securities with no return -- the mundane.  The structure of these securities is shown on the left of Fig. \ref{stable.fraud.fig}.  

There have been a much more colorful array of crypto-currencies.  They start with the bleached stable coin like Tether and USD Coin.  They are very low risk securities like the centralized currencies.  Their structure is shown on the left of Fig. \ref{stable.fraud.fig}. 

The next class are the mined, or proof-of-work crypto-currencies like Bitcoin.  These currencies have no liquid reserves and an unknown amount of tangible and intangible capital reserves.  There is no reporting of the use of the coin, so there is no way of knowing how many of the transactions are business transactions, and how much of the coin are held as transactional savings.  It is very likely that only a fraction of the coin is being used for business transactional purposes.  It is a misconception that because computation and therefore energy is expended in creation of Bitcoin, that Bitcoin is a liquid asset like gold.  Bitcoin has no utility value like gold.  You can not heat your home with the energy that was expended to create it.  Because of that, it has no liquid asset reserve like a gold coin.  The mining process is a method of monetary control.  When the value of a Bitcoin is greater than the cost of mining it, then Bitcoin will be mined and the supply of Bitcoin will be increased (decreasing the value) until the value is equal to the cost of mining.  The supply can never be decreased, even if the value is less than the cost of mining, because Bitcoin is never destroyed.  As discussed in Sec.~\ref{resilience.sec}, this is a poor method of currency control.  It is also a waste of a natural resource that has a significant utility value.  To date the mining of Bitcoin has wasted more than \$500 billion USD in energy, about 25\% of the world's annual power production.  A total waste of resources, when a Bitcoin could be created and destroyed programmatically for negligible cost.  The structure of Bitcoin as a security is shown in Fig. \ref{crypto.risk.fig}.  It is a very high risk, leveraged derivative security, despite the general belief that there is a significant liquid energy (information) reserve that mitigates its risk.  

Proof-of-stake crypto-currencies like Ethereum only make the situation worse by phase shifting the control signal (by integration), destabilizing the control system.

This is taken to the fraudulent extreme by infinitely leveraged coins like OneCoin which are backed with absolutely no assets.  The structure of these coins is shown on the right of Fig. \ref{stable.fraud.fig}.  These are infinitely leveraged derivative securities which are pyramid schemes -- financial frauds.  A sophisticated cousin of these pyramid schemes, that at first glance looks to be doing the risk tranching that we propose, is TerraLuna.  The problem is that it has no assets beyond a modest cash reserve.  the result is Terra being a high risk leveraged derivative security, despite it being advertised as a very low risk coin pegged to the USD.  Luna is a very high risk, infinitely leveraged derivative security, no better than OneCoin.  The structure of TerraLuna is shown on the left of Fig. \ref{stable.fraud.fig}.

None of the electronic currencies to date, have an investment policy or plan.  Except for the stable coins, crypto-currencies have been high-risk, leveraged derivative securities or frauds with no investments, in crypto clothing.  This has been overlooked by even very educated VC due their zealous greed, given the bull market.  It should not be a surprise that the crypto market has collapsed at the first sign of a bear market with the bankruptcy of firms like FTX.

What we propose if very different.  The structure of what we propose, as a security, is shown in Fig. \ref{assets.equity.fig}.  It shows how our electronic currency is a low risk, very stable, low return security.  It is also very easy to transfer.  All of this makes the electronic currency well suited to be a useful facilitator of economic activity.  The stock is a high risk, high return security.  Both are completely backed with assets.  This is by design, and as advertised.  There is an economically driven fiscal investment policy that is coordinated with the monetary policy.  It is designed to maximize the virtuous economic activity.

\section{Interaction of sub-economies}
\label{interaction.sec}
Up to this point we have examined in detail the theory and operation of a sub-economy.  Some hints have been dropped of how these sub-economies will interact.  Time has come to examine the interaction of sub-economies in detail.  We start with examining the issue of multi-objective optimization by multiple sub-economies.  In the theory that has been developed, each sub-economy, $i$, will develop its own reward function $R_i(q)$,  which will be its ``utility function'' and a basis vector in a Hilbert space $\left|\xi_i \right>$.  This is how members of the sub-economy value, that is weight, the different objectives.  The functional form of $R_i(q)$ is modified to the desired form by the leader (that is electronic currency management firm) of the sub-economy by a combination of strategic investments, subsidies, and penalties.     A set of interacting sub-economies forms a graph network on a manifold with basis vectors $\{\left|\xi_i \right>\}$ and coordinates $\{\xi_i\}$, that is an economy.  The metric, that is geometry and topology of the economy, will be specified by the matrix of currency exchange rates, $\Xi_{ij}$.  Furthermore, each sub-economy will be operated both efficiently and optimally to maximize the virtuous economic activity of that sub-economy $\int{R_i[q(t)] \, dt}$.  The free trade between sub-economies will be ``negotiated'', that is governed, by the free market exchange rate matrix $\Xi_{ij}$.  The market is free in the sense that there is a public currency exchange where any entity is free to exchange any currency they own to any other currency they can own.

Let us now consider the trade between a sub-economy that is operated to maximize exploitation (that is operated unaesthetically) and a sub-economy that is operated to maximize virtuous economic activity (that is operated aesthetically).  For the aesthetic sub-economy, it will take more real resources to compensate labor, and more real resources to develop the new technology and to engineer the aesthetic product that is beautiful and ecologically friendly.  The unaesthetic sub-economy will develop an inferior product based on the technology of the aesthetic sub-economy while giving less real compensation to their labor.  As a consequence, the unaesthetic sub-economy will find that their currency is worth significantly more than the currency of the aesthetic sub-economy.  The result will be that the relative price of the products will be the same, and the relative currency compensation of labor will be the same.  Given that there is free trade and free currency exchange, members of both economies will buy the product of the aesthetic economy, and the unaesthetic economy will have neither an internal or external market for their product.  The exchange rate is effectively imposing a import tax on the product of the unaesthetic sub-economy to pay for technology development done by the aesthetic sub-economy, ecological and beautiful design done by the aesthetic sub-economy, and proper compensation of the labor of the unaesthetic sub-economy.

The response of the leader of the unaesthetic sub-economy will be to restrict that freedom of trade by eliminating freedom of currency exchange.  The leader will do this by requiring that they are the sole authorized entity of exchange at a fixed exchange rate, much less than the free exchange market rate.  This will flood the market of the aesthetic sub-economy with cheap, ugly, and environmentally destructive product.  The response of the aesthetic sub-economy should be to erect a topological obstruction to trade with the unaesthetic sub-economy.  This is simply done by banning ownership of the currency of the aesthetic sub-economy by members of the unaesthetic sub-economy, and vice versa.  Both things are under the control of the leader of the aesthetic sub-economy.  The topological obstruction will require that all transactions between the two currencies go through an intermediate ``gateway currency''.  This is the topological obstruction.  The leader of the gateway currency will see the value of their currency go up since it is now a surrogate currency for the currency of the unaesthetic sub-economy.  This will damage the balance of trade and the sub-economy of the gateway currency.  The result will be the leader of gateway currency erecting a further topological obstruction to trade between their sub-economy and the unaesthetic sub-economy.  The unaesthetic sub-economy will quickly find itself isolated in a cluster of other unaesthetic sub-economies.  The economy will find itself having two disconnected clusters.  One of the clusters containing only aesthetic sub-economies with freedom, and the other containing only unaesthetic sub-economies with no freedom.  The unaesthetic sub-economies are subject to rebellion and revolution that will replace their leadership with aesthetic leadership due the desperation coming from the hardship of unaesthetic sub-economies.  Obviously, the sub-economy would then become part of the aesthetic cluster.

The result of the interaction of sub-economies is the evolution of any economy to a well connected cluster of only aesthetic sub-economies.  This is even more important with respect to the global economy where it is important that each state be led in an aesthetic manner.  The evolution outlined in the previous paragraph will lead to the trade embargo of an unaesthetic state by its major aesthetic trading partners.  The embargo will quickly spread to all aesthetic states so that they are not exploited as gateway currencies.

\section{Electronic currency as a religion}
\label{religion.sec}
With the background of the proceeding sections we now address the question, ``why is crypto a religion?''. Here when we say crypto, we mean any form of electronic currency whether it be centralized or peer-to-peer.

 The equation for the price of a crypto token, Eq.~\ref{r0.eqn}, is made up of two parts.  The first is a metric of social good.  The second is a long range Probability Density Function (PDF) that recognizes social good well into the future.  Other ways of saying this is that it is a belief in a future, or the existence of an ``auctioneer at the end of time'' \citep{buiter.09}, or that there is an eschaton \citep{dodd35} at infinity.  In religious terms, the auctioneer at the end of time is ``God'', and the belief in the future is equivalent to beliefs in reincarnation and an afterlife.  In contrast, the equation for the NPV, Eq.~\ref{npv.eqn}, is made up of two very different parts.  The first is a metric of personal greed.  The second is a denial of a future.  In religious terms, this is a denial of the existence of ``God'', and disbelief in reincarnation and an afterlife.  Because of this, faith-based investors and the metric of transactors are fundamentally religious with a belief in ``God'', and a fundamental value of the social good.

One can say that our capitalistic economy is now driven by atheist greed, and that turning to crypto is being saved by this true religion of faithful pursuit of social good.  This is why it is no accident, as will be described in Sec.~\ref{inspiration.sec}, that the crypto culture is described as a cult or a religion.  Historically, this also explains why most societies are under pinned by religious beliefs.  It also explains why those religious beliefs come into conflict with business motivations derived from the DCF optimization of Eq.~\ref{npv.eqn}.

There is a force of ``evil''.  It is the thermal force discussed in Sec.~\ref{nonlinear.sec}.  This force first leads the financial dynamics to a saddle, mountain pass, or x-point.  At this point, there is self-determination where small changes make a large long-term difference in the trajectory of the financial dynamics and the ultimate destination, that is equilibrium.  In order to make the saddle point the ultimate destination there need to be small incremental guidance given on an ongoing basis.  This guidance is given via the monetary value control system, the invisible hand of ``God'', like the ``evil'' hand of ``Satan'' coming from the thermal force.  If there is no hand of ``God'', the trajectory of the financial dynamics will move into one of the basin of attraction and ultimately end up at a valley center, stable equilibrium, or o-point.  This can be viewed as ``Hell''.  Once at this point there is very little self-determination.  It will take a great effort to climb out of the valley back up to the mountain pass.  This hopelessness leads to existential angst and Latin despair.  This religious landscape is illustrated in Fig. \ref{lexi.graphic}.

We return for a moment to economics, specifically to Adam Smith's (i.e., The Father of Capitalism) ``invisible hand'' that he talks about in his 1776 book, \textit{The Wealth of Nations} \citep{smith76}.  This invisible hand has several origins.  It is the force of the virtual individual of the collective, the thermal force leading to the stable equilibriums of maximum exploitation, the economic stimulation coming from investment, and the ponderomotive force coming from the monetary value control system which stabilizes the unstable equilibriums of maximum sustainable virtuous economic activity.  The last two are the fiscal and monetary policy that Adam Smith described as ``sensible government intervention to improve and optimize free markets''.  The invisible or virtual nature of these forces are hard for humans to comprehend.  Therefore, humans personify the virtual individual, the source of the force, with mythology and metaphor resulting in ``Gods''.  Economic and social leadership and guidance is given as doctrine (laws and commandments) such as the Golden Rule of collective behavior, which can be best summarized as ``the collective acts as one'' or ``local actions have global consequences''.  This forms a religious philosophy, that is really an economic or social philosophy.

There is a choice of following a Shepard that is a ``God'' or a ``Satan'', that is choose a religion, when a member of the Flock makes a choice of a currency to use.  This is the choice of a Shepard.  Obviously, if it becomes apparent that a Shepard is a ``Satan'', the member of the Flock will choose to change leader, unless the member of the Flock is not free to choose, which is often the case.

The belief in immortality, whether that be through reincarnation or by spending it in ``Heaven'', if one has looked out for the social good ensuring there is a virtuous future, gives a strong motivation of choosing the eskaton of virtuous economic activity over the eskaton of personal greed.

The slogan on the US dollar, ``In God we trust'', is a statement that the nation trusts that the government will be sheparded by the Federal Reserve Bank, the President, and Congress for the good of the Flock, that is nation.  When the statement is made that ``the US dollar is backed by the full faith and credit of the US government'', what is really meant is that the US dollar will be monetarily and fiscally managed to maximize the virtuous economic activity of the US, and that it is backed by that virtuous economic activity.

The foundation of religious philosophies have been:  (1) the existence of a virtual individual of the collective, that is God, (2) the Golden Rule of the collective, and (3) the antithetical nature of resistivity, such as usury (that is interest or riba) and oppression.  These are the foundation of Judaism, Islam, Buddhism, Hinduism, Confucius philosophy, and native African Ubuntu philosophy.  It is especially true of native American and Aboriginal philosophy that have a focus on sustainability, that is the common good and future.

For instance, consider Christian religious philosophy.  There is a belief in the Holy Trinity of Gods.  The Golden Rule of the collective appears in Matthew 7:13 as ``do onto others as you would have them do onto you''.  Prohibitions on usury were many.  The First Council of Nicaea in 325 forbade clergy from engaging in usury.  The Third Lateran Council of Rome in 1179 decreed that persons who accepted interest on loans could receive neither the sacraments nor Christian burial.  The Fifteenth Ecumenical in Vienne France in 1312 declared the belief in the right to usury a heresy, and condemned all secular legislation that allowed it.  The Fifth Lateran Council in 1515 gave a definition of usury ``when, from its use, a thing which produces nothing is applied to the acquiring of gain and profit without any work, any expense or any risk''.  Saint Anselm of Canterbury (1093-1109) led the shift in thought that labeled charging interest the same as theft.  Saint Thomas Aquinas (1225-1274), the leading scholastic theologian of the Catholic Church, argued charging of interest is wrong because it amounts to ``double charging'', charging for both the thing and the use of the thing.  Aquinas said this would be morally wrong in the same way as if one sold a bottle of wine, charged for the bottle of wine, and then charged for the person using the wine to actually drink it.

Despite these numerous prohibitions on usury, the Medici Popes (Leo X and Clement VII) instituted the Chest of Indulgences as the church sponsored method of finance in the 16th century.  Martin Luther, whose primary interest was economics and social welfare \citep{lindberg.16} and who believed that social ethics flowed from Christian philosophy (love), felt that the practice of indulgences was fundamentally destructive to social welfare.  Elimination of the Chest of Indulgences and its replacement with the Community Chest \citep{luther23} was the primary reformation of the Protestant Reformation that Luther led.

\begin{figure}
\noindent\includegraphics[width=12pc]{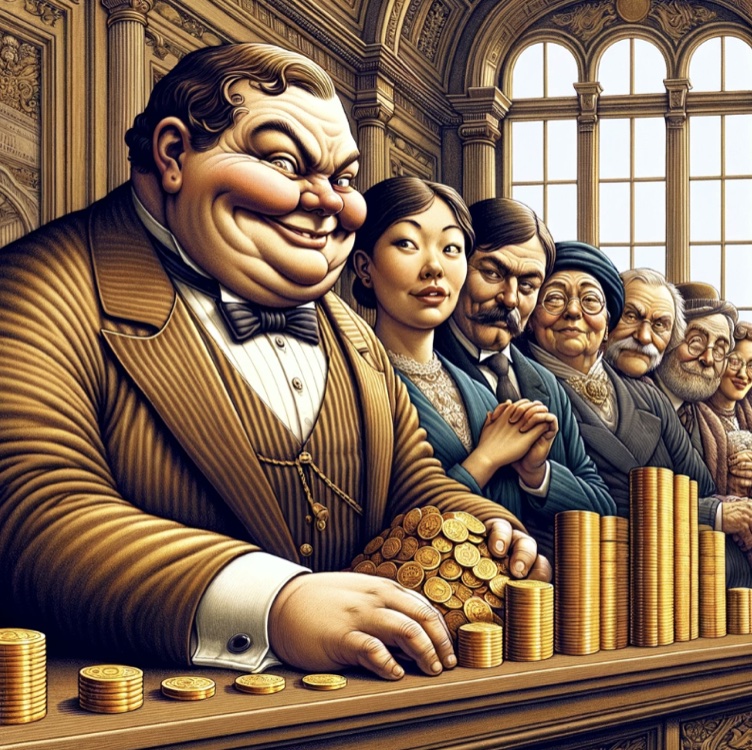}
\caption{\label{homage.fig} Homage being given to Mr. Greed with his pile of money.  Image generated by OpenAI's DALL-E.}
\end{figure}

This can be expressed as follows.  A pile of money is not entitled to homage or periodic indulgence, that is additional money being added to the pile, as shown in Fig. \ref{homage.fig}.  A vendor is not entitled to multiples of what they have invested or will invest in a good or service, in compensation for that good or service.  Generative Artificial Intelligence (genAI) is about better forecasting and control of collective systems, leading to better decisions, and ultimately social (including economic) good — enlightenment, not entitlement.

It is interesting to note that Lorenzo de' Medici \citep{parks05}, known as \emph{Lorenzo il Magnifico} (1449-1492), truly wanted to be a benevolent leader who valued and developed the simple beauty of society, art and science.  This was the Renaissance that he inspired.  It was unfortunate that he had to contend with the resistance, that is the evil, of the Pazzi and Savonarola.  The ultimate irony is that his son Giovanni became Leo X, and his nephew Giulio became Clement VII -- the Popes that established the Chest of Indulgences and inspired the Protestant Reformation of Martin Luther.  The Medici operated the largest bank for most of the 15th century out of Florence, based on debt financing, but making most of its income from currency exchange futures associated with international trade.  It fell victim to the fractional reserve banking risk, caught stealing from the public dowry fund for needed reserves, and went out of business in 1494, shortly after the death of Lorenzo, being founded in 1397 by Lorenzo's great grandfather, Giovanni di Bicci.

\section{``Stargate'' and ``Mr. Robot'' inspiration}
\label{inspiration.sec}
Like Newton being inspired by watching an apple fall from a tree, there were two sources of inspiration for this work.  The first was the movie and television show ``Stargate''.  The analogy is as follows.  The territories leading to the mountain passes and the valley centers can be identified as universes.  The mountain passes or x-points are the beings called the Asgard, are points at which small changes can move one from one universe to another.  The Asgard are at the Stargates between universes.  If the system is in the universe of an Asgard, that system will, via thermal forces, move to the Stargate. The valley centers or o-points we will call the Goa'uld.  If the system moves from the universe of an Asgard, at the Stargate, to the universe of a Goa'uld, that system will, via thermal forces, move to the Goa'uld.  The Goa'uld is a long distance from the Stargate, and it will take a large amount of energy to reach the Stargate.

\begin{figure}
\centering
\includegraphics[width=20pc]{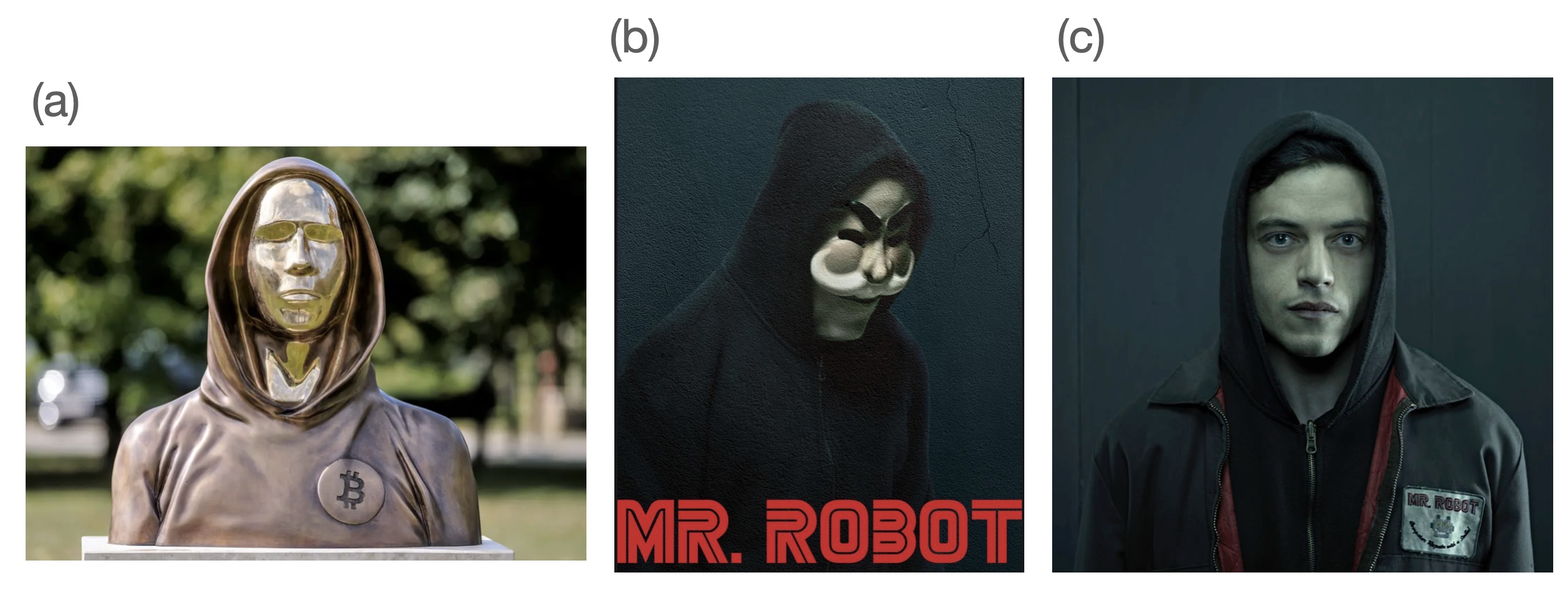}
\caption{\label{fig:crypto_culture}The similarity between the mythical founder of the crypto culture  and the characters of the TV series ``Mr.\ Robot'':  (a) the statue in Budapest of the mythical founder of the crypto culture, Satoshi, (b) one of the members of \texttt{fsociety}, (c) Rami Malek as the lead character, Elliot Alderson, of ``Mr.\ Robot''.}
\end{figure}

There is another aspect of electronic currency that leads to the second inspiration of this work.  It is the crypto culture \citep{roberts20}.  The origins of crypto are those of a religion or cult.  The cult leader is a mythical leader called Satoshi.  Governments and the global finance industry are viewed by this cult as inherently evil, manipulating international finance to their interest and to the detriment of society.  Crypto currency was created to be a currency beyond the control of governments and the global finance industry.  This is exemplified by the USA Network television series, ``Mr.\ Robot'', where a group of computer hackers endeavor to destroy E-corp (an obvious reference to Evil Corp or Enron, the logo for E-corp is the crooked E of Enron).  ``Mr.\ Robot'' is the second inspiration for this work.  The members of \texttt{fsociety}, the group of hackers, even wear the hoodies pulled over their heads and the white masks that appear on a statue of Satoshi in Budapest, Hungary (see Fig. \ref{fig:crypto_culture}).  This is closely related to the Robinhood investor movement behind the Game Stop play, the subject of the MSNBC documentary ``Diamond Hands: The legend of WallStreetBets''.  This was a crowd of common people that banded together to form a virtual consortium to out maneuver the global finance industry.

\section{Discussion and Conclusions}
\label{conclusions.sec}
In conclusion, what has been presented is a new economic paradigm for economics, banking and finance, as well as a new paradigm for strategic and operational business decision-making.  These new paradigms were inspired by ``Stargate'' and ``Mr. Robot'', after some initial irrational crypto exuberance.  At the kernel is the Golden Rule of collective behavior -- ``the collective acts as one'', the virtual individual.  It is both the economic objective, and at the foundation of the innovative genAI based collective system control -- fiscal (investment), monetary, capital planning, and operational.

We realized that current theory for microeconomics, banking and finance was based on an invalid local assumption (in terms of the business graph and time) that leads to a greedy maximization of economic exploitation, and a constrained global minimization of virtuous economic activity.  We needed to return to the fundamentals of macroeconomics, and develop a new theory of microeconomics which brings the global perspective to the local economy.  It is based on a local maximization of $m_e S_0 R_0$ rather than NPV.  There is a coordinated monetary and fiscal policy that is matched to the structure (topology) of the economy.  The proper investments are made, locally, that maximize the local virtuous economic activity.  The electronic currency supply is then adjusted to financially support this investment.  The businesses are operated using the principles of decision analysis which identify the business options (that is alternatives), then evaluates the options according to the metric of long-term virtuous economic activity (that is $m_e S_0 R_0$), not short-term greedy economic exploitation (that is NPV).

This is \textbf{a new unified economic theory} \citep{glinsky.24a}.  We mean by ``new unified economic theory'' is that there is a unification of the monetary, that is how the money supply is determined and controlled, with the fiscal investment policy, that is how investments are made in the economy.  Traditionally, the monetary policy has been governed by Monetarism \citep{fisher.11,friedman.59,friedman.62,desai.81} that has been uncoordinated with the fiscal policy. With Monetarism, you have a central bank that is controlling the money supply via a discount or prime rate at which the central bank loans out money.  This is Autocratic Capitalism with a central profit or resistivity or interest based monetary policy and control. Traditionally things have been handled from the fiscal investment side with the concepts of Keynesianism \citep{keynes.36}.  That is where you borrow and spend in order to be able to make the investment by the central government.  Since you are borrowing, there is an interest rate associated with the borrowing by the issuing of something like T-bills by the US Government or GILTs by the British Government.  This is an autocratic.  There is a central capitalistic government.  This means the government is profit based with this interest based fiscal policy and control.  From the firm's perspective, the optimization is made in terms of profit.  But, these two things, the monetary and fiscal policy, end up being uncoordinated.  

Then there are the separate macroeconomics, the way that the overall economy is controlled based off of the Fundamental Equation of Monetary Policy;  and microeconomics, the firm perspective which maximizes DCF under the constraint that the NPV of the opportunity needs to be greater than zero.  

What we have done is to bring both sets together -- Monetarism and Keynesianism, and macroeconomics and microeconomics.  In fact, we examined the firm like an economic collective comprised of that firm and its suppliers, its supplier’s suppliers and so on.  We then approached the overall economic system as a multiscale web, mathematically a graph, of economic collectives.  This resulted in a distributed constrained optimization at each one of the nodes of this multiscale web, each one of the the corner points, where $m_e S_0 R_0$ is maximized, under the constraint that there is no inflation.  This is equivalent to a revenue maximization or a GDP maximization of the sub-economy or a maximization of the money supply.  There is not just one central currency, but there will be individual currencies, that is transactional equities, associated with each one of the nodes in the graph of this web of economic collectives, with coordinated Ubuntu (print/invest) Financing.  The Ubuntu Financing coordinates the monetary and fiscal policy by printing the transactional equity needed to make the investment.  This brings Monetarism together with Keynesianism.  Both the macro economy and the micro firm are approached as a constrained optimization of their respective GDPs.  This brings macroeconomics together with microeconomics.

This new unified economic theory is stable Democratic Socialism, that is African Ubuntu philosophy, which: (1) relaxes the capitalistic diffusive approximation made for control so that there is coordinated, yet distributed, monetary and fiscal policy made to maximize GDP, not DCF, and (2) replaces viscous forecasting and control with genAI forecasting and control.  This replaces stable Autocratic Capitalism, that is fascist philosophy, which has uncoordinated viscosity-based centralized Monetarism (for monetary policy) and Keynesianism (for fiscal policy).

There is no separation of the social (that is economic) and the religious philosophy of the State.  The State has a philosophy.  The best philosophy is Ubuntu (that is, Democratic Socialism), not Fascism (that is, Autocratic Capitalism).  Society has searched for a metaphor for the Puppet Master or Wizard of Oz, that is the virtual individual exerting that ``invisible hand'' or force on members of society.  The metaphor or mythology becomes a God.  This person does not exist.  It is the natural coordination of all members of the society, through their interactions, that coordinates their motions as they follow the easiest path (the geodesics) determined by the ``invisible hand'', that is force, created by their coordinated motion.  This is the simple beauty.  The symmetry of the coordination creating the force, that creates the coordination.  Fascism comes about when the society identifies a real person with this virtual individual -- a false God like the Pope, dictator, or monarch.

Another way of understanding what we have done is to recognize that previous economic theory has been built from advanced propositions, not from theorems proved from basic principles, that is not built up from a foundation.  These propositions have been either neoclassical capitalism (Keynesianism for fiscal demand-side policy, and Monetarism for monetary supply-side policy) relying on resistive control, or socialism (Marxian Economics, Marxism, Modern Monetary Theory, Complexity Economics, and the Economics of Mutuality) with no method of control.  Note that resistivity controls investment, that is allocates capital, by commercial banks loaning money with interest, giving a cost to capital.  Resistivity controls the money supply by the central bank loaning money with interest to commercial banks.  We have built up a new unified economic theory from a foundation of basic principles, without resorting to resistivity to control the economic collective as neoclassical economics does.  Instead, we have used genAI to control the economic collective, yielding a well controlled and high performing socialism.  In the case that resistive control is used, a low performing neoclassical capitalism is the result.

It is instructive to examine the two inequalities that are the constraints on the optimizations.  For existing microeconomics the constraint is
\begin{equation}
\label{npv.constraint.eqn}
\begin{split}
    \text{NPV} &= \int{\text{e}^{-\nu t}R(t) \, dt} - \int{\text{e}^{-\nu t}E(t) \, dt} \\
    &= \left( \frac{T_I}{m_e S_0}\right) \left(\frac{R_T-E_T}{R_T} \right) \left( \frac{\text{e}^{-\nu T_I}}{\nu T_0} \right) \, R_{mSR} - I \\
    &= \text{DCF} - I \ge 0,
\end{split}
\end{equation}
where $(R_T-E_T)/R_T$ is the profit margin,
\begin{equation*}
    R_T \equiv \int_0^{T_0}{R(t) \, dt} = T_0 \, R_0,
\end{equation*}
\begin{equation*}
    E_T \equiv \int_0^{T_0}{E(t) \, dt} = T_0 \, E_0,
\end{equation*}
\begin{equation*}
    I \equiv \int_0^{T_I}{I(t) \, dt} = T_I \, I_0,
\end{equation*}
and
\begin{equation*}
    R_{mSR} \equiv \left( \frac{m_e S_0}{T_I} \right) \, R_T = m_e S_0 R_0 \, \left( \frac{T_0}{T_I} \right).
\end{equation*}
Note that this constraint is that the short-term return less all of the short-term expenses is greater than zero.  This is in contrast to the proposed constraint that brings macroeconomics to microeconomics
\begin{equation}
\begin{split}
    \frac{T_0 \Delta M}{T_I} &= \frac{m_e S_0}{T_I} \int{ R(t) \, dt} - I \\
    &= \left( \frac{m_e S_0}{T_I} \right) \, R_T - I \\
    &= R_{mSR} - I \ge 0.
\end{split}
\end{equation}
This constraint is that all of the returns multiplied by both the network and the savings multipliers ($m_e S_0 / T_I \sim 10$) less some of the short-term costs (the ones not covered by short-term returns) be greater than zero.  The constraint $\Delta M>0$ is that the increase in the demand for the electronic currency be greater than zero.  Just so that it is greater than zero, there will be an increase in the value of the currency and no inflation.  The difference between the two constraints is stark -- short-term returns less all short-term costs versus all of the returns times $m_e S_0 / T_I$ less some of the short-term costs being greater than zero.  Much more investment is justified by the $\Delta M>0$ inequality than by the $\text{NPV}>0$ inequality.  Looking at the three multipliers in the second line of the expression for NPV in Eq.~\eqref{npv.constraint.eqn} is very instructive.  The term for the DCF, or the upper limit on investment, is the total return $R_{mSR}$ (the upper limit on investment in our theory) times three factors.  The first factor is a small factor given by the inverse of the network and savings multipliers $T_I / m_e S_0 \sim 0.1$.  Remember that these two multipliers come from the fact that an increase of the yearly revenue of a member of the sub-economy is multiplied by both the network effect and the savings effect.  The second factor is a term that favors projects with a large profit margin.  The third factor is a function that favors projects with a short life time $T_0 \lesssim 1/\nu$, and quick payback $T_I \lesssim 1/\nu$.  These are all false incentives, with the first term a false reduction in the potential investment because the network and savings multipliers are not recognized by NPV.  

The combined effect, a focus on immediate profit, causes investment to be preferentially given to firms that are currently generating a profit.  These are the firms that need investment the least.  Whether it be because a firm is in a startup phase or because a firm is in a business cycle low, that is the time they need investment the most, but qualify for the least investment.  For example, this results in banks egregiously having loan covenants that call a line of credit of a firm who is experiencing a business cycle low.  It also results in firms required to have significant revenue when they IPO.

The main difference between the objective of $m_e S_0 R_0$ and NPV is a belief in the future. The use of NPV assumes that instead of investing in four companies with a 25\% chance of success so that one of these companies will likely succeed and will return the reward of increased self-sustaining economic activity, that all of the companies will succeed but there is a 75\% chance that the world will end before they return the investment.  One has no belief in a future nor an ``auctioneer at the end of time'' nor an eskaton of religious philosophy.  One has no religious faith.  Another difference between the objective of $m_e S_0 R_0$ and NPV is a belief that what is good for others is always good for oneself.  The use of NPV is a belief that what is good for others is always bad for oneself.

The introduction of viscosity into the economic systems is motivated by several things.  The first is the simplicity and resulting practical solutions of the theory.  There are both analytic solutions and numerical viscosity solutions resulting from the regularization (removal) of singularities in the theory by the viscosity.  Second, it is much easier to control systems with viscosity.  Conservative systems are like a tube of tooth paste.  When they are squeezed in one direction they squirt out in another direction.  It is like herding cats.  It is much easier to herd the cats if weights are tied around the necks of the cats.  Finally, exerting a viscous force on the system is a convenient way of destructively measuring the system.  This takes many forms as was discussed in Sec.~\ref{quantum.sec}.  It ranges from the US Federal Reserve Bank demanding interest payment on their loans of newly printed currency to measure the money supply, to banks demanding interest on loans to measure the free cash flow of borrowers, to bank depositors demanding interest payments on their deposits to measure the reserves of the banks, to some crypto currencies demanding proof of work to measure the supply of the crypto currency, to other crypto currencies demanding payments be made for proof of stake to measure the supply of those crypto currencies, to bond holders demanding coupon payments to measure the free cash flow of the issuers of bonds, and to stock holders demanding dividends and stock buybacks to measure the free cash flow of the firms who issued the stock.  These destructive measurements are then used to feedback control the economic systems.

The equity of the business is risk tranched into two securities -- the electronic currency (that is, the transactional equity) and the stock (that is, the common equity) of the local economy.  The electronic currency is the low risk (that is, small fluctuations), low return, easy to transact (that is, trade) security, and the stock is the high risk (that is large fluctuations), high return, more difficult to transact security.  The fluctuations of the electronic currency are further reduced, and the equilibrium stabilized by arbitrage trading of the electronic currency.  This arbitrage trading transfers the fluctuations (that is, risk) from the electronic currency to the stock.  The fluctuations in the electronic currency are driven by a lack of liquidity in the electronic currency, that is the counter party to the trade not being there at the time of the trade.  The arbiter steps in and makes the trade, then makes the opposite trade when the counter party arrives.  As the local economy gets larger, the liquidity increases and the required liquid reserve ratio decreases, as one over the square root of the size of the local economy.  The arbiter buys for a small amount less than the true currency price and sells for a small amount more than the true currency price.  The difficulty is knowing the true currency price.  When the arbitrage trading is done according to Eq.~\eqref{stable.pondermotive.eqn}, the arbiter will always buy low and sell high, creating the financial heat pump described in Sec.~\ref{solve.hjb.sec}.  The risk tranching is a consequence of the backing (that is, securitization) of the electronic currency by tangible assets whose ``replacement'' costs can be accounted by transactions (or equivalently the transactional monetary demand given in Eq.~\eqref{ec.value.eqn}), and the backing of the stock by predominately intangible assets whose value can not be accounted for (that is, things like know-how, show-how, value-in-use, goodwill, and brand value).  It is essential that when these two securities are traded by the public, that the public have complete real-time on-line visibility of the finances of the local economy.  This allows the public to correctly value the securities.

While much improvement can be made by using simple estimates of the long-term virtuous economic activity without the exponential discount factor, significantly more improvement can be made by using a more realistic multi-scale model of risk for both the economic evaluations and the arbitrage trading system (that is currency value control system).  These multi-scale models of risk (that is system response) are based on a geometric, topological analysis.  They have much in common with recent developments in AI.  These multi-scale models of risk are the second new paradigm that are super charging the new economic paradigm.  The first is the use of $m_e S_0 R_0$ instead of NPV.

This new paradigm is very different from current systems of monetary and fiscal control.  Sovereign currencies have an uncoordinated political fiscal policy at solely the global level, and a system of monetary control based on debt. This method of control has a phase lag that leads to an exponential instability.  Existing centralized electronic currencies (PayPal, Venmo, Zelle, etc.) and the stable cryptos (Tether, and other stable coins) are 100\% liquid asset backed with no financial leverage or associated fiscal investments.  Proof-of-work cryptos have a single scale control system with no liquid asset reserves, large financial leverage and no fiscal investments.  Proof-of-stake cryptos have an unstable single scale control system with no liquid asset reserves, large financial leverage and no fiscal investments.  Then there are the cryptos that are either highly leveraged derivative securities or infinitely leveraged pyramid schemes dressed in crypto clothing.

Freedom (economic, personal, legal, and intellectual) is essential to the success of this new paradigm.  It will protect the economy from exploitation, and maintain the conservative dynamics of the economy.  While the maximization of virtuous economic activity will modestly increase the wealth of the sovereign of the local economic commune, it will significantly increase the wealth of the members of the commune. ``A rising tide floats all ships.''. This will significantly decrease social stratification and the associated social unrest and totalitarian abuse.  Finally, it also eliminates the ``Research Valley of Death'' and the ``Innovator's Dilemma'' with respect to technology.  The use of the capitalistic (that is short-term profit-based) business metric and constraint leads to a focus on selling existing technology and products at a profit, rather than the development of technology that will increase the virtuous economic activity by enabling better business decisions and giving better options from which to choose, including better technology and products.

The economic system presented in this paper resolves the conundrum that all sovereign and family funds find themselves.  Being caught between the goals of wealth preservation and growth, and the social good of their societies, whether that be their sovereign responsibility or comes from philanthropic tendencies.  This conflict is aggravated by financial counselors who advise them that wealth is best preserved and grown by maximizing returns, that is profits.  As we have demonstrated, short-term profit maximization is counter to the maximization of virtuous economic activity, that is the social good or aesthetic.  For the presented economic system, the growth and preservation of the sovereign is a logical outcome of maximization of the social good of their societies.

This economic system is resilient to economic collapse.  Economic collapse is initiated by a failure of the economic control system, many times leading to run-away inflation. This triggers a panic when the banks get caught in the ``risk squeeze'' that sees them short on capital reserves on deposits.  The resulting run on bank deposits puts a ``liquidity squeeze'' on the bank, leading to immediate bank failure due to a cash flow crisis.  This economic system has immunity to all three phases from: (1) a stable control system with currency value insurance, (2) securitization of loans based on the true risk, (3) different currency loaned than saved, (4) currency is committed to the sub-economy, (5) floating value with no guaranteed exchange, and (6) significant liquid reserves and complete capital reserves.  It simply \textbf{solves the fractional reserve banking problem}, thereby eliminating the humongous risks in our current banking system.

The essence of this paper has been to understand what electronic currency is from the perspective of macroeconomic theory based on the fundamental equation of monetary policy $\text{GDP}=M\,V$.  This leads to the maximization of long-term virtuous economic activity, not the microeconomic maximization of local short-term economic exploitation (that is profits).  This optimisation is matched to the structure of the economy with different currencies for each market and each country, region and locality.  What is good for the United States is not necessarily good for the United Kingdom, New Mexico or Santa Fe.  What is good for the energy industry is not necessarily good for the micro-electronics industry.  As the saying goes ``horses for courses''.  The outcome is benevolent monopolies that are the result of the combination of ``Communism \& Freedom'' or a Free Market Communism.

The bottom line is that electronic currency leveraged financing allows more effective operation of sub-economies and more appropriate levels of fiscal investment.  This is supercharged by approaching monetary and fiscal policy as a question of financial systems control based on innovative genAI methods of system characterization (i.e., risk estimation) and system control.  The economy is operated at the point of maximum sustainable virtuous economic activity, that is at the point of economic and social prosperity.  This is in contrast to conventional resistive control.  In this case, the supply of capital is controlled by the central bank interest rate (raise the rate to reduce the supply and lower the rate to increase the supply), and the distribution of capital is rationed by the cost of capital, the interest rate at which it is loaned.  The economy is operated at the point of maximum short term profit, that is at the point of maximum economic and social exploitation.

The embedding of exploitation into the economy leads to an assumed form of economic activity that goes to zero in a couple business cycles 
\begin{equation*}
    R(q,\tau) = R(q) \, \text{e}^{-\nu_b \tau} \xrightarrow[\nu_b\tau \to \infty]{} 0
\end{equation*}
where $\nu_b=1/\tau_b$ and $\tau_b$ is the length of a business cycle which is about 5 to 10 years.  This is a self fulfilling prophesy.  Assume that there will be an End of the World, and it will come to pass.  The assumption and embedding of dissipation into the economy and optimization of profit and NPV, ensures that businesses are managed to extinction.  Prime examples are petroleum companies that are returning 10's of billions of dollars annually to their shareholders instead of investing in their transition to sustainable energy, and Sears that shut down their catalogue sales and distribution the very same year that Jeff Bezos founded Amazon, bringing catalogue sales to the internet.

The current methods of economic control are resistive.  It is like ``going full gas'' and ``riding the brakes'' to control the speed of a bicycle or automobile or motorcycle.  It is effective control but wasteful of economic energy (revenue) and leads to much less performance as shown in Fig. \ref{full.gas.fig}.  For an economic collective, this effectively stabilizes and controls the economic collective, but constrains the economic collective from reaching the maximum level of sustainable economic activity, and robs it of revenue needed for sustenance and growth.  What is even worse is that when an acceleration (stimulation) of the economy is needed, the acceleration is accomplished by letting up on the brakes with a commitment to ride the brakes even harder in the future.  Eventually, the economy is so damped it will grind to a halt.  It is much better for the currency management firm to sell electronic currency for liquid assets, to increase investment with electronic currency, to decrease repayment of previous investment with electronic currency, and/or to buy assets using electronic currency.  This is pushing on the accelerator, rather than letting up on the brakes.

Given this resistive method of control, when there is inflation that needs to be reduced, that is when a deceleration of the economy is needed, the interest rate $\nu_\text{prime}$ must be increased to slow down the economy and reduce the money supply.  This significantly increases the resistivity in the economy.  The effect on the topology is to significantly lower the economic activity, to decrease the value (total energy) of the economy, to decrease compensation, and to increase unemployment.  If the increase of interest rates is large enough, it can destroy all the local minimums and cause the economy to crash, that is spiral down to very little economic activity -- a deep depression.  All these things are very detrimental to the economy, and avoidable.  It is much better for the currency management firm to buy electronic currency using liquid assets, to decrease investment with electronic currency, to increase repayment of previous investment with electronic currency, and/or to sell assets for electronic currency.  This is like a hybrid car that decelerates by running the electric motors as electric generators and storing the energy for later accelerations.

With this being said, what is proposed by this paper is a more proactive continuous, not reactive, control of the economy to have the maximum sustainable amount of virtuous economic activity, that is sustainable economic and social prosperity.

The answer to the question, ``Why is crypto a religion?'', can now be given.  For when a person or economic entity chooses a crypto (that is electronic currency) to use for their economic activities, they are choosing a ``God'' to shepard them!  The originators of crypto endeavour to save the world from the tyranny of ``Freedom \& Capitalism''.

It is very beautiful how conservative dynamics foliates the renormalization space into two sets of planes.  One set of planes are specified by the conservative sustainable dynamics of the sub-economy determined by the energy $H$ and parameterized by the adjoint or canonically conjugate time $\text{Ad}(H)=\tau$.  The other set of perpendicular planes specify the way that the external economy can influence the sub-economy determined by the adjoint of the energy which is the time $\text{Ad}(H)=\tau$ and parameterized by the adjoint of the adjoint which is energy $\text{Ad}(\text{Ad}(H))=H$, where i's and $\omega$'s have been omitted for clarity.  These are complementary motions (that is $H$ and $\tau$, or $H+\text{i}\omega\tau$), technically canonically conjugate, or a complex analytic continuation.

It is beautiful for this inherent sustainability and structure, and for the cooperative amplification of the network.  This profoundly beautiful structure and the resulting ``min-max dilemma'' of game theory is quite fascinating.  What is the ``min-max dilemma''?  It comes from the fact that your opponents costs in a game are your rewards and vice versa.  The dynamical trajectory is determined by local cost minimization, that is operational efficiency; but those costs are also the rewards that someone wants to maximize.  The conservative system will follow a trajectory of operational efficiency, that is a trajectory of locally minimum costs or action; but that should be a trajectory of maximum energy or total reward.  The problem is that this trajectory is not stable, that is not sustainable without intervention.  The objective in playing a game is to stabilize this trajectory, that is guide the moves of your opponent so that the move that minimizes your opponent's local costs or action, that is the most efficient move, is also the move that will maximize their total reward or energy.  In other words, you want to make your moves so that there is an alignment of your opponents short-term and long-term interests, so that the short-term minimization of costs leads to a long-term maximization of rewards.  This is very counter-intuitive.  The best strategy for winning a game is making moves so that your opponent will be guided to winning the game.  It will not make a difference if your opponent is playing the game locally or with a global strategy.  Their choice of moves will be the same.  A complex dynamical system that is being controlled will always follow the trajectory of locally minimum costs or action.  It has no ability to think strategically, or globally.  This philosophy of control is the reason that good players of a game do not like playing against poor players.  The poor player keeps leading them into playing a poor game.  This is also the reason that poor players like playing against good players.  The good players lead the poor player into playing much better.  This is why a player plays up to or down to the level of their competition.  The control forces for an economic system are applied by the investment and coordinated monetary policy, that is decisions, that inject the economic stimulus (that is economic energy) required to be on the trajectory of maximum economic energy or activity.  The control force is also applied by the ponderomotive arbitrage trading and costs levied for taking undesirable actions, required to sustain (that is stabilize) the trajectory of maximum virtuous economic activity.

This perspective on game theory leads to an even more enlightened perspective on leadership, ethos (the ethical or aesthetic or beautiful versus the unethical or unaesthetic or ugly), politics, education, and intelligence.  Let us start with leadership.  Leadership can take many forms from the conducting of an orchestra, to leading a research effort, to quarterbacking an American football team, to playing a game of chess, to investing in businesses.  Building on the game theory concepts of the previous paragraph, leadership is an issue of system control.  The control is exercised by the leader exerting a force on an opponent or the sub-economy that aligns the direction of minimum local action with the direction of global desired ethos.  The leader has the strategic vision to see the long-term implications of the short-term actions.  The opponent or members of the sub-economy can then just follow the modified direction of minimum local action in a reflective or reactive way.  They may also improve the efficacy of the leader by refining the direction in which they will move (that is the action they will take) by looking for a slightly better small change in direction, given the same ethos.  

The leader has a choice of two ethos: the aesthetic or the unaesthetic.  The choice is between a point of global aesthetic maximum where everyone wins or a point of global aesthetic minimum (the unaesthetic) where only a few people win and everyone else looses.  This is a choice of the future or the religious eskaton.  There are two ways that a strategic player can play a game against a reactionary opponent.  The strategic player can lead his opponent to play the game very poorly leading to quick and decisive defeat of the opponent where the strategic player always wins.  This is a choice of the unaesthetic, an ugly boring game with the misery of the opponent maximized.  The strategic player will feel that the game is boring and unrewarding, too.  In contrast, the strategic player can lead his opponent to playing well resulting in a closely matched interesting game where there is a significant chance that the opponent will win.  This is a choice of the aesthetic, a beautiful engaging game with the enjoyment of both participants, as well as the spectators, maximized.  This is the beautiful game of football of which Brazilian fans speak, the ``Joga Bonito''.  It is not whether one wins or looses, it is how one plays the game.  The best football match ever played was the 2022 World Cup final between Argentina and France.  It does not matter that Argentina won the match.  France could have easily won with a slightly different trajectory of the ball off of the foot of Mbappé late in the match.  What made it the most beautiful match ever was both Messi and Mbappé playing the best football of their career against each other.  They both elevated the play of each other to a perfectly matched level of performance never seen before.

Intelligence is the ability to ``see'' the future, that is understand the long-term consequences of short-term actions.  It is not simply taking the reactionary path of least resistance.  It is knowing and taking the path that will ultimately lead to the desired ethos.  Without intelligence or ethical leadership, the path of least resistance will lead to an unethical, ugly future.  In order to ``see'' the future, an intelligent entity must be educated, that is it must learn.  Without education, an intelligent entity will follow the path of the ignorant, the path of least resistance.  An intelligent entity learns by retrospective studies of the past, that is by studying history.  This is because the curvature of the dynamical manifold does not change, that is history repeats itself.  The intelligent entity also learns by study of logical patterns, that is the theory of knowledge or epistemology.

Now let us examine politics and government, that is social leadership.  What is paramount is all intelligent entities having the choice of leadership or, equivalently, sub-economy.  This can be viewed as freedom of religion, that is a freedom to choose leadership and more importantly the ethos of the leadership.  It is also paramount that all intelligent entities are educated so that they can make an informed choice of leadership.  It is naive to think that there is not an ethos of the uber system, that is the state.  There is a state religion and that state ethos must be ethical or aesthetic.  The state must zealously protect all freedom, and above all the freedom to choose leadership.  The state must also ensure that all intelligent entities are well educated, that is have the ability to choose.  An educated entity is one who never would choose an unethical political leader.  Being unethical should be disqualifying for being a political leader.  It is hallmark of a state in grave political peril to have a major political party that is unethical, that is for the state to have a significant population of the uneducated.  Once in power, that political party will eliminate freedom, and above all the freedom to choose leadership.  That political party will also endeavour to destroy education, especially the teaching of history and epistemology.  The result will be a society dominated by unethical behavior, crime, poverty and ugliness.  If this take over by the unethical does not happen, civil war is inevitable.  At least, it will lead to a terrorist rebellion or a political revolution.  The two ethos can not co-exist at the level of the state.

The choice of an ethos is an all or nothing choice.  If a choice is made of the unaesthetic, it means that everything will be ugly.  There will be low levels of economic activity, unemployment, poverty, crime, power and wealth concentrated in the hands of a few, loss of freedoms, poorly designed ugly products that have short lifetimes and no ability to be repaired, destruction of the environment, an even worse life for our descendants, financial fraud, corruption, unethical behavior, lack of compromise, ignorance (lack of education and critical thinking), intolerance, discrimination, propaganda or dogma, a caste or class system that limits economic mobility, and possibly an end to the earth as we know it with catastrophic climate change and the accompanying loss of the atmosphere.  It is antithetical to think that power and wealth can be concentrated in the hands of a few, and that poverty, crime and revolution will not ensue.  Unfortunately, the consequences of the unaesthetic ethos is not immediate.  It will take time before the House of Medici is reduced to a family of subterranean rats traversing Florence in a labyrinth of tunnels to avoid the poverty and crime.  It will take a generation before the daughter-in-law of Louis XIV is beheaded.  It will take a generation or two before the earth losses its atmosphere.

The very good news was given in Sec.~\ref{interaction.sec}.  This section discussed the interactions of economic sub-systems and sovereign states.  It started by describing how the economic system or global community accommodates multi-objective and multi-party optimization.  This led to the concept of the geometry and topology of the economic system that is captured by the metric of exchange rates, $\Xi_{ij}$ given the discrete set of basis vectors $\left|\xi_i \right>$ forming a low dimensional manifold on a Hilbert space.  A natural evolution first will lead to a separation of topology into two disconnected network graph clusters, one of the aesthetic sub-economies and another of the unaesthetic sub-economies.  Whether by migration of the members of the sub-economy, if there is enough freedom, or by replacement of the leaders of the unaesthetic sub-economies, usually by armed revolt; the economic system will then evolve to one aesthetic cluster.

History has shown us that the unaesthetic has dominated, though.  There is a paradox of tyranny.  The one person that must have an aesthetic ethos is the leader.  The paradox arises from the fact that the leader is the one person who will benefit from an unaesthetic ethos.  Unless a super majority of the community is educated, the leader can propagandize enough of the community to follow the greedy path of exploitation, so that the leader can stay in power.  The leader can also change the rules so that the leader will always win the game.  So, education, that is enlightenment, is fundamental for evolution to an aesthetic society.  The additional fundamental ingredients are electronic currencies and genAI based monetary, investment, and operational control.

The paradox of tyranny has been reinforced by the tyranny of money.  Monetary and financial systems have been operated and controlled based on usury, that is based on an unaesthetic ethos.  The choice of ethos has been imposed on the leader.  While that choice is good for the leader, it is bad for society.  The new economic and financial theory of money, shown in this paper, liberates economies to operate and be controlled based on an aesthetic ethos. An economy so operated and controlled will have far superior performance, aesthetic, beauty, and growth of wealth (that is return) for everyone including the leader, both over the short-term and the long-term.  Businesses will not have to worry about cash flow, will have the proper level of investment, will grow much more quickly, and will produce better products and provide better services that sustain a beautiful environment.  That is exemplified by the example presented in Sec.~\ref{valuation.sec}, shown in Fig. \ref{valuation.fig}, and detailed in App.~\ref{econ.model.app}.  This is why the economic growth rate of 40\% of the New Energy sub-economy, the seven fold increase in profit of Target Energy, and the three fold increase in production of Target Energy seem unbelievable.  This is what happens when a sub-economy is liberated to reach its full potential.

The beautiful structure of conservative dynamics should be embraced and enhanced.  The rewards can be modified by fiscal policy that rewards creation of virtuous and sustainable economic activity.  There also exist natural points of safe harbor from external forces ($\beta^*$ of $H(\beta)$), that can be further fortified by monetary policy.

Instead, society has capitulated to the forces of external exploitation (ugliness) by building that exploitation (resistivity) into the dynamics of the sub-system (via debt financing), destroying the topological beauty of nature.  This is the cause of both the existential angst of Jean-Paul Sartre in ``Les Jeux Sont Faits'' \citep{sartre47}, and the profound acquiescence to fate in Latin American literature of authors such as Gabriel Garcia Márquez in ``Cien Años de Soledad'' \citep{marquez67}.

In summary, this paper presented a new unified economic theory that relaxed the capitalistic diffusive approximation made for control so that there is coordinated, yet distributed, monetary and fiscal (print/grant investment) policy made to maximize GDP, not DCF.  This is a socialist, not a capitalistic metric, where the economic system is treated as a web (multiscale graph) of economic communes.  The local microeconomics is treated like the global macroeconomics.  The viscous forecasting and control is replaced by a Generative Artificial Intelligence (genAI) forecasting and control.  The control is distributed (that is, democratic), not centralized (that is autocratic).  Monetarism argues for only an autocratic (centralized) capitalistic (profit based) monetary control, and Keynesianism argues for only an autocratic capitalistic fiscal (borrow/spend investment) control.  This new theory argues for coordinated democratic socialist monetary and fiscal control.  It does have philosophical connections to African Ubuntu, the Economics of Mutuality, Modern Monetary Theory, Marxian Economics, Complexity Economics, and religious philosophy.  It also makes the connection between genAI and the physics of economic collective systems.  The result is an economic system that maximizes the good of society, not an economic system that maximally exploits society.

We leave you with a final thought.  Our current economic system is one large resistor.  It needs to be transformed into a much more ``conservative system'' (that is Hamiltonian system), an economic superconductor, with significant savings and storage (that is potential energy).  Creation of this capacity for storing economic energy will take significant financing.  It can not and will not be financed with resistive financing (that is debt).  A new method of ``conservative financing'' must be created (that is transactional equity based Ubuntu Financing).  That is what this paper is proposing -- \textbf{true economic conservatism}.

\begin{acknowledgments}
First of all, we would like to thank Ted Frankel and Michael Freedman and for giving a deep appreciation for the primal importance of geometry and topology to the sciences, including economics.  To the Santa Fe Institute and Murray Gell-Mann, for a stay where the richness of complex systems and how to approach them from a scientific perspective was learned.  To the Institut des Hautes Etudes Scientifique and Stephane Mallat, for another stay where the methods of topological analysis were inspired.  To Rev.\ Benjamin Larzelere III for intriguing and useful discussions about religious philosophy.  Last, but not least, to Robert Baird who gave a deep love and understanding of economics.
\end{acknowledgments}

\bibliography{crypto_religion_refs.bib}

%
\onecolumngrid
\appendix
\section{Economic model with electronic currency}
\label{econ.model.app}
What will be presented in this appendix builds on the traditional way that principles of money, banking, and finance are presented in texts \citep{ritter89,welsch81}.  The presentation will be extended to include electronic currency.  A sequence of transactions will be described.  After each set of transactions are described the double entries into the accounting T-accounts (books) for each financial entity are made.  One thing to note is how the assets and liabilities are reversed with respect to New Energy Inc.\ similar to either what happens with the accounting books of a bank or to the books of a parent company when it invests in a subsidiary.  Be sure and note the savings levels in both capital assets and EC\$ of each of the entities and how this ensures that there will never be a cash flow crisis for any entity.  Other things to note are how the electronic currency leverages the cash reserves, how profits from the hydrocarbon operations generate cash that is subsequently leveraged with electronic currency and re-invested, how the New Energy Strategic Hydrocarbon Reserve significantly increases the profit of the hydrocarbon operations, and how the businesses repay the invested electronic currency when or if they can (and how much they can, which can be significantly more than the initial EC\$ investment).

The following treatment of the accountancy deviates from Generally Accepted Accounting Principles (GAAP) in the treatment of intangible assets.  GAAP has evolved to focus on an accurate accounting of cash flows, profit, and loss.  Ultimately, this enables a good estimate of Discounted Cash Flows.  GAAP also focuses on the tangible assets compared to the debt, to understand the Capital Asset Pricing Model (CAPM) of the entity.  Intangible assets, such as Goodwill and Value-in-place, have been seen as distorting, volatile, and not quantitative.  Therefore, they are ignored by GAAP.  Although the changes in intangible assets obfuscate the cash flows, they are essential for accurate estimates of asset value.  For this reason, the intangible assets are included in the following treatment.

\begin{figure}
\noindent\includegraphics[width=30pc]{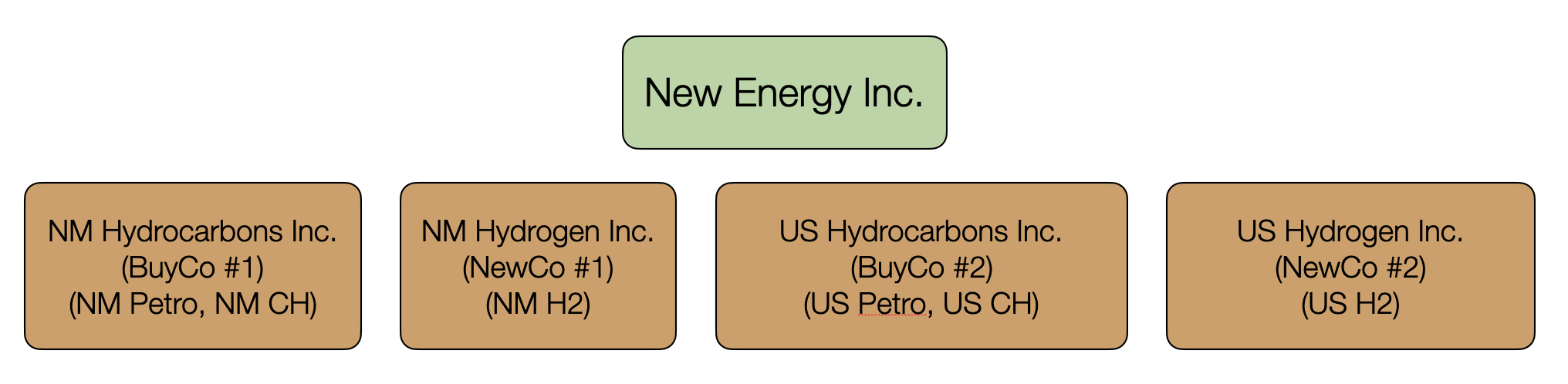}
\caption{\label{org.chart.fig} Chart showing the corporate organization of the New Energy Group of companies.}
\end{figure}

\begin{figure}
\noindent\includegraphics[width=\columnwidth]{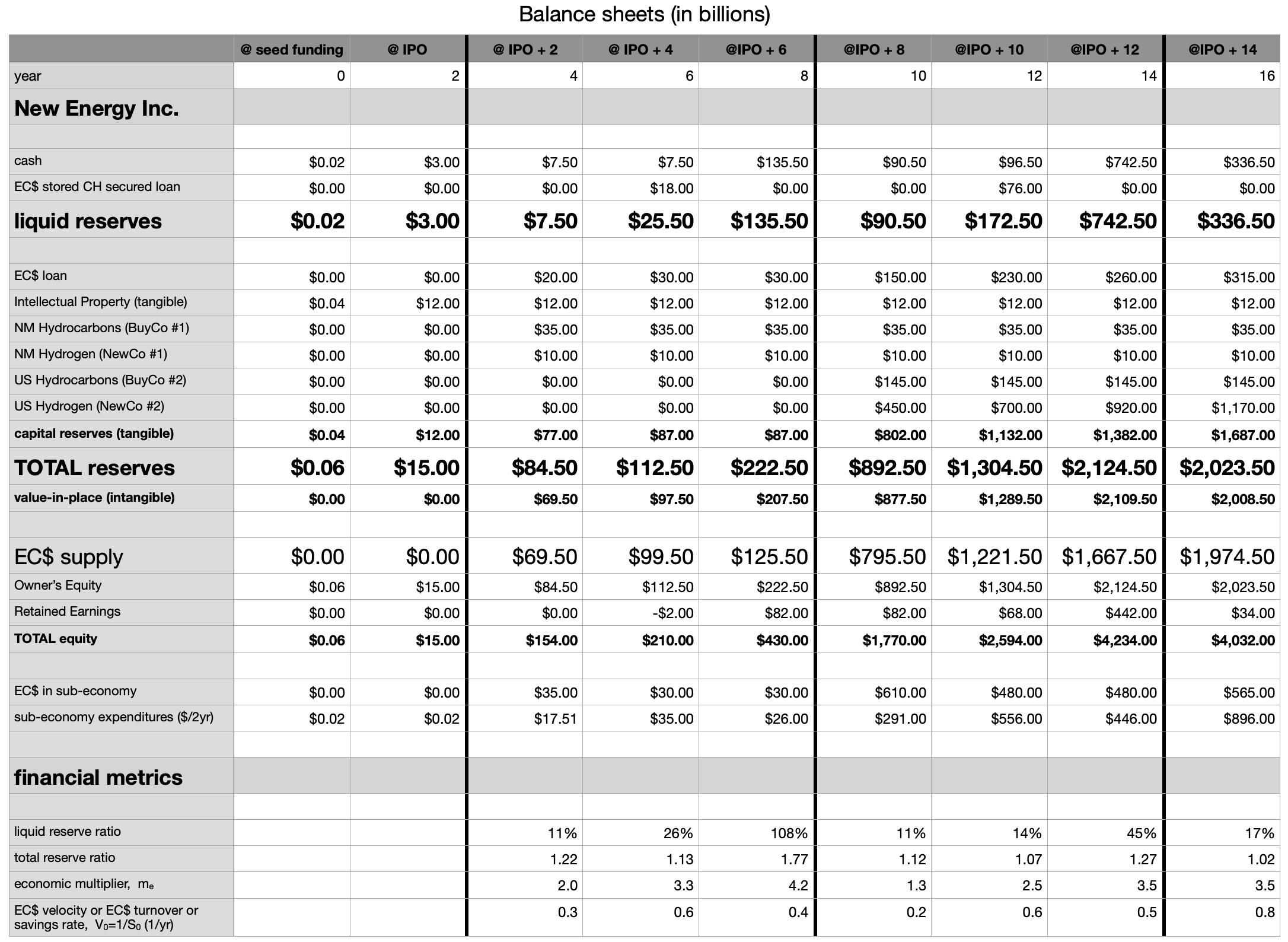}
\caption{\label{balance.sheet.fig} Shown is a summary of the New Energy Inc.\ balance sheets, along with other financial metrics.  The spreadsheet can be found online at this \href{http://tiny.cc/new_books}{Link} (Apple Numbers format) and this \href{http://tiny.cc/new_books_xlsx}{Link} (Microsoft Excel format along with Fig. \ref{valuation.log.fig} and Fig. \ref{valuation.fig}).  The journal entries for the changes can be found in the formulas, along with the mathematical expressions for the other financial metrics.  This spreadsheet also contains the complete accounting T-accounts and formal balance sheets.}
\end{figure}

\begin{figure}
\noindent\includegraphics[width=30pc]{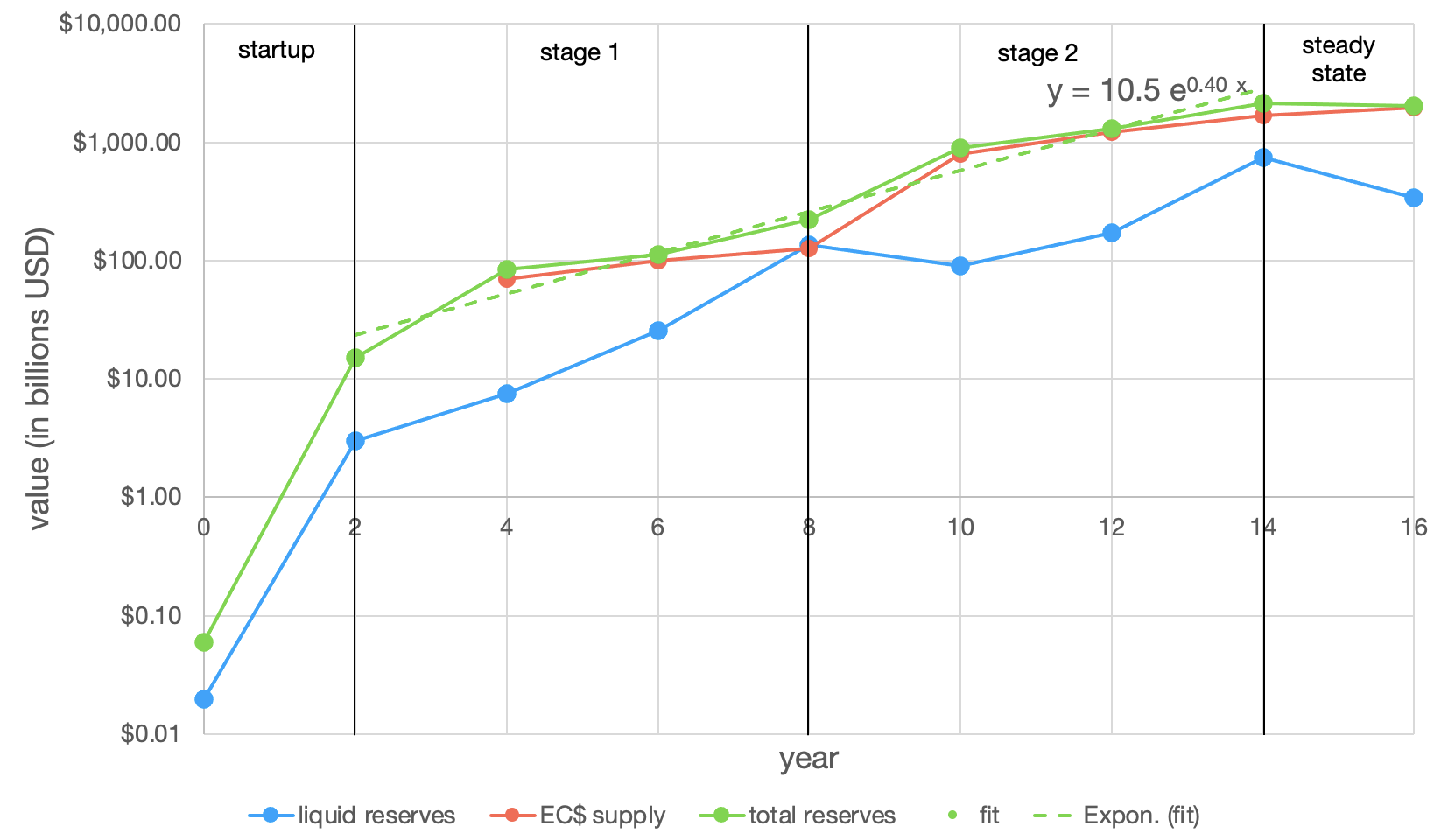}
\caption{\label{valuation.log.fig} Plot of the value and reserves of New Energy Inc.\ versus time.  The time between each event is roughly two years.  The ordinate is the $\text{log}_{10} \text{USD}\$$ value.  Shown are the value of the: (1) liquid reserves as the blue circles, (2) EC\$ supply as red circles, (3) total reserves or the value of the sub-economy or equivalently liquid assets + capital assets (which includes the EC\$ savings of the sub-economy) as the green circles.  The different stages of growth are indicated, and the value of the sub-economy is fit to an exponential.  The fit has a slope (that is growth rate of the sub-economy) of 40\% per year.  The return on investment is therefore about 40\% per year, and the dividend is about 14\% per year when steady state is reached at a value of \$2 trillion USD. Some of the dividend is paid to the stockholders (4\%) and the rest of the dividend (10\%) is paid to society, the ultimate stakeholders in the sub-economy, in the form of $\text{CO}_2$ sequestration.  The value of the initial seed investment grows about 30,000x and the value of the IPO investment grows about 100x.}
\end{figure}

There will be two stages of investment.  The second is about 10x larger than the first.  The cash raised by the IPO is leveraged with electronic currency to fund the NM energy transition.  This is the first stage.  The cash generated by operating the NM energy transition is leveraged with electronic currency to fund the US energy transition.  This is the second stage.  The cash generated by operating the US energy transition is returned to the shareholders as a dividend and the sub-economy reaches steady state.  In practice, not all the cash may be returned to the shareholders, and instead re-invested in the sub-economy to maintain the economic activity as some of the businesses become no longer viable, such as the hydrocarbon business.  It may also be used for socially aesthetic purposes such as $\text{CO}_2$ sequestration.  Additional investment in both New Mexico and the United States may also be needed.  For simplicity in this example, at each stage there is one acquisition (BuyCo) and one new startup (NewCo), as shown in Fig. \ref{org.chart.fig}.  These two stages will be proceeded by a setup stage and followed by a steady state stage.  The balance sheet as a function of time is shown in Fig. \ref{balance.sheet.fig}, along with other important financial metrics, and plotted in Fig. \ref{valuation.log.fig}.

The example is initiated by the seed funding of New Energy Inc.\ with \$20 million USD of cash at a valuation of \$60 million, to develop and mature the technology.

\begin{figure}
\noindent\includegraphics[width=\columnwidth]{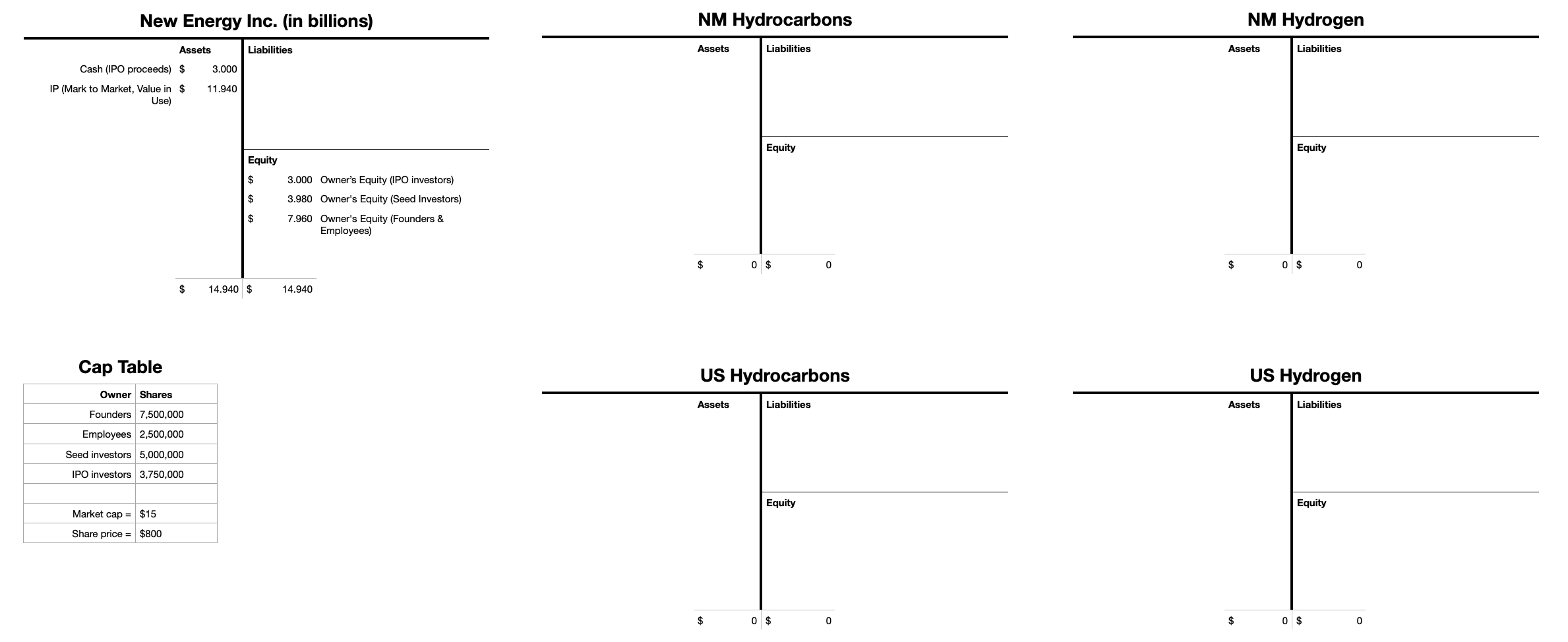}
\caption{\label{balance.0.fig} Shown are the T-accounts for New Energy Inc.\ and its subsidiaries at year 2.  Also shown is the capitalization table.}
\end{figure}

After two years, New Energy Inc.\ has spent \$14 million USD on the development of the technology and preparation for the IPO.  At this time they have a successful IPO, raising \$3 billion USD at a valuation of \$15 billion USD.  This is the start of the first stage.  The T-accounts at this time are shown in Fig. \ref{balance.0.fig}.

\begin{figure}
\noindent\includegraphics[width=\columnwidth]{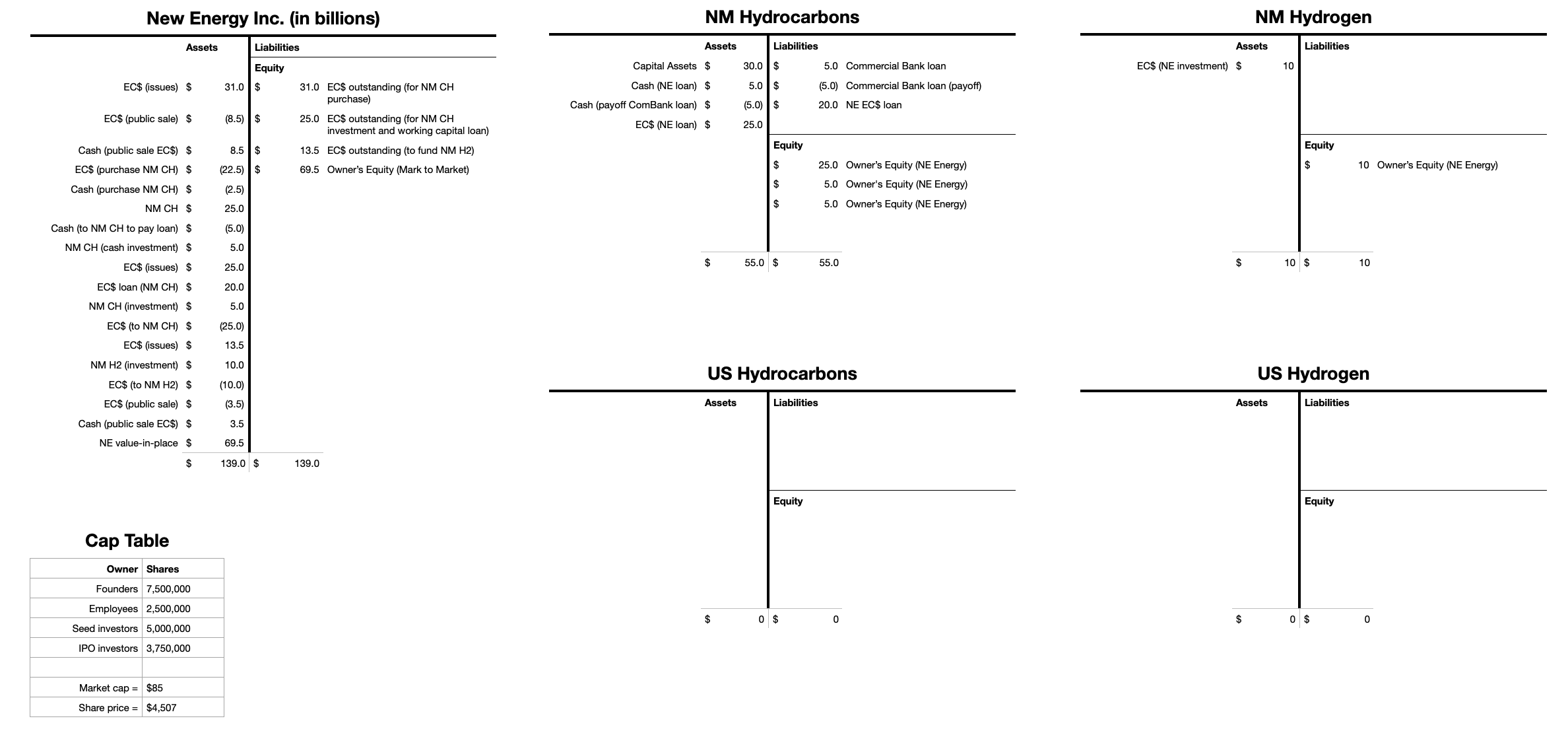}
\caption{\label{balance.1.fig} Shown are the T-accounts for New Energy Inc.\ and its subsidiaries at year 4.  Also shown is the capitalization table.}
\end{figure}

Over the next two years New Energy Inc.\ issues 31 billion EC\$ (throughout this calculation we will assume an constant exchange rate of EC\$ = USD) and sells 8.5 billion of those EC\$s.  It subsequently acquires a company operating hydrocarbon assets in the Delaware Basin with \$5 billion USD in debt for \$25 billion USD, 22.5 billion in EC\$ and \$2.5 billion in cash.  After the acquisition, New Energy Inc.\ sets up the NM Hydrocarbons subsidiary, pays off the \$5 billion USD in debt, and issues another 25 billion in EC\$ and invests that in NM Hydrocarbons to fund the New Energy Strategic Petroleum Reserve and for working liquid assets.  Over this same period it founds a new company, NM Hydrogen to develop Blue Hydrogen production facilities close to NM Hydrocarbons methane production and future $\text{CO}_2$ sequestration facilities.  New Energy Inc.\ issues another 13.5 billion in EC\$, investing 10 billion EC\$ in NM Hydrogen to develop technology, to construct the facilities, and sells 3.5 billion EC\$ for cash reserves on the EC\$.  The T-accounts at this time are shown in Fig. \ref{balance.1.fig}.

\begin{figure}
\noindent\includegraphics[width=\columnwidth]{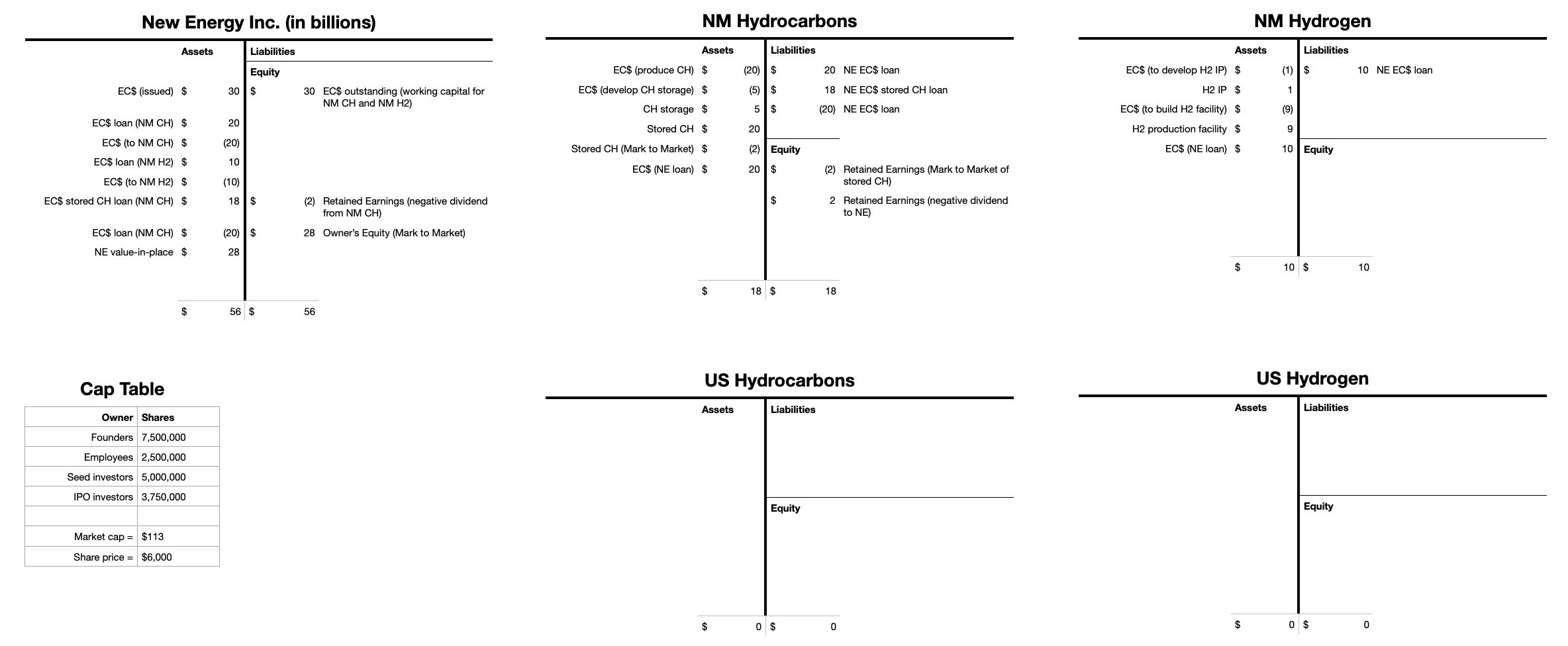}
\caption{\label{balance.2.fig} Shown are the T-accounts for New Energy Inc.\ and its subsidiaries at year 6.  Also shown is the capitalization table.}
\end{figure}

Over the next two years NM Hydrocarbons operates the assets, minimizing methane emissions, using 20 billion EC\$ to produce hydrocarbons with a current market value of \$18 billion USD and storing those hydrocarbons in the New Energy Strategic Petroleum Reserve.  It uses another 5 billion EC\$ to develop the New Energy Strategic Reserve, to develop the $\text{CO}_2$ sequestration facilities, and to build the pipelines to and from the NM Hydrogen facilities.  Meanwhile, NM Hydrogen spends 1 billion EC\$ on technology development and 9 billion EC\$ on the construction of the hydrogen production facility.  During the later half of this period New Energy Inc.\ issues another 30 billion EC\$ investing 20 billion EC\$ in NM Hydrocarbons and 10 billion EC\$ in NM Hydrogen to refresh their working capital.  No further cash reserve is necessary due to the \$18 billion in the New Energy Strategic Petroleum Reserve.  The T-accounts at this time are shown in Fig. \ref{balance.2.fig}.

\begin{figure}
\noindent\includegraphics[width=\columnwidth]{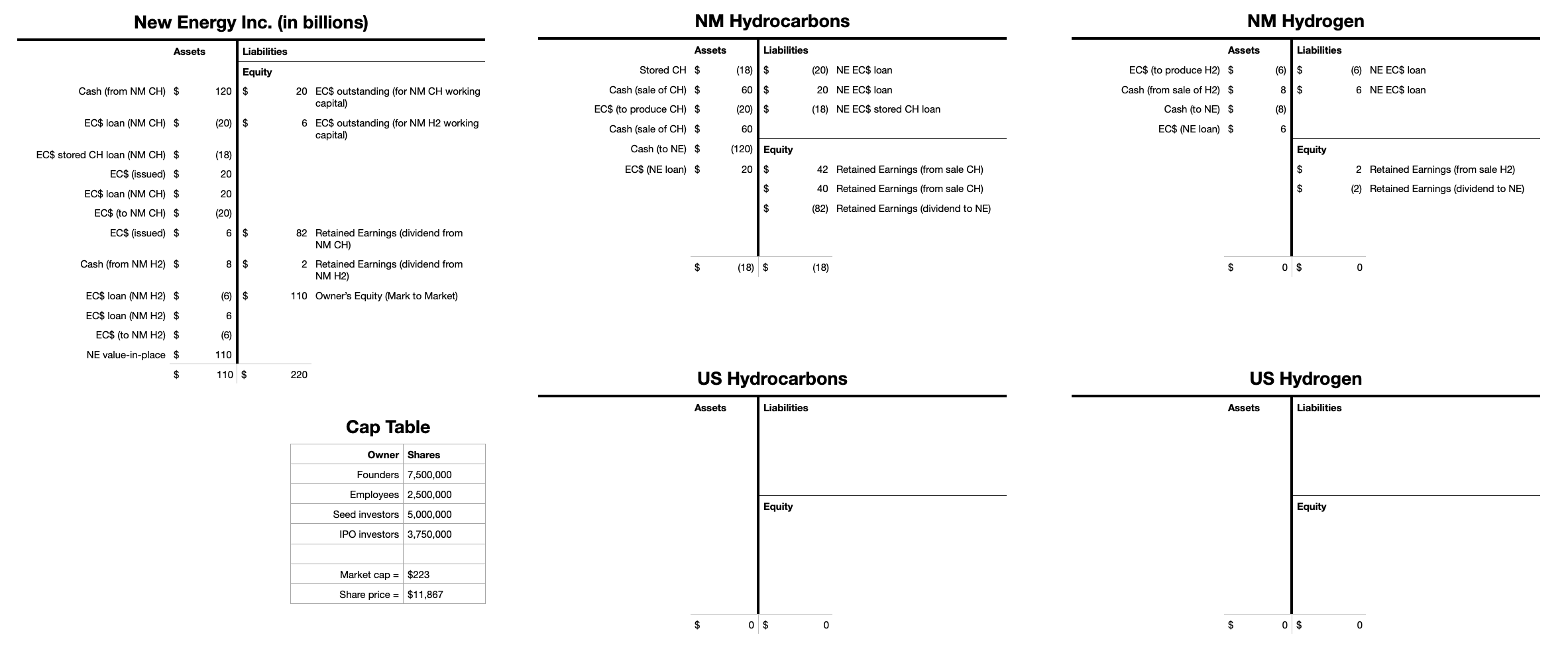}
\caption{\label{balance.3.fig} Shown are the T-accounts for New Energy Inc.\ and its subsidiaries at year 8.  Also shown is the capitalization table.}
\end{figure}

The next two years finds a favorable price environment and NM hydrocarbons sells the New Energy Strategic Reserve for \$60 billion USD, produces additional hydrocarbons for 20 billion EC\$, and sells them for \$60 billion USD.  Subsequently it gives \$120 billion USD to New Energy Inc.\ for 20 billion newly issued EC\$.  Meanwhile NM Hydrogen is now profitable and generates \$8 billion USD after spending 6 billion EC\$.  Subsequently it gives \$8 billion USD to New Energy Inc.\ for 6 billion newly issued EC\$.  Note that New Energy Inc.\ has now received a \$120 billion return on the 40 billion EC\$ ``loan'' that it has given NM Petroleum with an outstanding balance of 20 billion EC\$.  New Energy Inc.\ has also received an \$8 billion USD payment on the 10 billion EC\$ ``loan'' that it has given NM Hydrogen with an outstanding balance of 10 billion EC\$.  New Energy Inc.\ is now ready for the next stage of investment with a cash reserve of \$135.5 billion USD, \$87 billion USD in capital reserves including \$13 billion USD in intellectual property.  It has issued a total of 125.5 EC\$.   The value of New Energy Inc.\ (that is the sub-economy) is \$222.5 billion USD.  The T-accounts at this time are shown in Fig. \ref{balance.3.fig}.

At this point it is very instructive to examine how NM Hydrocarbons is operated, in contrast to how it would be traditionally operated.  We have based NM Hydrocarbons on a typical unconventional hydrocarbon producer we call Target Energy Inc.\ which has a market cap of \$30 billion, debt of \$10 billion, annual expenditures of \$4 billion, and liquid savings of \$200 million.  Their total cost is \$40 per barrel.  The price over the first two years would then be \$36 per barrel.  The price over the second two years is \$120 per barrel.  Note that we have operated NM Hydrocarbons with no debt, increased their annual expenditures by 3x, and increased their savings by 100x compared to Target Energy Inc.  Therefore, the velocity of money for NM Hydrocarbons is 0.5/year compared to 20/year for Target Energy Inc.  This means that Target Energy turns over its savings every 18 days and NM Hydrocarbons turns over its savings every two years.  Why has NM Hydrocarbons operated the asset so differently than Target Energy?  The answer lies in the fact that Target Energy maximizes the DCF and NM Hydrocarbons maximizes the value of the EC\$ and its production.  Target Energy also has to service their debt.  In order to maximize the DCF, Target Energy must minimize their assets and savings while only developing the most efficient petroleum reserves.  In contrast, NM Hydrocarbons needs to operate with increased production (expenditures) and savings (both increase the value of the EC\$).  Another way of looking at this is that macroeconomic evaluation encourages savings in assets (capital, inventory, cash, and EC\$) and spending, while microeconomic evaluation discourages the same.  Increased savings also builds resiliency into the system, by reducing the cash flow risk.  The increased spending leads to more economic activity and growth.  Practically, the increased savings of NM Hydrocarbons is ultimately used to store the hydrocarbons, and some of the savings is also needed to support the increased expenditures.  It is very interesting to note that NM Hydrocarbons has generated \$45 billion in economic activity over the four years compared to \$16 billion for Target Energy (3x more).  What is even more remarkable is that NM Hydrocarbons has generated \$75 billion in profits compared to \$11 billion for Target Energy (7x more).  This is because NM Hydrocarbons is producing more, selling at a higher average price, and not paying interest on its debt.  The pursuit of DCF has put Target Energy into a Nash Equilibrium which is a local minimum (o-point, valley center, stable equilibrium), compared to the ``local maximum'' of NM Hydrocarbons (x-point, saddle point, mountain pass, dynamically stabilized unstable equilibrium).

Furthermore NM Hydrocarbons, with the consistent and increased expenditures, will be having consistent and increased employment.  In addition to the obvious benefit of the increased employment to the social aesthetic, it will be easier to recruit employees if they perceive increased job stability and their stress level will be reduced and their productivity increased, also improving the social aesthetic.  There also will be consistent and increased technology development and deployment.  The consistency will improve the productivity of the technology development and deployment, and the increase in technology level will increase the efficiency and quantity of the production.  NM Hydrocarbons will also be leveraging the \$75 billion in profits over the next years by a factor of 10 and reinvesting in the energy transition and offsetting $\text{CO}_2$ sequestration.  In contrast, Target Energy returns much of their \$11 billion in profits to their investors.

With respect to NM Hydrogen, the current private financial system would not make this investment based on DCF valuation.  NM Hydrogen has generated \$16 billion in economic activity and the associated employment over these four years, along with \$2 billion in profits which will be leveraged by a factor of 10 and reinvested in the energy transition.  This economic activity is critical to address climate change.  All of this, improves the social aesthetic.

\begin{figure}
\noindent\includegraphics[width=\columnwidth]{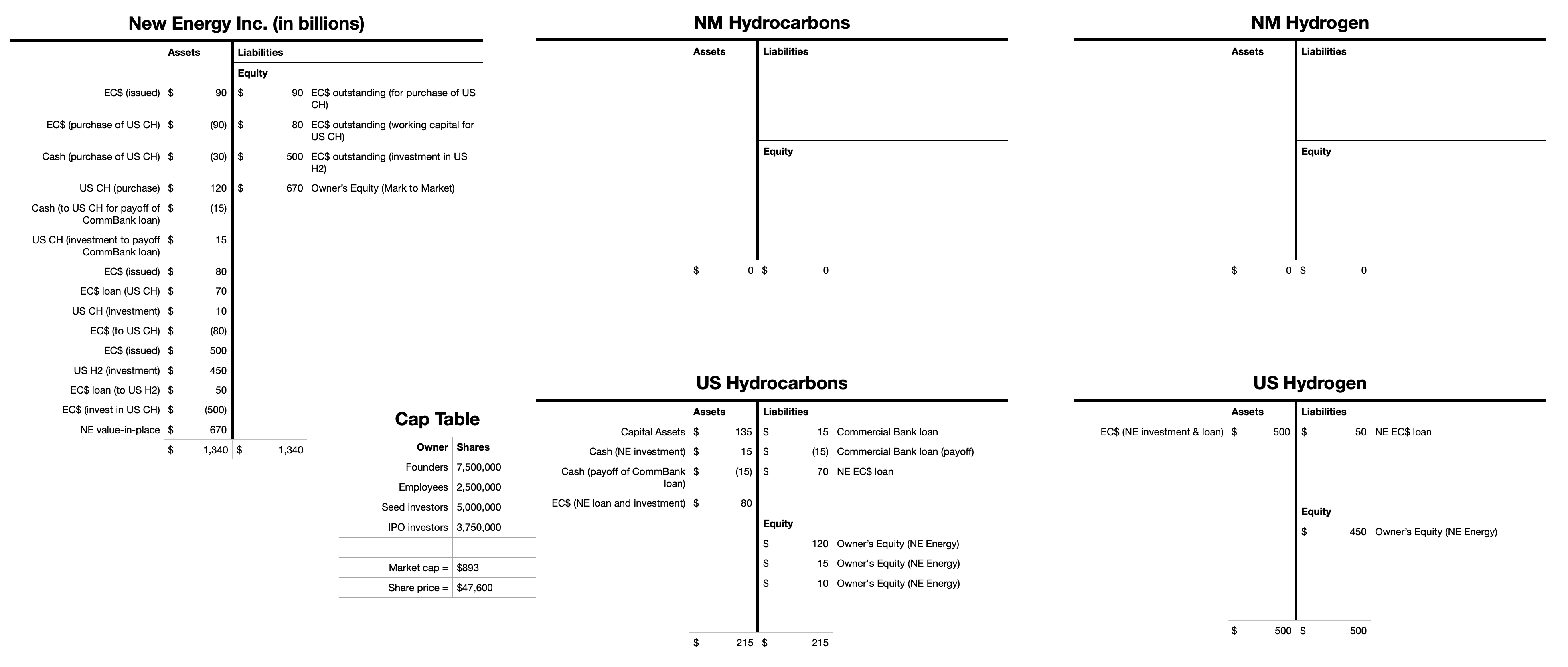}
\caption{\label{balance.4.fig} Shown are the T-accounts for New Energy Inc.\ and its subsidiaries at year 10.  Also shown is the capitalization table.}
\end{figure}

The New Energy sub-economy is now positioned to expand to the US in scope at a scale about 10x larger than NM.  Over the next two years, New Energy Inc.\ acquires a major US based international petroleum company for 90 billion newly issued EC\$ and \$30 billion USD.  It pays off \$15 billion USD in debt, and invests 80 billion newly issued EC\$ in this US Petroleum subsidiary.  Over the same period it founds US Hydrogen and invests 500 billion newly issued EC\$ in it.  The T-accounts at this time are shown in Fig. \ref{balance.4.fig}.

\begin{figure}
\noindent\includegraphics[width=\columnwidth]{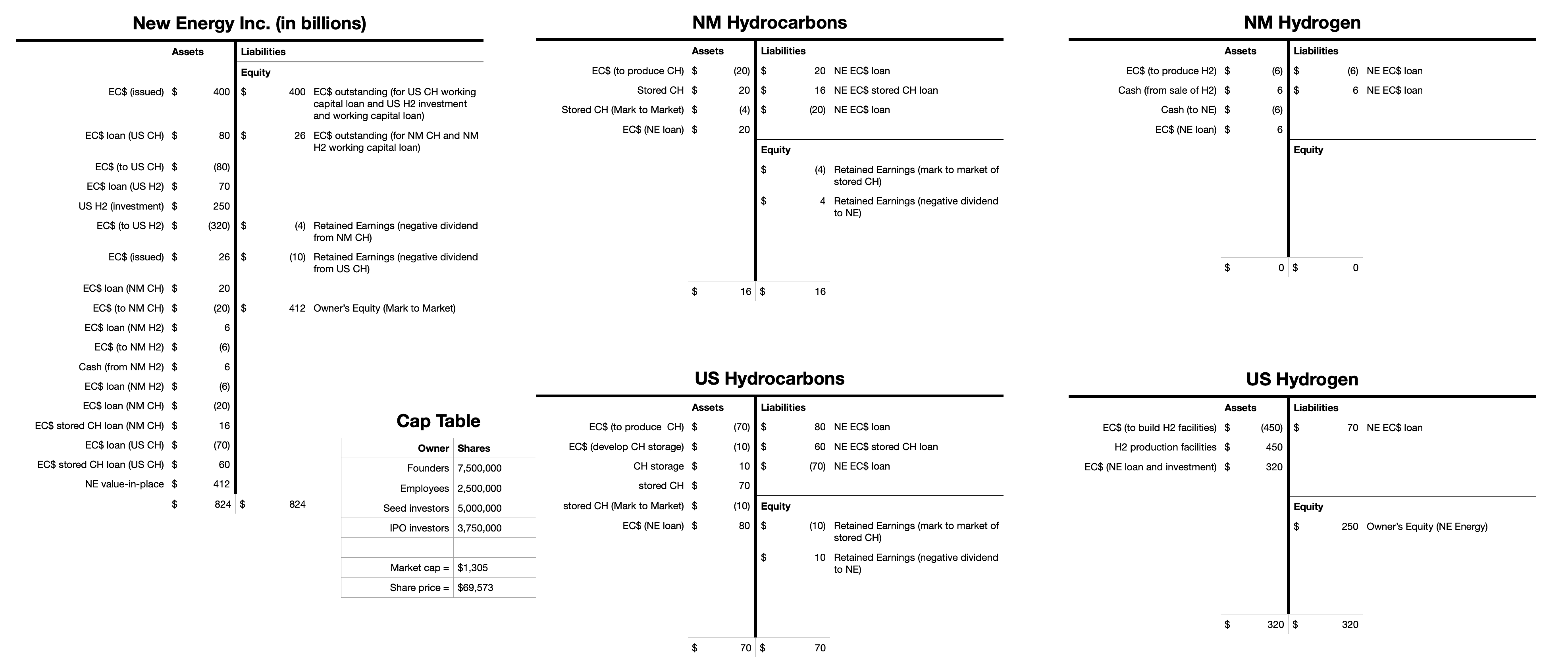}
\caption{\label{balance.5.fig} Shown are the T-accounts for New Energy Inc.\ and its subsidiaries at year 12.  Also shown is the capitalization table.}
\end{figure}

Over the next two years, US Hydrocarbons operates its assets, minimizing methane emissions, using 70 billion EC\$ to produce hydrocarbon with a current market value of \$60 billion USD and storing those hydrocarbons in an expanded New Energy Strategic Petroleum Reserve.  It uses another 10 billion EC\$ to develop the New Energy Strategic Reserve, to develop the $\text{CO}_2$ sequestration facilities, and to build the pipelines to and from the US Hydrogen facilities.  Meanwhile, US Hydrogen spends 450 billion EC\$ on the construction of the hydrogen production facilities.  During the later half of this period New Energy Inc.\ issues another 400 billion EC\$ investing 80 billion EC\$ in US Hydrocarbons and 320 billion EC\$ in US Hydrogen to refresh their working capital.  No further cash reserves are necessary due to the \$60 billion in the New Energy Strategic Petroleum Reserve.  Meanwhile NM Hydrocarbons continues to operate its assets, using 20 billion EC\$ to produce hydrocarbons with a current market value of \$16 billion USD and storing those hydrocarbons in the New Energy Strategic Petroleum Reserve, and NM Hydrogen continues to operate, generating \$6 billion USD after spending 6 billion EC\$.  Subsequently NM Hydrogen gives \$6 billion USD to New Energy Inc.\ for 6 billion newly issued EC\$, and New Energy Inc.\ issues and invests 20 billion EC\$ in NM Hydrocarbons.  The T-accounts at this time are shown in Fig. \ref{balance.5.fig}.

\begin{figure}
\noindent\includegraphics[width=\columnwidth]{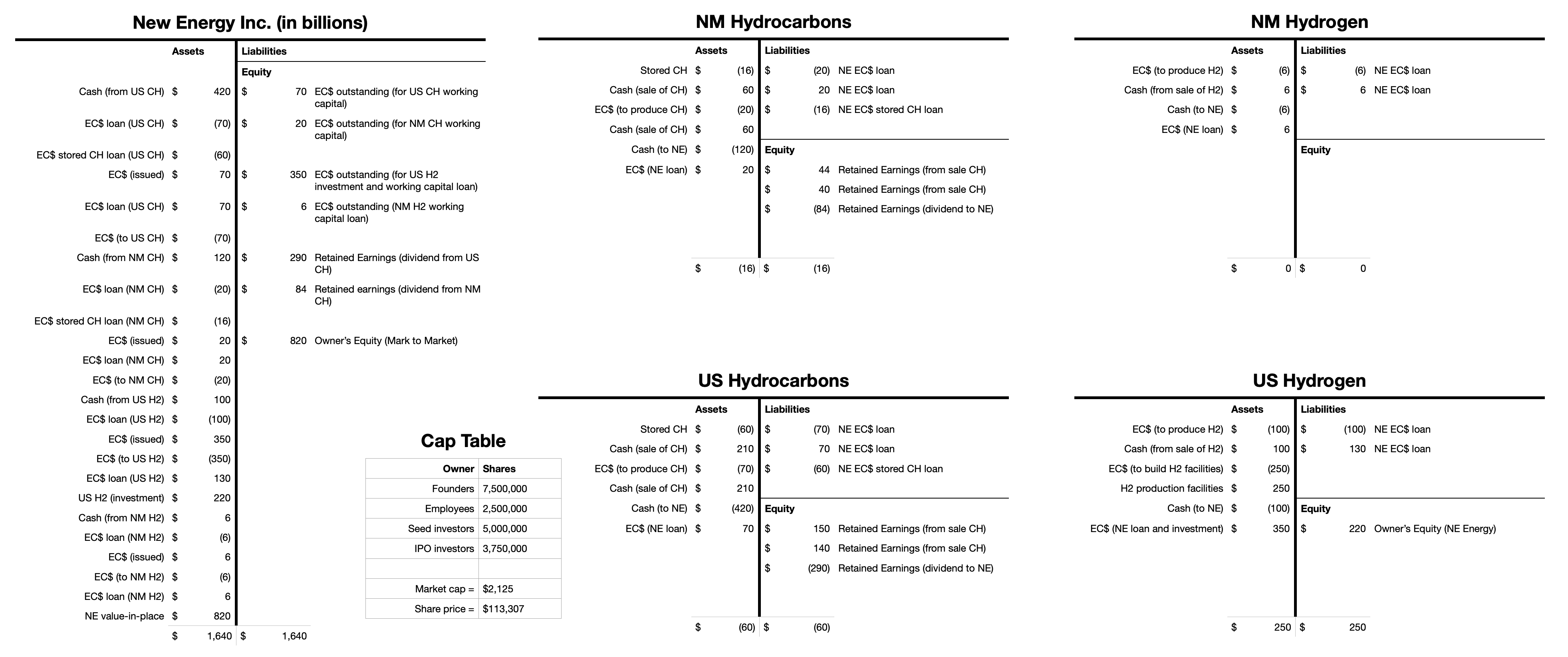}
\caption{\label{balance.6.fig} Shown are the T-accounts for New Energy Inc.\ and its subsidiaries at year 14.  Also shown is the capitalization table.}
\end{figure}

The next two years find the maturation of New Energy Inc.  There is a favorable price environment so that US Hydrocarbons sells its deposits in the New Energy Strategic Petroleum Reserve for \$210 billion USD and produces additional hydrocarbons for 70 billion EC\$ that it sells for another \$210 billion USD.  Subsequently it gives \$420 billion USD to New Energy Inc.\ for 70 billion EC\$.  At the same time, NM Hydrocarbons sells its deposits in the New Energy Strategic Petroleum Reserve for \$60 billion USD and produces additional hydrocarbons for 20 billion EC\$ that it sells for another \$60 billion USD.  Subsequently it gives \$120 billion USD to New Energy Inc.\ for 20 billion EC\$.  Meanwhile US Hydrogen is now operational and generates \$100 billion USD after spending 100 billion EC\$.  It also invests 250 billion EC\$ on new hydrogen production facilities.  Subsequently it give \$100 billion USD to New Energy Inc.\ for 350 billion newly issued EC\$.  Meanwhile NM Hydrogen generates \$6 billion USD after spending 6 billion EC\$.  Subsequently it gives \$6 billion USD to New Energy Inc.\ for 6 billion newly issued EC\$.  The T-accounts at this time are shown in Fig. \ref{balance.6.fig}.

This leaves New Energy with \$742.5 billion USD in cash, \$1.382 trillion USD in capital reserves including \$13 billion in intellectual property.  It has issued a total of \$1.6675 trillion EC\$.  The value of New Energy Inc.\ is now \$2.1245 trillion USD.  It now can return up to \$575 billion USD to its investors and stakeholders as a dividend over the next two years which would be an average of 14\% per year, or continue to invest all or part of that in the transition to a sustainable energy economy.

\begin{figure}
\noindent\includegraphics[width=\columnwidth]{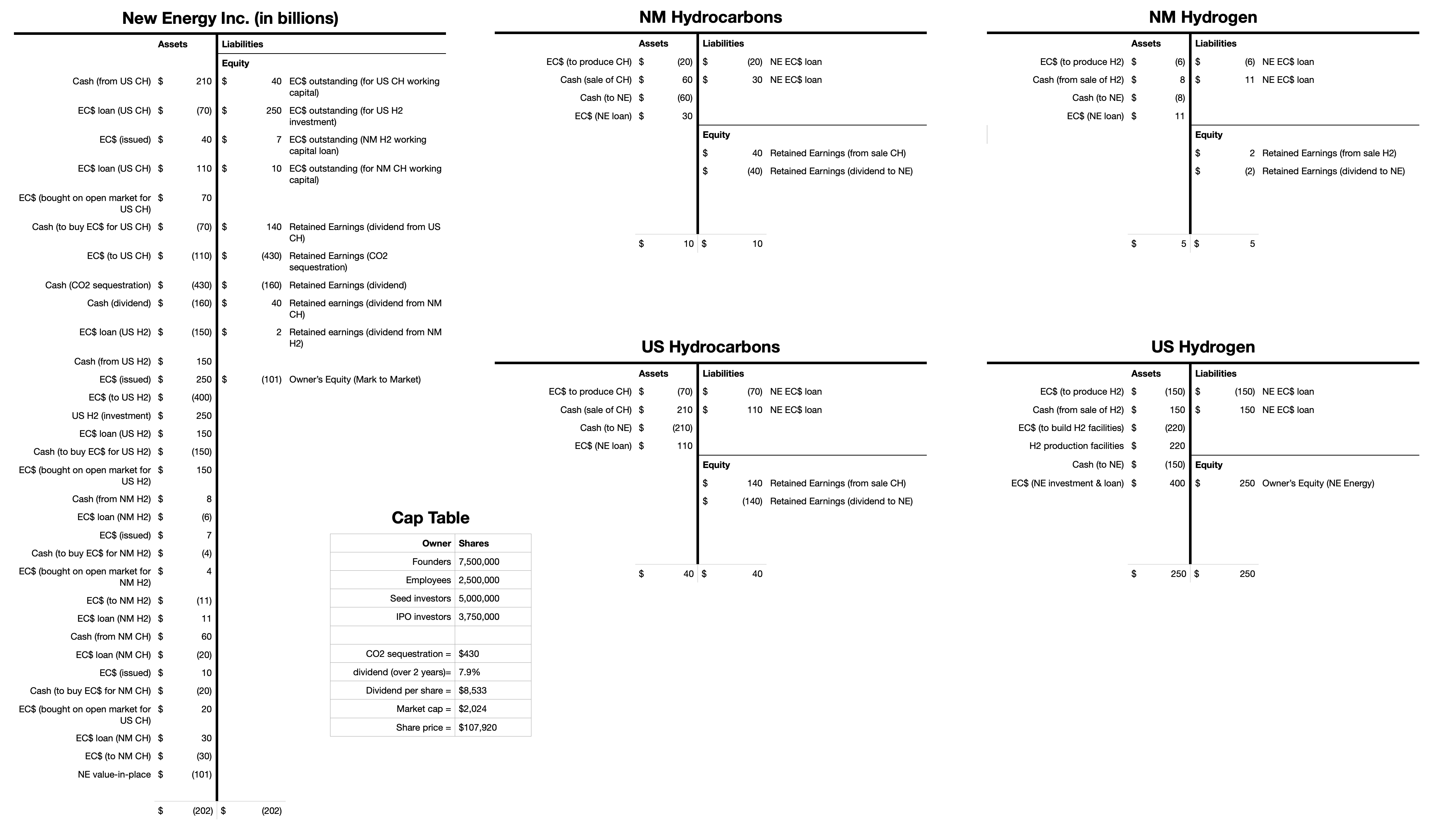}
\caption{\label{balance.7.fig} Shown are the T-accounts for New Energy Inc.\ and its subsidiaries at year 16.  Also shown is the capitalization table.}
\end{figure}

Over the next two years New Energy Inc.\ transitions to steady state.  The price environment remains favorable.  US Hydrocarbons produces hydrocarbons for 70 billion EC\$ that it sells for \$210 billion USD.  Subsequently it gives \$140 billion USD to New Energy Inc., buys 70 billion EC\$ on the open market via New Energy Inc.\ in order to mitigate the inflation of the EC\$ as the sub-economy reaches steady state, and gets 40 billion newly issued EC\$ from New Energy Inc.  New Energy Inc.\ spends \$430 billion USD on $\text{CO}_2$ sequestration and issues a dividend of \$160 billion USD (approximately 4\% per year).  US Hydrogen generates \$150 billion USD after spending 150 billion EC\$, and invests 220 billion EC\$ in new facilities.  It buys 150 billion EC\$ on the open market via New Energy Inc.\ and gets 250 billion newly issued EC\$ from New Energy Inc.   Meanwhile NM Hydrogen generates \$8 billion USD after spending 6 billion EC\$.  It buys 4 billion EC\$ on the open market via New Energy Inc., gives \$4 billion USD to New Energy Inc., and gets 7 billion newly issued EC\$ from New Energy Inc.  NM Hydrocarbons produces hydrocarbons for 20 billion EC\$ that it sells for \$60 billion USD.  It buys 20 billion EC\$ on the open market via New Energy Inc., gives \$40 billion USD to New Energy Inc., and gets 10 billion newly issued EC\$ from New Energy Inc.  The T-accounts at this time are shown in Fig. \ref{balance.7.fig}.

After this transition to steady state,  the resulting return on investment to the stockholders is about 40\% per year, and the dividend is about 14\% per year when steady state is reached at a value of \$2 trillion USD. In this example, some of the dividend is paid to the stockholders (4\%) and the rest of the dividend (10\%) is paid to society, the ultimate stakeholders in the sub-economy, in the form of $\text{CO}_2$ sequestration.

All of these transactions are summarized in Fig. \ref{balance.sheet.fig}, plotted in Fig. \ref{valuation.log.fig} which shows the time evolution of the value and reserves of New Energy Inc., and shown diagrammatically in the ``financial hydrodynamics'' in Fig. \ref{hydrodynamics.fig}.  Additional information regarding the supply of EC\$, the liquid and capital reserves, the reserve ratios, the economic multiplier $m_e$, and the savings rate $1/S_0$ is shown in Fig. \ref{balance.sheet.fig}.  Note that the liquid reserve ratio (cash and petroleum reserves) never falls below 10\%, and the total reserve ratio (capital assets, cash and petroleum reserves) never falls below 100\%.  The primary EC\$ savings of the New Energy Inc.\ subsidiaries in steady state is 565 billion (about 2 years operating expenses on average) compared to the 2 trillion in circulation, giving a reasonable economic multiplier of 3.5.

\end{document}